%% file: newthesis.tex
\documentclass[12pt,oneside,english]{book}
\usepackage{epsfig}
\usepackage{float}
\usepackage[T1]{fontenc}
\usepackage[latin1]{inputenc}
\usepackage{babel}
\usepackage{setspace}



\newpage
\hbox{}

\centerline{\Huge Decoherence and the appearance of classicality} \vspace{20 pt}

\centerline{\Huge in physical phenomena}
 
\bigskip  \bigskip \bigskip \bigskip \bigskip

\centerline{\bf  \Large Daniel Bar}

\bigskip \bigskip \bigskip 

\centerline{\bf \large Department of Physics} 

\bigskip   \bigskip   \bigskip   \bigskip   \bigskip   \bigskip \bigskip

\bigskip

\centerline{\Large Ph. D. Thesis}

\bigskip \bigskip \bigskip \bigskip \bigskip \bigskip  \bigskip 

\bigskip \bigskip \bigskip \bigskip \bigskip

\centerline{\large   Submitted to the Senate of Bar-Illan University, Ramat-Gan, Israel}

\bigskip \bigskip \bigskip 

\centerline{\bf October 2002}

\begin{document}

\newpage
\thispagestyle{empty}
\hbox{}

\bigskip \bigskip \bigskip \bigskip \bigskip \bigskip \bigskip 

 \centerline{\LARGE Several parts of this thesis,} \vspace{28 pt}

 \centerline{\LARGE mainly the quantum ones, were} \vspace{28 pt}

 \centerline{\LARGE worked in consultation  with} \vspace{28 pt}

\centerline{\LARGE Professor  L. P. Horwitz}  \vspace{78 pt}

{\LARGE Department of Physics \hspace{40 pt}  Bar-Ilan University}

\bigskip \bigskip \vspace{6 cm}

\newpage

\newpage
\thispagestyle{empty}
\hbox{}

\bigskip \bigskip \bigskip \bigskip \bigskip  

\centerline{\huge Acknowledgement}

\bigskip \bigskip \vspace{2 cm}

\noindent

\noindent {\Large I wish to thank Professor L. P. Horwitz}.  \vspace{20 pt}
 
 \noindent {\Large  I wish, especially, to thank my brother Moshe without
his help}  \vspace{20 pt}

\noindent  {\Large  not only this thesis but also all my academic studies would not} 
\vspace{20 pt}

\noindent  {\Large  be
possible.} 

\bigskip \bigskip \vspace{4 cm}

\noindent {\large Where Science has progressed the farthest, the mind has but regained
from nature that which the mind has put into nature. }

\bigskip  \bigskip 

{\Large Arthur. S. Eddington}

\hspace{2 cm}  {\large in} 

{\large Space Time and Gravitation}

\newpage 

\tableofcontents{}
BIBLIOGRAPHY . . . . . . . . . . . . . . . . . . . . . . . . . . . . . . . . . .
. . . . . . .78
\include{listoffigures}

\pagestyle{empty}
\section{\large Abstract}

\footnote{Due to certain limitations imposed upon the permitted 
number of the abstract 
lines we introduce here an abridged version of it. The full  abstract, as
represented in the  thesis submitted to the Bar-Ilan University, is shown in 
Appendix $A$.} 

\bigskip \bigskip \bigskip \bigskip \bigskip \bigskip \bigskip 
{\large  It is accepted  that among the ways through which a quantum 
phenomenon decoheres and becomes a classical one is   what is termed 
in the literature the Zeno effect. This effect,  named after the  
ancient Greek philosopher Zeno of Elea (born about 485 B.C),     
 were used in 1977 to analytically predict  
that  an initial 
quantum state may be preserved in time  
 by merely repeating a large number of times,  in a 
finite total time,  the experiment of checking its state. Since then this effect 
has been experimentally validated and has become an established physical fact. 
It has been argued by Simonius  
that the Zeno  effect   must be  related not only  to quantum phenomena 
but also to many macroscopic 
and classical effects.  Thus, since it operates in both quantum and classical
regimes it must cause to a more generalized kind of decoherence than the
restricted one that ``classicalizes'' a quantum phenomenon. We show that 
this generalized decoherence,  
{\it obtained as a result of  dense measurement},  
not only gives rise to new phenomena that are
demonstrated through new responses of the densely
interacted-upon system  but also may physically {\it establish} them.      
   For that matter  
we have found and established the  analogous  {\it space Zeno effect} 
 which  leads to the necessity  
of an ensemble of related observers (systems) for the remarked physical 
validation  of  new  
phenomena. As will be shown in Chapters 3-5  of this work  the new phenomena 
(new responses of the
system) that result from  the space Zeno effect may be of an 
unexpected nature. We use quantum  field theory  in addition to 
the more conventional methods of analysis and  also corroborate our analytical 
findings by numerical simulations.}     
   
\newpage 

 \pagestyle{headings}
 
 \chapter{\label{Chapter1} Introduction}
\section{\label{sec1} Static and dynamic Zeno effects}
We wish to discuss  in this work, as its name implies, the detailed stages 
of the decoherence process, where by this term we mean also the mechanism that
not only physically validates and establishes a new-encountered phenomenon
\cite{Bar0} 
but also, as will be shown, may initially give rise to  it \cite{Bar0}. 
We note that one 
generally 
finds  in the literature (see, Giulini {\it et al} in \cite{Zeno} and
references therein) this term as meaning the process that ``classicalizes'' a
quantum phenomenon so that its former wavy character disappears \cite{Franco}. In this work 
we also take this term  more generally  to mean the process that may first 
give rise to  
a new phenomenon, in a manner to be fully described in this work, and then   
physically establish it.    The involved  process 
includes the stages of first trying   \cite{Bar1},  
especially through mathematical
expressions, 
to explain this phenomenon   and then  of
validating (or refuting) the suggested expression by experiments. 
  The last stage, in which one tests the assumed mathematical 
relation to
see if it conforms to the experimental findings, is the most important one since
by this one may decide   the status of the assumed theory  to be either elevated
to the physical level  or to be refuted. We note that since the advent of
quantum mechanics there is in the literature a long and continuous effort that
tries to clarify and understand the problem of measurement (see, for example,
\cite{Wheeler,London,Rosenfeld,Adriana}). \par We wish  to
describe by various examples  the means through which a  new phenomenon
 is first discerned as such and then decoheres to become an established 
 physical fact. We do this by various
 methods that, although appear to be  different at first sight,
 nevertheless, they  all yield the same  result that what may first initiate  
 and then 
physically establish a new phenomenon (and the theory that explains it) 
 is not only
the experiment one performs in order to test it  but the   multiple 
repetitions of it (or of similar versions of it) in a finite total time 
\cite{Bar1,Bar2} as will be explained. Moreover, we show that the degree of  validity depends
upon the number of these repetitions,  that is, the more
large is this number   in the   total alloted time the more
physically established
 the phenomenon and its proposed theory will be \cite{Bar1}.  \par 
  These  repetitions are regarded by many authors
  \cite{Zeno,Itano,Aharonov,Facchi,Harris} as an important factor in giving rise
  to 
  the remarked restricted decoherence through which a {\it quantum} phenomenon
 becomes  ``classicalized''. This appearance of classicality in the quantum regime is
  termed  the quantum Zeno effect
\cite{Zeno,Itano,Aharonov,Facchi,Harris,Simonius} and it denotes the mechanism 
by which the initial state of some
quantum system is preserved in time by only repeating a large number of times,
in a finite total time, the experiment of checking its state. The last result is
termed the quantum static Zeno effect \cite{Facchi}   to differentiate it from 
 the quantum dynamic Zeno effect \cite{Facchi} which  is not
composed of repeating  the same experiment but of performing a very large
number of different experiments each of them reduces the system to different
 state so that it proceeds through a specific path of
consecutive states. Thus,  if the  different experiments that reduce the 
system to this path  are done in a dense manner  this results in ``realizing''
 this path, 
  from the large number of possible different ones,  as has been demonstrated
  theoretically \cite{Aharonov,Facchi} and experimentally \cite{Itano}. This
  ``realization'' is effected through the probability  to proceed along the
  different possible paths of states which tends to unity  for the one that dense
  measurements were performed along it and to zero for the others. \par 
  It has been
  argued \cite{Simonius} that the Zeno  effect is not restricted only to quantum
  events but it may be found also in many macroscopic and classical phenomena.
  That is, even the classical phenomena are established  as a result of the
  Zeno effect. In other words, this
  process may be the source that causes,  through the remarked repetitions, the
  ``realization'' of many physical phenomena and not only to their
  ``classicalization'' from any former quantum stage they may be in. We show in
  the following that this effect may indeed establish  the
  physical character (and not only the classical properties) of many phenomena. 
  \par 
   We show that since these repetitions are an important factor in constituting 
  the physical aspect of  real phenomena then the reality of the latter 
  do   not depend only on their specific nature but also on this dense
  measurement. That is, we expect to find this Zeno
  effect demonstrated in many  disciplines of physics, as well as  other
  scientific regimes,   as has been argued
  in \cite{Simonius}. This effect has indeed been experimentally found  in diverse phenomena
  \cite{Itano} including chemistry \cite{Harris}.               
     We have shown 
   its possible existence  in both
  quantum \cite{Bar0,Bar5,Bar6,Bar7,Bar8} and classical phenomena 
  \cite{Bar10,Bar11,Bar111} 
  and also in chemical reactions \cite{Bar12}. \par 
  One may argue that all the real phenomena, physical, chemical, biological etc 
  do not seem to obtain their reality from any repetitions at all, so how and in
  what manner these supposed iterations become effective in the remarked
  physical validation ?. That is, our physical laws and phenomena do not appear
  to result from any dense measurement as described here. The answer is that the
  relevant repetitions that may  establish the physical character of the real
  phenomena are not the time repetitions one usually connects
  \cite{Zeno,Itano,Aharonov,Facchi,Harris,Simonius} with the Zeno
  effect but a space version of it. That is, in order to appropriately discuss
  the process of physically establishing the new-encountered phenomena we have
  introduced a new kind of Zeno effect which we call space Zeno effect
  \cite{Bar1,Bar7,Bar8,Bar11,Bar3}. In the last effect, to be fully discussed in the
  following  chapters, the remarked dense repetitions are done in
  space and not in time so that the relevant ensemble is of related observers
  (systems) in space 
  and not of repeated experiments in time. This  effect stands in the basis of
  the remarked physical validation  of  new phenomena as we later show in
  Chapters 3-5   when we discuss the effect of the large ensemble of related
  observers. That is, the repetitions effected in the physically establishing
  process are not the conventional ones of the time Zeno effect
  \cite{Zeno,Itano,Aharonov,Facchi,Simonius} that are performed serially in time
  but those that are done by a large ensemble of observers.  \par 
   Moreover, the number of observers in the
  ensemble does no have to be large in order to accomplish this establishing
  process. This may be seen clearly from  Section 4.3 (see also \cite{Bar7}
  which is partly shown in
  Appendix $C_2$) where we discuss the one-dimensional multibarrier potential of
  finite range which constitutes a quantum example of the space Zeno effect in
  which the barriers in the finite spatial axis represent a one-dimensional 
  version of observers (systems). One may realize
  from
  Figures 2-5 in \cite{Bar7},  which are shown also in Appendix $C_2$, 
   that  for both cases of $e>v$ and $e<v$ and for either a constant or a variable
  length of the finite section,  along which the barriers are arrayed,   it is
  possible to obtain a significant transmission of the passing wave even for a
  30 potential barriers. This may be seen also for the analogous classical
  one-dimensional multitraps in a finite section as seen in the 
  relevant papers \cite{Bar11,Bar111} (see these articles, and especially their 
   graphs in
  Appendices $C_4$-$C_5$). This significant, and unexpected,  transmission of the
  particles (either quantum or classical) through the barriers or traps
  constitutes the new response (phenomenon) that not only comes into being as a 
  result of this multi-measuring process but also may be established by repeating
  these kinds of experiments. In summary, one does not have to take the limit 
  of a very large number of systems (observers) for obtaining  this new response
   and for physically establishing it.       
            \par \section{Time and space Zeno effects}
   The remarked repetitions,  of either the same experiment or along a large
   set of  different ones, that characterize the static and dynamic Zeno 
   effects respectively are
   performed in time. Thus, an important element of these repetitions is, as
   remarked,  that
  they  must be  done in a finite total time $T$ so that both Zeno effects are obtained in
   the limit in which the number $N_t$ of repetitions in $T$ satisfies $N_t\to
   \infty$. It has been shown \cite{Bar1,Bar3,Bar4} that the same effect is obtained
   also when these experiments  are performed on a large number of 
   similar systems
   occupied in a finite region of space $R$  instead of repeating it 
     a large number of times,    in a finite
   total time $T$,  on the same system. The corresponding space Zeno effect
    is obtained \cite{Bar1,Bar3,Bar4} in the limit when the number of systems $N_r$
    ( observers) in $R$ satisfies $N_r \to \infty$.  
    We note that what characterizes both kinds of the Zeno effect is the
    continuous and uninterrupted experimentation either in time or 
    space so there
    is no time (in the total time)  or region (in the total spatial volume) 
     that the
    system is not interacted upon. Thus, this kind of uninterrupted 
    experimentation  in
    which the system is not left to itself causes it to behave differently, even
    unexpectedly as will be shown, especially, in Chapters 3-5. Moreover, 
    as remarked,
    the dense
    measurement condition not only gives rise to  this new behaviour of the
     system but may also physically  establish 
    it.   \par 
      We must note that  we do not 
regard 
each separate experiment as constituting an experiment on its own, only the whole 
array of all these   similar experiments, all confined 
to be done 
in a finite region of space, is considered as an experiment.   
 The  conclusion  obtained  is that in such a 
limit, when the magnitude of each experimental set-up becomes 
very small whereas the total volume (in which all these experiments are 
performed) does not change,   we get actually, as remarked,  a field 
of such probes (see Section 1.4 of this work,  Bixon in \cite{Harris}, and \cite{Davies}).  
  In such cases 
it is meaningless to discuss these fields 
 in terms of the separate  points  as it is meaningless to treat the 
 electromagnetic field pointwise. The same is true 
 also for the time Zeno effect.  That is, 
  it is necessary to look upon 
 the whole array of these elements of measurements,  
 and from such a perspective the Zeno effect is obtained not only theoretically 
 but also experimentally, as has been done for the time Zeno effect 
 by  Itano {\it et al}  \cite{Itano}.   Kofman and Kurizki  in  
\cite{Itano}  show the  existence  of 
  this effect   with regard to the excitation decay of 
 the atom in open cavities and waveguides  using a sequence of pulses on the 
 nanosecond scale;  see also Raisen in  \cite{Itano}  for another way of showing 
experimentally 
 the Zeno effect, this time in quantum tunnelling.     
 \par    
 Thus,   we  may regard the whole process, 
composed of the large 
number of repetitions of the same experiment, as one inseparable process that 
should not be decomposed as we show in the following. 
  This view is related to that adopted, for example, by
  Gell-Mann-Hartle-Griffiths \cite{Hartle,Isham,Griffiths}   
  in their histories formalism (see Section 2.2 of this 
work).  In this formalism  only the 
complete  history is regarded as a physical process and the separate 
parts of it  are not considered to be  reduced physical entities on their own. 
  It has been shown in 
\cite{Bar0}, by discussing in terms of Feynman paths 
\cite{Feynman2,Schulman,Gert} the three processes of 
the EPR paradox \cite{Einstein}, 
 the  
Wheeler's 
delayed 
choice experiment \cite{Wheeler}, and the teleportation phenomenon 
\cite{Bennett},  
that the 
paradox  in these 
 processes arises only from discussing  them  
before they are complete. For example, in the EPR process,  we assume that one particle 
of the two involved must  always have some definite  direction for its spin 
even before we 
measure 
the spin of the second particle \cite{Einstein};  but the EPR experiment is 
complete only after the latter measurement is performed.   Thus,  the Zeno 
 effect 
should   be 
discussed  on the basis of the entire ensemble of repetitions without 
 considering  the 
particular experiment 
that is repeated. This has been shown in \cite{Pati} by using the geometric 
structure of the Fubini-Study metric defined on the projective Hilbert space of 
the quantum system. With the help of this projective geometry a  
quantum Zeno effect has been predicted for many types of  systems even those described by 
nonlinear and nonunitary evolution equations, that is, even the linear Schroedinger 
equation is not a necessary condition.   
\par 
    \section{Single and ensemble of observers} The differences between
    the time Zeno effect and its space analog constitute, 
      as remarked, the 
     important  differences between the case 
when the  remarked dense experiments  are  
done by one observer or by  a large number of them  \cite{Bar1} (where in the 
last case no one
has to repeat his specific experiment). That is, the time Zeno effect
corresponds to the first case and its space analog to the second. In the first
case  one may establish his tested theories   
 {\it for himself}  but,  as
  should be obvious,  this  will not be common to other unrelated observers.
   When, however,   a large ensemble of 
 observers is involved the relevant new 
 phenomenon and its  theory will be physically  established, for all of them, 
  without
 having each one  repeating his experiment so long as they  are {\it related} 
 to 
 each other in the sense that the results of any
 specific experiment done by any one of them are  valid also for all the others. 
 That is, although only one,  from a large number of observers,
   does his specific experiment, nevertheless, the relationship
   among the ensemble members, to be discussed in details in Chapters 3-5 of this
   work,  ensures that any  other
   observer that repeat the same experiment under the same conditions obtains
   the same results.  We show that the more large is
   the ensemble of related observers that perform the experiments the more 
   physically established for all of them will be the new 
   phenomenon and its theory  \cite{Bar1}.  \par
    \section{Field realization of the Zeno effect}  The physical 
    validation  of new phenomena due
    to densely  experimenting   with  a large ensemble of  
    systems that are
    confined in a finite region  of space necessitates, as remarked in Section
    1.2,   to discuss these systems, especially 
    in the limit of a large number
    of them,   in terms of fields. That is,
    when the number of systems increases 
     whereas 
    the total volume does not change 
     the magnitude of each system  becomes very small and we 
     get actually, as remarked,  a field of them.  
    The known physical fields,  such as the electromagnetic field, 
 can be regarded as such fields of probes as has been done by several authors. 
Bixon in \cite{Harris} has shown  that the stabilization 
of the localized Born-Oppenheimer states \cite{Schiff} is due to the surrounding medium 
composed from such a field 
of probes. In his article Bixon himself regards this stabilizing effect of the 
surrounding field as a manifestation of the Zeno effect, although he regards it as 
the conventional time Zeno effect and not its space analog.    Davies \cite{Davies} has 
likewise treated the electromagnetic field as a 
field of 
probes and show that the interaction with it causes the localized state to acquire 
lower energy than the extended one,  thus stabilizing it.  It seems, therefore,  
appropriate to use field formalism  in order to discuss this effect. We exploit in
the following both quantum  \cite{Mahan,Mattuck} and classical field 
\cite{Mattis,Masao,Namiki1} methods for demonstrating  the possible existence of the Zeno effect
in various different phenomena.  \par \section{Numerical simulations 
 as corresponding to Zeno processes}
  An important example in which the effect of the large  ensemble of related
  systems (observers) is sharply
  pronounced is, as shown in \cite{Bar2,Bar001},  that of  computer numerical simulations.  
  In this respect we 
   point out  the striking similarity of the remarked  repetitions that lead to
  establishing  and validating   of  real phenomena  to the corresponding
 numerical repetitions of many  computer  simulations especially those concerned
 with finding numerical solutions of physical problems. For example, the
 numerical solution of any differential equation, including those that govern
 the evolution of physical systems, is obtained only after repeatedly updating
 the given differential equation. Moreover, the larger is the number of these
 iterations the larger will be the number of samples and the better is the 
 resulting
 statistics.    Thus, there is a strong correspondence  between the remarked stages of
    physically establishing a new  phenomenon to the mechanism of numerical
    simulation \cite{Bar2}. An important example of numerical simulations,  
    discussed in \cite{Bar2,Bar001}, 
  is related to Internet webmastering \cite{Bar2,Bar001}, where by this term we
  mean the stages of first writing the software sources of the websites (by HTML, Java
  or other script) then  running these  codes to  show the relevant
  websites on   the
  computer screen. We extensively discuss in \cite{Bar2,Bar001} the  stages of 
  these  numerical 
  simulations and especially those  related to their code-writing  
   and conclude from the obtained
  results about the corresponding processes of 
  real  phenomena (see especially \cite{Bar2}).   \par 
  \par \section{Scope of this
  work}
 This work is composed of five  chapters, the first of which is the present
 introduction,
 that are connected by the unifying principle of the Zeno effect. Each chapter  
   is constructed from several sections that may serve as preliminary
   introductions to the main calculations and results that have been published
   in articles (see list of publications at the end of this work). The relevant
   papers are shown completely or partly in special Appendices in this work. 
   The common result  demonstrated
 in the five chapters and their relevant articles in the Appendices 
  is  that the  factor that causes a physical
 system to first responds in a different,  even unexpected, manner and then 
 physically establish  this new response (new phenomenon)  is, as remarked, 
  the repetitions
 of experiments embodied in  the time and,  especially,  in the space Zeno 
 effect.    \par
  In Chapter 2  we discuss the time Zeno
  effect in the framework of quantum field theory.  
  Section  2.1    discusses  both kinds of the static and
  dynamic Zeno effects by using the  quantum field examples analyzed in 
  \cite{Bar13} (which is completely shown in Appendix $A_1$). We show by 
  these examples that experimenting in an uninterrupted
  manner on the system results in a new response of it that occurs only because
  of this kind of experiment. Thus, all one have to do in order to reconstruct
  this new response of the system is to perform again dense measurement. In
  other words, this kind of experimentation not only gives rise to  this new result
  but also physically establishes and validates it. 
   We also show that the Zeno process  is responsible for the occurence
  of additional effects that do not appear in the absence of the relevant
  repetitions \cite{Bar13}. Section  2.2  
   discusses  the histories formalism
  \cite{Hartle,Isham,Griffiths} in conjunction with the Flesia-Piron quantum 
  extension 
  \cite{Horwitz} of the Lax-Phillips theory \cite{Lax}. We  show, using  the 
  results of 
  \cite{Bar14} which  is  shown in Appendix $A_2$,  that the histories evolution 
  is stationary if the corresponding
  elements of the sequence $(h_{t_1}, h_{t_2}, h_{t_3},\ldots)$, which denotes
  the relevant history in a Lax-Phillips sense \cite{Horwitz,Lax}, are 
  determined by the  
  Schr\"oedinger  evolution between infinitesimally close neighbours in this
  sequence. That is, the stability of the real histories evolution corresponds to the
  dynamic Zeno effect \cite{Aharonov,Facchi}. We also show that the static Zeno
  effect is obtained if all the elements of the former sequence are identical so
  that $h_{t_k}=h_{t_0}$ for all $k=1, 2, 3, \ldots$.  \par 
    As remarked,  the Zeno 
    effect is not restricted only to
  quantum events but is effective also in the classical regime. This may be
  realized from \cite{Bar2} in which  we  apply  the stochastic quantization (SQ)
  method of Parisi-Wu \cite{Namiki1,Parisi1} to the programming process that  
  simulates physical phenomena.  
   We use in \cite{Bar2} the   path integral method 
   \cite{Feynman2,Schulman,Gert} 
  to show the effect of  
   the remarked repetitive element. For this  method the 
   repetitions
   are effected    through 
  introducing into the actions $S$ of all the path integrals of the relevant 
  ensemble the {\it same}  expression that determines the involved  process. In
  \cite{Bar001}  we  discuss the Internet  websites 
   without referring to any particular site. We  use for that
  purpose  a
  statistical machanics approach \cite{Reichl,Huang,Ursell,Mayer} that allows us to
  consider large clusters of mutually linked sites \cite{Bar001}. We obtain 
  similar results to
  those obtained from the SQ method,  discussed in  \cite{Bar2},  and from the Feynman diagram
  summation \cite{Feynman2,Schulman,Gert} of Chapter 2.  
    In all  these methods one obtains, as will be shown, in the limit of a 
    very large
  number of these repetitions, the known physical, or numerical, 
    equilibrium situations.  \par In 
   Chapter 3  we apply   the Zeno effect to the  general
  reversible classical chemical reaction  $A_1+A_2+\ldots \leftrightarrow
  B_1+B_2+\ldots$. We show that in the limit of repeating this reaction  a large
  number of times one remains with the initial constituent particles as if the
  reaction never happened. That is, one obtains the remarked static Zeno effect.
  We have also demonstrated   in Section 3.3 the possible existence of the
  dynamic Zeno effect in classical chemical reactions. These results were also 
  numerically   demonstrated    in Section 3.4   using the double circular billiard model
  \cite{Bar10,Bar12}.     This numerical billiard model was first introduced in
  \cite{Bar10}  to show the classical effects of dense measurement and in
  Section 3.4 we use this model for simulation of classical reactions. 
  The paper
  \cite{Bar10} is completely shown in Appendix $B_1$. \par  
  In Chapter 4 we discuss  the space Zeno effect in both its quantum and 
  classical realizations. We do this with the help of the Appendices $C_1$,
  $C_2$,  $C_3$ and $C_4$ which show the papers or the relevant parts of them 
    that
  discuss this aspect of the Zeno effect. 
  In  Section 4.2 we  show, using  \cite{Bar3} which is partly shown in 
  Appendix $C_1$,  that  the
  relevant expression for the space Zeno effect may
  be derived analytically \cite{Bar3} from that of 
    the time Zeno
  effect by  replacing, and doing the appropriate changes, 
   the time variable by the space one. In Section 4.3  we 
   discuss two systems that although appear similar at first sight they are very
   different fron each other since one  is quantum and the other is
   classical. We show for both systems the existence of the space Zeno effect.
   The first one which is discussed in the papers \cite{Bar7} and \cite{Bar8}, 
   which are shown  in
   Appendices $C_2$ and $C_3$ respectively (only Sections 1-2 of \cite{Bar7} is
   shown in $C_2$), is 
     the quantum  one-dimensional
   multibarrier potential of finite range  and the second, discussed 
   in \cite{Bar11,Bar111},  is the
   classical one-dimensional array of imperfect traps (the  papers
   \cite{Bar11}, \cite{Bar111}  are 
   shown in the Appendices $C_4$ and $C_5$ respectively). we have 
   demonstrated  for
   the first quantum  system also chaotic \cite{Bar7,Bar8} and phase transition
   \cite{Bar15} effects.    The existence of the space Zeno effect 
    for each system and for both cases of finite
   and infinite number of barriers or traps  in the finite interval is
   demonstrated. That is,  
    we show  analytically and numerically,  for both  
    systems, 
    the somewhat unexpected
   result \cite{Bar7,Bar8,Bar11,Bar111} that the  larger is the number of barriers 
    or traps the value  of either the probability amplitude of the quantum 
    wave after passing the barriers or 
      the density of
   the classical diffusing particles after going through  the traps  tends 
    to the 
   initial value    before traversing  either system.  \par 
     We have discussed,
   especially, these quantum and classical systems since they both exhibit the
   noted characteristics of the Zeno effect.  That is,  an uninterrupted
   experimentation, in a spatial sense as will be explained in Chapters 4-5, on either
   one causes them to behave differently and unexpectedly  compared to their
  known  behaviours in the absence of these spatial dense measurement.  That is,
  their known and expected physical behaviour changes to an  unexpected
  one not because of changing the value or the condition of any physical
  parameter related to the experiment but only because of repeating it in a
  dense manner. In other words, the dense   repetitions of these  known
  interactions  change the known response of the relevant systems,
  quantum or classical, and not only the manner by which this response is
  accomplished. Moreover, as remarked, if we wish to reconstruct this new
  response any number of times then all we have to do is to perform each time 
  this kind of
  experiment.  Thus, these  spatial  dense measurement, may be
  thought of as processes  that  establishes and validates the physical aspect
 \cite{Bar0} of the new responses (new  phenomena). 
      \par 
 In Chapter 5   we discuss in details 
  the effect of performing 
experiments by an ensemble of {\it related}  
observers  compared to the ensemble of unrelated ones or to the 
case in which one observer is involved. 
In  Section 5.2   we show the quantum effect of the related ensemble using 
the 
 Feynman path method 
\cite{Feynman2} and  in Section 5.3 we use for that purpose 
 the relative state approach  of Everett
 \cite{Everett,Graham,Bar16}.
 In Section  5.4 we  show the classical effect of the related ensemble using
thermodynamical  considerations \cite{Bar1}.        
 \newpage

\pagestyle{myheadings}
\markright{CHAPTER 2. \ \ \ QUANTUM FIELD THEOTY AND THE HISTORIES......}

 \newpage

\chapter{\label{Chap2}  Quantum field theory  and the histories formalism}

\protect  \section{\label{sec1}Quantum field theory and dense measurement}  
 
 We have remarked more than once in Chapter 1  about the importance of acting 
ceaselessly, through repeatedly experimenting,  on the physical system which 
results in a response that is different from the one obtained when the 
experiments are not repeated in such a dense manner. This kind of repeated 
experimentation and its unique results have been discussed in details in   
\cite{Bar13}  which is represented  in Appendix $A_1$.  In \cite{Bar13} 
 we have exploited 
the powerful  quantum field theory method of summing Feynman diagrams 
\cite{Mahan,Mattuck}  to all orders and adapt it for discussing quantum Zeno effects. 
This is done by summing to all orders not the Feynman diagram of the relevant 
process but the $n$ times repetitions of it where the limit of $n \to \infty$ 
is taken. This generalized summation have been applied in \cite{Bar13} for both 
kinds of the Zeno effect static and dynamic. Thus, in Section 2 of 
\cite{Bar13}  we use the quantum field example of the bubble process 
\cite{Mahan,Mattuck} for discussing the static Zeno effect. It has been shown in 
\cite{Bar13} by taking the summation to all orders of the $n$ times repetitions 
of  the bubble process that the probability to remain with  the initial state 
tends to unity when $n \to \infty$. By taking this limit  we ensure that the system
is interacted upon all the time without being left to itself even for  a very
short time and this is so not only  for the higher order terms of the 
relevant Feynman
diagrams but also for the 
lower order terms.  That is,  we  first sum the basic Feynman diagrams 
of the bubble 
interaction   $n$ times, where $n \to \infty$,  and then 
we sum to all orders the Feynman 
diagram of this
$n$-times repeated process.
By these summations we   
cause the system to behave
differently and unexpectedly  compared to its known behaviour in the absence
of these $n$-times repetitions. 
 In other
words, the dense experimentation entailed by the $n$ times repetitions of some
experiment, over a finite time interval, assigns to the system, 
in the limit $n \to \infty$,  new
characteristics   that causes it to behave
differently from its behaviour under the same experiment performed once or
repeated not in a dense manner. Moreover, this new behaviour of the system
includes not only the former unity value for the probability to remain with the
initial state but a much more unexpected response of it. This response arises
when  we express the former probability in the $(k,w)$ representation   in which
case it may be written as $\lim_{n \to
\infty}(\frac{1}{(w+i\delta)-\epsilon_k})^n$ (see Eq (21) in
\cite{Bar13} and in Appendix $A_1$) where $\epsilon_k$ is the excited energy.
That is, one obtains from the last expression that there exists a pole for each
value of $w$ that satisfies $|(w-\epsilon_k)|<1$ and not only for the Hartree 
value of $(\epsilon_k+V_{klkl}-i\delta)$ \cite{Mattuck} which is obtained when
the bubble process is summed to all orders ($V_{klkl}$ is the potential related
to this process \cite{Mattuck}). In other words, the summation of the $n$
times repetitions of the bubble interaction to all orders results, in the limit of
$n \to \infty$,  in a large number of poles along a cut. \par
In Section III of \cite{Bar13} (see also Appendix $A_1$) we use the quantum
field example of the open-oyster process \cite{Mattuck} for discussing the
dynamic Zeno effect \cite{Aharonov,Facchi}. We have shown there that if we take
the summation to all orders of the $n$ times repetitions of the open-oyster
process then the probability to begin at a given initial state and end at
another given one tends to unity in the limit of $n \to \infty$. That is, as in
the formerly discussed bubble process the dense experimentation entailed by the
$n$ times repetitions of it, over a finite time interval, assigns to the
system, in the limit of $n \to \infty$, new characteristics that cause it to
behave differently from the behaviour expected when the same experiment is
performed once or repeated not in a dense manner. Moreover, it has been shown in
Section 3 of \cite{Bar13} (see also Appendix $A_1$) that since in the
version of the open-oyster process discussed there there is no element of
repetition whatever,  as explained after Eq (32) in \cite{Bar13}, then there is
also no poles at all (see Eq (32) in \cite{Bar13} and in Appendix $A_1$). This
is in accordance with the former discussion of the bubble process where we find
that the mere repetitions of this process results in a continuum (cut) 
 of poles.  \par 
 Note, as remarked, that if the bubble process itself is discussed
 without the $n$ times repetitions of it, as in \cite{Mattuck}, then it results
 with the Hartree single pole $(\epsilon_k+V_{klkl}-i\delta)$.   
 Also it is known \cite{Mattuck} that if the open-oyster process is discussed in the
 version in which the energy of the leaving particle is the same as that of the
 entering one then it yields also \cite{Mattuck} the single pole 
 $(\epsilon_k+V_{lkkl}-i\delta)$ (note the difference  between 
 the potential 
 $V_{lkkl}$ and that of the bubble  which is $V_{klkl}$ (see the discussion in
 \cite{Mattuck} and in Appendix $A_1$)). Thus, since the kind of the open-oyster
 process discussed in Section 3 in \cite{Bar13} (Appendix $A_1$) has no element
 of repetition at all then it has also no pole whatever. In other words, we see 
 that the poles which are so intimately related to the occurence of physical
 resonances and quasi-particles \cite{Mattuck} are enabled through these
 repetitions. That is, if the process itself has an  element of repetition such 
 as the bubble procedure or the open-oyster one in which the energies of the
 leaving and entering particles are equal then there exists pole in these
 systems. If these processes are discussed in the version in which they are 
 repeated $n$ times and the Feynman diagram of these repetitions are summed to
 all orders then in the limit of $n \to \infty$ one finds a whole cut of poles
 as found for the bubble processs of section 2 in \cite{Bar13} (Appendix $A_1$).
 If, on the other hand, there is no element of any repetition whatever 
 as in the version
 of the open-oyster process discussed in Section 3 of \cite{Bar13} in which the
 energies of the leaving and entering particles are different then there is also
 no pole in the system. Thus, we see that the same system may respond
 differently not only to different experiments but also to the presence or
 absence of repetitions of the {\it same experiment}.       
     \par  In summary, we see, using the field examples of
the bubble and open-oyster processes,  that summing to all orders the $n$-times
repetitions of the relevant Feynman diagrams one may establish not only a given
initial state (static Zeno effect \cite{Zeno,Itano,Harris,Simonius}) or a 
specific Feynman path of states (dynamic
Zeno effect \cite{Aharonov,Facchi}) but also may result in other new responses that are not obtained in
the absence of these repetitions. Moreover, as remarked, these iterations also
physically  establish these new processes in time. In Chapters 4-5 we show
that  these  processes are established also in space
\cite{Bar7,Bar8,Bar11,Bar3}. 
\section{\label{sec2}The histories formalism and dense measurement}
  We discuss in this section a field of histories where by the term history we
 mean the sequence into which a physical process may be subdivided into its
 parts (see Appendix $A_2$).  The subdivision may be coarse grained or fine grained \cite{Hartle}
 where a real physical history must be fine-grained.  
  In \cite{Bar14}, which is shown in Appendix $A_2$,  we use the histories 
 formalism
 of Gell-Mann-Hartle-Griffiths (GMHG) \cite{Hartle,Isham,Griffiths} together with the Flesia-Piron
 \cite{Horwitz} extension  of the Lax-Phillips theory \cite{Lax} which discuss
 resonances and semi-group evolutions (i. e., irreversible processes) for the
 classical scattering of electromagnetic waves on a finite target.  A Lax-Phillips generalized
 state is defined as \cite{Lax} the sequence $(h_{t_1},h_{t_2},h_{t_3},\ldots)$ where each
 element satisfies $h_{t_k} \subset H_k$,  and $H_k$ is the corresponding
 Hilbert space at $t_k$. We consider the Flesia-Piron generalization
 \cite{Horwitz} of the Lax-Phillips theory and refer to the former sequence as 
 a
 GMHG history \cite{Hartle,Isham,Griffiths} projected as a result of  a large
 set of corresponding experiments. We show in  \cite{Bar14} (see also  
  Appendix $A_2$) that the
 equilibrium state of this history is obtained only in the limit in which it is
 finely-grained and any two infinitesimally close element of which are related
 by the Schr\"odinger evolution. We note that  the real physical histories 
  must
 satisfy these conditions of fine graininess and Schr\"odinger evolution in
 order to be real (a coarse-grained history is only a virtual approximation to
 the real fine-grained one).  Thus, we see that the real histories 
  behave as if they were formed from a dynamic Zeno-type process
 that characterizes a dense set of experiments which realizes
 \cite{Aharonov,Facchi} the specific resulting  path of states. 
  The static Zeno effect is obtained when all the
 elements of the relevant history are identical.  \par 
  Moreover,  it is shown
 in \cite{Hartle,Griffiths} that the real physical histories must be 
   only the consistent ones \cite{Hartle,Griffiths} that are
 exclusive in the sense that the problematic superposition principle
 \cite{Wheeler,Griffiths} is excluded.  Thus,  real physical phenomena are
 constructed only from single sequences of consecutive states and not from the
 superpositions of such sequences.  We have  shown in \cite{Bar14} that
 these consistent histories become stationary  in the limit
 of dense measurement in the sense of \cite{Aharonov,Facchi}. That is, also 
 with respect to the consistency condition of GMGH the
real  consistent   histories correspond to the dynamic Zeno effect in which a 
path of states (history) becomes ``realized'' as a result of performing  densely  
 all the specific experiments that reduce the system to these states. 
  In other words,
 regarding any real phenomenon as composed and assembled from a large number
 of consecutive states  we see that its physical reality 
  is obtained as if it has
 been established through densely performing the specific experiments that
 reduce the system to these states. Thus, we see, as remarked, that these
 repetitions (along this specific path),  which are the essence of the Zeno 
 effect, correspond to the physical realization of  real phenomena.


\pagestyle{myheadings}
\markright{CHAPTER 3. \ \ \ CHEMICAL REACTIONS AND THE CIRCULAR BILLIARD .....}

 \newpage
\bigskip \noindent 
\chapter{\label{Chap3}Chemical reactions  and the circular billiard model } 

\protect \section{Introduction \label{sec1}   }

\footnote{This chapter
 was later published with some changes, after the thesis's submission, 
 under the title 
 "The effect of increasing the rate of repetitions of classical reactions" 
 in IJTP, {\bf 43}, 1169-1190 (2004)}

\noindent In this chapter we show the general character of the Zeno effect
\cite{Simonius} which may be present not only in physical quantum events but 
 also in  chemical classical reactions. \par 
 The general reversible reactions  
 $A_1+A_2+\cdots+ A_r\leftrightarrow B_1+B_2+\cdots +B_s$, where 
$r$ and $s$  are two arbitrary positive 
natural numbers,  have
been studied by many authors (see, for example, \cite{Havlin} and references
therein). These studies discuss, 
especially,  the effects of the single reactions, or, in case they are repeated $N$
times,  the effect of these repetitions where
the general total time  increases proportionally to $N$. We
can, however, imagine a situation in which the {\it rate} of these repetitions
increases and discuss the effect of this increase upon the reaction. Such an
effect has been studied in \cite{Kac} with respect to random walk and it was
shown that when the rate of repeating the random walk becomes very large one
obtains a Brownian motion. It has also been shown \cite{Bar1} with respect to a
one-dimensional  array of imperfect traps \cite{Havlin,Smol} that as their 
number along
the same finite interval,  becomes very large,  the survival probability 
\cite{Havlin} of the 
particles that pass through them tends to unity. We show in this work that if 
 either direction of the   reaction is repeated  a
large number  of times $N$ in a finite total time $T$, then in the limit of very
large $N$, keeping $T$ constant,  one remains with the  initial reacting 
particles of the repeated
direction 
only.  \par We  use   quantum theory methods and   terminology  
      as done by  
many authors that use quantum formalism  for analysing
classical reactions (see for example \cite{Mattis} and annotated bibligraphy
therein).  The most suitable method is the coherent state one
\cite{Klauder,Glauber} since 
 it allows us  to define simultaneously, 
  as has been remarked in \cite{Swanson}, 
the expectation values of the coherent
state conjugate variables $Q$ and $P$, so that they both may have nonzero
values. Thus, this formalism resembles \cite{Swanson} the classical one and 
 is appropriate
 for using  it in classical reactions.  The use of the coherent
states formalism, together with  second quantization methods, for classical
systems have been studied by Masao \cite{Masao}.  \par Now,   since the
described phenomena and,   especially,  the particles participating in the 
reactions 
 are classical we  represent them 
 by real 
coherent states. 
That is,   we shall represent
    the reacting  and  product particles 
by  the real coherent states denoted  
$|z\!>= e^{-\frac{1}{2}|z|^2}\sum_{n=0}^{n=\infty}\frac{z^n}{(n!)^{\frac{1}{2}}}|n\!>$,
where $z$ is the real number 
$z=\frac{q+p}{(2)^{\frac{1}{2}}}$, $q$ and $p$ are two arbitrary real $c$ 
numbers (the masses of the reacting and product particles are assigned, for 
convenience the unity value), and $|n\!>$ are number representation eigenstates  
\cite{Klauder}. We note that although the mathematical entities and "operators" 
involved in this method  do not conform, as will be shown, to the known 
quantum operator formalism, nevertheless we follow, except for the differences,  the 
 conventional definitions and methods of the last theory. The results  
 obtained 
by applying this  
real coherent state formalism for classical reactions 
are exactly the same as those previously obtained \cite{Bar6} 
by applying the complex coherent state methods for 
 quantum optics reactions. 
Moreover, although the real coherent states formalism entails  an evolution operator 
(see the following Eq (\ref{e3.3})) which is nonunitary and unbounded nevertheless, 
we show that the results we obtain are valid also for this kind of operator. That is, 
we obtain for the classical reactions  the same results that were 
 obtained \cite{Bar6}  for the 
quantum particles which are represented by the complex coherent states 
\cite{Klauder,Glauber} that entail unitary bounded evolution operators. 
Needless to say that one may, obviously, use the conventional 
complex coherent state formalism \cite{Klauder,Glauber} for discussing also classical 
reactions as has been remarked. 
Thus, the 
  scalar product  
with the conjugate 
$<\!z|z\!>$ is interpreted  \cite{Masao,Mikhailov} (in accordance with the
conventional interpretation of quantum machanics) as 
the probability to find the
system in the state $|z\!>$.  \par 
 We assume that 
we have some set of $N$ identical particles so that the configuration in
which the $i$th particle is located at $q_i$ $(i=1,2,\cdots,N)$ is defined as 
a
state of the system and denoted, in Dirac's notation, 
$|q_1,q_2,\cdots,q_N\!>$ 
    ($|q^N\!>$) (for distinguished sets of particles we partition the total 
    $N$ system to 
    $N_1, \ N_2, \ \ldots$ subsystems). Thus,  when representing, in the 
    following, classical particles by
states we mean that they are elements of some configuration of the whole 
system.
Following this terminology we may calculate the probability to find the set 
of particles in some definite state $|q^N\!>$ as \cite{Masao} 
$$F^{(N)}(q_1,q_2,\cdots,q_N;t)=\sum_{all  \ permutations \ of \ q_i}
f^{(N)}(q_1,q_2,\cdots,q_N;t), $$ where  $f^{(N)}(q_1,q_2,\cdots,q_N;t)$  is some
normalized distribution function. To this probability one assign, as done in 
  \cite{Masao},  
a ``state''  $|F(t)\!>=\sum_{N=0}^{\infty}\int dQ^N
F^{(N)}(q_1,q_2,\cdots,q_N;t)|q^N\!>$, where $\int dQ^N=\frac{\int dq^N}{N!}$ (the
division by $N!$ is necessary \cite{Masao} so as not to overcount the state $|q^N\!>$ $N!$ 
times). Thus the former probability to find the system in the state $|q^N\!>$ 
may be written as \cite{Masao} $F^{(N)}(q_1,q_2,\cdots,q_N;t)=<\!q^N|F(t)\!>$. 
\par
In Section 3.2 we discuss the special cases of $r=s=1$ and $r=s=2$, that is, the
reversible reactions $A \leftrightarrow B$ and $A_1+A_2 \leftrightarrow 
B_1+B_2$, and show that repeating either direction of 
each   
 a large number $N$ of times  in a finite total time $T$ results, in the
limit of very large $N$,   in a unity probability to remain with  the initial 
reacting particles only.   The generalization to any finite $r$ 
and $s$ follows.  As remarked, exactly similar results were obtained also 
\cite{Bar6} for quantum reactions and particles that are represented by 
 complex coherent states \cite{Klauder,Glauber}. This has been explicitly 
shown  \cite{Bar6} for the case that these states represent both the optical 
point source and point detector. It was shown \cite{Bar6} that the 
cross-correlation \cite{Klauder} between these points becomes maximal in the 
limit of repeating a very large number of times the experiment of detecting the light 
from the point source by the point detector. Similar results were obtained 
also for the more realistic case of an extended light source that emanates 
light from many points which is detected by an array of point detectors. \par 
Note that similar results were obtained also for the bubble process 
discussed in Chapter 2.  That is, although the bubble process is a typical
quantum field phenomenon and the former reactions are   of the 
 classical chemical
type, nevertheless,  under the dense measurement condition they are both  
typical examples of the static Zeno effect. \par
We  note that since we discuss  
  the probability to remain with  the initial reacting 
 particles the product particles of such reactions are not relevant (as the 
 reacting ones) to our
 discussion. In Section 3.3 we discuss the more general and natural case in which
 the product particles are relevant. That is, 
 we assume that the  particles of the  ensemble  interact  at
 different places and  times and that they begin from some given initial
 configuration of reactions and end at a final different one. 
  We calculate the probability that 
   a specified system of 
 reacting
 particles evolves along a prescribed path of reactions, from a large number 
 of different
 paths that all begin at the given initial configuration, end at the  final one
 and are composed of intermediate consecutive different configurations of 
 reactions. 
  We note that such paths of
``states'' for the diffusion controlled reactions have been discussed in
 \cite{Masao,Namiki1,Mikhailov} where  use was made of   quantum field 
 theory methods 
 \cite{Mahan,Mattuck}, including 
 Wick's theorem \cite{Mahan,Mattuck}, to derive the classical Feynman diagrams. 
   These methods were
 also used in \cite{Mikhailov} for chemical kinetics. We show that
 taking the limit of a very large number $N$ of  reactions  
 along  the prescribed path and performing them in a dense manner one obtains 
 a unity value for the 
 probability of evolution along  that path. \par 
   Note that similar results were 
 obtained also for the open-oyster process 
discussed in Chapter 2.  That is, although the last process is a typical
quantum field phenomenon and the reaction discussed in Section 3.3 is of the 
 classical chemical
type, nevertheless,   they are both,   under the dense measurement condition,  
typical examples of the dynamic Zeno effect.
 \par 
 In Section 3.4 we use a numerical 
 model that has been used in \cite{Bar10} for showing the effect of dense
 measurement on classical systems.  This is 
 the model of two dimensional concentric billiard \cite{Bar10} 
  that is used here to  numerically 
 simulate  the reversible  reaction 
 $A+B \leftrightarrow A+C$.  
  The two possible modes of reflections inside the billiard, either between 
  the two
 concentric circles or between points of the outer circle, represent the two
 directions of the reaction. We note that nuclear and radioactive reactions are
 well simulated by billiards in which the stationary scattering circles
 represent the interactions between particles (see, for example,  \cite{Bauer}
 in which a model of a rectangular billiard with a circle inside was used to
 discuss the decay law of classical systems).    
 We show that if either direction of the reaction  $A+B \leftrightarrow A+C$  
     is repeated a large 
 number  of times $N$  in a finite total time then in the limit of very large 
 $N$  the  result obtained,  as will be explained, is as if   
  no repetition is involved at all. That is, the very large number of 
 repetitions in the same 
 direction of the 
 reaction, where the opposite direction occurs at some prefixed lower rate, 
 has an effect 
 as if the high rate repeated reaction 
 did not occur.   This  is exactly what we obtain 
 {\it analytically}  in 
 Sections 3.2 and 3.3  where  the large number of repetitions of either direction 
 of
 the general reversible reaction $A_1+A_2+\cdots A_r \leftrightarrow
 B_1+B_2+\cdots +B_s$ results in remaining,  with a unity probability,  with the
 initial reacting particles only as if the repeated reaction did not occur at
 all.  The paper \cite{Bar10} in which we discuss the concentric billiard model
 in relation to dense measurement is shown in Appendix $B_1$. Note that in this
 paper we take the fundamental trajectory as the one that begins from some 
 point, reflected at a second point and ends at a third one (that may be
 identical to the first). Here this trajectory is taken to be that between two
 consecutive reflection points thus it is, due to the elastic reflection, 
  only half in length compared to the first. The final outcome do not depends on
  such difference.  
 
 \protect \section{The reversible reaction  $A_1+A_2+\cdots A_r \leftrightarrow B_1+B_2+\cdots +B_s$ \label{sec1}}
We discuss first the specific case of $r=s=1$ and note, as we have 
remarked, that since we calculate the probability to remain with the initial reacting
particles  the product part of the reaction is not essential to the following
discussion,  as will be
shown.  Nevertheless, we take in this section 
the specific examples of $r=s=1$ and $r=s=2$ and begin with the first reaction,
that is, $A \leftrightarrow B$  where $A$ and $B$ are, as noted, represented  by the two coherent states \cite{Klauder}
\begin{eqnarray}
&&|z_A\!>=e^{-\frac{1}{2}|z_A|^2}\sum_{n=0}^{n=\infty}\frac{z_A^n}{(n!)^{\frac{1}{2}}}|n\!>
\label{e3.1} \\
&&|z_B\!>=e^{-\frac{1}{2}|z_B|^2}\sum_{n=0}^{n=\infty}\frac{z_B^n}{(n!)^{\frac{1}{2}}}|n\!>
\nonumber  \end{eqnarray}
 Using the
following general equation for any two operators $X$ and $Y$  
$$e^YXe^{-Y}=X+[Y,X]+\frac{1}{2!}[Y,[Y,X]]+\cdots, $$ where $[Y,X]$ is 
the commutation 
$[Y,X]=YX-XY$,  one obtains 
\begin{equation} \label{e3.2} U(q,p)(\alpha P+\beta
Q)U^{-1}(q,p)=\alpha(P+p)+\beta(Q+q)  \end{equation}
 The  $\alpha$,
$\beta$ are arbitrary parameters,  $U(q,p)$ and $U^{-1}(q,p)$ are given 
respectively by $U(q,p)=e^{pQ-qP}$, $U^{-1}(q,p)=U(-q,-p)$, and $Q$,
$P$ are the coherent state operators that satisfy $[Q_i,P_j]=\delta_{ij}$. 
That is, $U(q,p)$ translates the operators $Q$ and $P$ by $q$ and $p$
respectively.  Now, since the coherent states has been defined, as remarked, in
terms of the number representation eigenstates (see Eq (\ref{e3.1})) we 
 write the time evolution operator of the relevant states 
 as $e^{Nt}$,  where $N$ is the number operator
\cite{Klauder} (note that since we discuss in this paper real coherent states 
 the evolution operator is real  also).  
$$N=a^{\dagger}a=(\frac{Q-P}{\sqrt 2})(\frac{Q+P}{\sqrt 2})=
\frac{1}{2}(Q^2-P^2+1)$$ 
$N$ is defined analogously to the corresponding operator of the complex coherent 
state formalism \cite{Klauder,Glauber} but without the complex notation $i$ in the 
middle expression. Note that the operator $N$ is not positive definite and this 
to remind us, as remarked, that the real coherent state formalism discussed here 
does not conform to the conventional quantum operator process. Nevertheless, as 
remarked, the final results obtained here are exactly the same as those accepted 
in \cite{Bar6} from the quantum complex coherent state formalism. 
The commutation  
$[Q_i,P_j]=\delta_{ij}$ has been used in the last equation. 
 Applying
the operator $N$ on the coherent state $|z\!>$ from Eq (\ref{e3.1}), 
and taking the scalar product of the
result with the conjugate  state  
$<\!z|$  one obtain (using
$<\!n|e^{Nt}|m\!>=e^{nt}\delta_{nm}$, since in the number representation the
operator $N$ is diagonal) 
\begin{eqnarray}
&& <\!\grave z|e^{Nt}|z\!>=\exp(-\frac{1}{2}|z|^2-\frac{1}{2}|\grave z|^2)
\sum_{n=0}^{n=\infty}\frac{(\grave
ze^tz)^n}{n!}=\nonumber \\ &&=\exp(-\frac{1}{2}|z|^2 -\frac{1}{2}|\grave z|^2+\grave
ze^tz) =<\!\grave z|e^tz\!>=\label{e3.3} \\ &&=<\!\grave z|(\cosh t+\sinh t)z\!>=<\!\grave q,\grave
p|q_t,p_t\!> \nonumber \end{eqnarray}   
  The last result is
obtained by writing $z$ in terms of $q, p$ in which we have  
\begin{equation}  q_t=q(\cosh t+\sinh t), \ \ \ p_t=p(\cosh t+\sinh t) 
\label{e3.4} \end{equation}  
  We, now,  calculate, using Eq (\ref{e3.3}),  the probability $p(|q_A,p_A\!>)$ to remain with the  initial particle $A$
 after the reaction $A \to B$ where the particle $B$ is represented by the 
 coherent state $e^{Nt}|z_A\!>$.  This is given by
\begin{eqnarray} 
&& <\!z_A|e^{Nt}|z_A\!>=<\!q_A,p_A|q_{A_t},p_{A_t}\!> 
=\exp(-\frac{1}{4}(q_A+p_A)^2- \nonumber \\ && -\frac{1}{4}(q_{A_t}+p_{A_t})^2)
\sum_{m,n=0}^{m,n=\infty}\frac{(q_A+p_A)^m(q_{A_t}+p_{A_t})^n}{2^{\frac{m+n}{2}}
(m!n!)^{\frac{1}{2}}}<\!m|n\!> =\nonumber \\ && 
=\exp(-\frac{1}{4}(p_A+q_A)^2- \frac{1}{4}(p_{A_t}+q_{A_t})^2)
\sum_{n=0}^{n=\infty}\frac{(q_A+p_A)^n(q_{A_t}+p_{A_t})^n}{2^nn!}
=\nonumber \\ && =\exp(-\frac{1}{4}(p_A+q_A)^2 -\frac{1}{4}(p_{A_t}+
q_{A_t})^2+ 
\frac{1}{2}(q_A+p_A)\cdot \label{e3.5} \\ && \cdot (q_{A_t}+p_{A_t}))
 =
\exp(-\frac{1}{4}(p_A+q_A)^2-\frac{1}{4}(p_A+q_A)^2(\cosh t+\sinh t)^2+
\nonumber \\ && +
\frac{1}{2}(q_A+p_A)^2\cdot (\cosh t+\sinh t))
 =\exp(-\frac{1}{2}(p_A+q_A)^2(\frac{1}{2}+\nonumber \\ && +\frac{1}{2}(\cosh t+\sinh t)^2 
 -(\cosh t+\sinh t))) \nonumber \end{eqnarray}
 Note that since we discuss coherent states the
interpretation \cite{Klauder} of the expression $<\!z_A|e^{Nt}|z_A\!>$ is 
the probability to
find the {\it mean}  position and momentum of the coherent state $e^{Nt}|z_A\!>$,  
which represents $B$, equal to those of $z_A$, which represents $A$,  and this 
probability is equivalent 
in our discussion to remaining with the particle $A$. 
From Eq (\ref{e3.5}) one obtains  the probability to remain with the initial 
particle $A$ after a 
single reaction $A \to B$. If it is repeated $n$ times in a finite total
time $T$ one obtains (using $n=\frac{T}{\delta t}$, where $\delta t$ is the 
duration of each reaction)
\begin{eqnarray} && p^n(|q_A,p_A\!>)=\exp(-\frac{T}{2\delta t}
(p_A+q_A)^2(\frac{1}{2}+\frac{1}{2}(\cosh \delta t+\label{e3.6} \\ && +
\sinh \delta t)^2-
(\cosh \delta t+\sinh \delta t))) \nonumber \end{eqnarray}
In the limit of very large $n$ (very small $\delta t$) we expand the
hyperbolic functions in a Taylor series and keep terms up to second order in
$\delta t$.  We obtain 
\begin{eqnarray} &&
p^n(|q_A,p_A\!>)=\exp(-\frac{T}{2\delta t}(p_A+q_A)^2(\frac{1}{2}+
\frac{1}{2}(1+2\delta t^2+2\delta t)-\nonumber \\ &&-(1+\frac{\delta t^2}{2}+\delta t))
=\exp(-\frac{T}{4\delta t}(p_A+q_A)^2\delta t^2)= \label{e3.7} \\ && =
\exp(-\frac{T}{4}(p_A+q_A)^2\delta t \nonumber
\end{eqnarray}
Thus, we obtain in the limit $n\to \infty$ ($\delta t\to 0$) 
\begin{equation} \label{e3.8} \lim_{n \to \infty} p^n(|q_A,p_A\!>)=\lim_{\delta t \to 0}
\exp(-\frac{T}{4}(p_A+q_A)^2\delta t)=1 \end{equation}
Now, although we refer in the former equations to  the direction $A \to B$ all 
our discussion remains valid also for the opposite one $B \to A$.  
That is, repeating either side of the  reaction $A \leftrightarrow B$ a large 
number of times $n$ in a finite
total time $T$ results, in the limit of very large $n$,  in remaining 
(with probability 1) with the initial 
particle of the repeated direction of the reaction.\par
We, now, discuss the reversible 
reaction $A+B \leftrightarrow C+D$ in which we have two
 reacting particles.   We continue to use the number evolution operator $N$ 
  and take
into account that the initial particles $A$ and $B$ interact. Thus, representing
these particles as the coherent states $|q_{A},p_{A}\!>$ and 
$|q_{B},p_{B}\!>$ we write, for example,   the left hand side direction 
of the former reversible 
reaction $A+B \to C+D$ as 
 \begin{equation} \label{e3.9} \exp((N_A+N_B+P_AP_B+Q_AQ_B)t)|q_A,p_A\!>|q_B,p_B\!>=
 |q_C,p_C\!>|q_D,p_D\!>,  \end{equation}
 where the terms $Q_AQ_B$ and $P_AP_B$ represent,  as for the boson particles 
 discussed in    \cite{Klauder},  
 the interaction of the
 particles $A$ and $B$, and $N_A$, $N_B$ are the number operators for them. 
 Note that, as for the reaction $A \leftrightarrow B$ (see the discussion after
 Eqs (\ref{e3.4}) and (\ref{e3.5})), the operation of the evolution operator, which
 is now more complicated due to the interaction between $A$ and $B$, on the
 coherent state $|q_A,p_A\!>|q_B,p_B\!>$  is represented by 
 $|q_C,p_C\!>|q_D,p_D\!>$. 
 We calculate, now, the probability that the reaction $A+B \to C+D$ results 
 in remaining with the initial particles $A$ and $B$ only  (we denote this
 probability by $p(|q_B,p_B\!>|q_A,p_A\!>)$). 
\begin{eqnarray} 
&& p(|q_B,p_B\!>|q_A,p_A\!>)= \label{e3.10} \\ && = 
<\!q_B,p_B|<\!q_A,p_A|\exp((N_A+N_B+P_AP_B+Q_AQ_B)t)|q_B,p_B\!>|q_A,p_A\!>
 \nonumber \end{eqnarray}
 Using Eqs (\ref{e3.1}), (\ref{e3.3})  and the following coherent states 
 properties \cite{Klauder}
 $<\!q,p|Q|q,p\!>=q$, \ \ \  \\ $<\!q,p|P|q,p\!>=p$ (derived  by using the operator $U$ 
  from Eq (\ref{e3.2})
 and the relation  $N|0,0\!>=0$)
we obtain 
\begin{eqnarray}
&&
p(|q_B,p_B\!>|q_A,p_A\!>)=\exp((q_Aq_B+p_Ap_B)t)\cdot \nonumber \\&& \cdot 
<\!q_B,p_B|<\!q_A,p_A|q_{B_t},p_{B_t}\!>
|q_{A_t},p_{A_t}\!>= 
\exp((q_Aq_B+p_Ap_B)t)\cdot \nonumber \\ && \cdot \exp(-\frac{1}{4}(q_A+p_A)^2-\frac{1}{4}(q_B+p_B)^2
-\frac{1}{4}(q_{A_t}+p_{A_t})^2- \frac{1}{4}(q_{B_t}+p_{B_t})^2)\cdot \nonumber
\\ && \cdot 
\sum_{m,n=0}^{m,n=\infty}\frac{(q_A+p_A)^m(q_{A_t}+p_{A_t})^n}
{2^{\frac{m+n}{2}}(m!n!)^{\frac{1}{2}}}<\!m|n\!>\sum_{s,r=0}^{s,r=\infty}
\frac{(q_A+p_A)^s(q_{A_t}+p_{A_t})^r}
{2^{\frac{s+r}{2}}(s!r!)^{\frac{1}{2}}}\cdot \nonumber \\ && \cdot 
<\!s|r\!>=
\exp((q_Aq_B+p_Ap_B)t)\exp(-\frac{1}{4}(q_A+p_A)^2-\frac{1}{4}(q_B+p_B)^2
-\nonumber \\ && -\frac{1}{4}(q_{A_t}+p_{A_t})^2-\frac{1}{4}(q_{B_t}+p_{B_t})^2)
\sum_{n=0}^{n=\infty}\frac{(q_A+p_A)^n(q_{A_t}+p_{A_t})^n}
{2^nn!}\cdot \label{e3.11} \\ &&  \cdot \sum_{r=0}^{r=\infty}
\frac{(q_A+p_A)^r(q_{A_t}+p_{A_t})^r}
{2^rr!} =\exp((q_Aq_B+p_Ap_B)t)\exp(-\frac{1}{4}(q_A+p_A)^2-\nonumber \\ && -
\frac{1}{4}(q_B+p_B)^2
-\frac{1}{4}(q_{A_t}+p_{A_t})^2 -\frac{1}{4}(q_{B_t}+p_{B_t})^2+ 
\frac{1}{2}(q_A+p_A)(q_{A_t}+p_{A_t})+\nonumber \\ &&+ 
\frac{1}{2}(q_B+p_B)(q_{B_t}+p_{B_t}))=
\exp(\frac{1}{2}(\cosh t+\sinh t)((q_A+p_A)^2+(q_B+p_B)^2)-\nonumber \\ && -
\frac{1}{4}(1+(\cosh t
+\sinh t)^2)\cdot  ((q_A+p_A)^2+  (q_B+p_A)^2)+ (q_Aq_B+p_Ap_B)t) 
\nonumber \end{eqnarray}
This is the probability to remain with the original particles $A$ and $B$ after
one reaction. Repeating it a large number of times $n$ in a finite total time $T$,
where $n=\frac{T}{\delta t}$ ($\delta t$ is the time duration of one reaction) 
one  obtains for the probability to remain with  $A$ and $B$. 
\begin{eqnarray} && P^n(|q_B,p_B\!>|q_A,p_A\!>)=\exp(\frac{T}{\delta t}
((q_Aq_B+p_Ap_B)\delta t+\nonumber \\ && + 
(\frac{1}{2}(\cosh \delta t+\sinh \delta t)
-\frac{1}{4}(1+(\cosh \delta t+\sinh \delta 
t)^2))\cdot \label{e3.12} \\ && \cdot ((q_A+p_A)^2+(q_B+p_B)^2))) 
\nonumber \end{eqnarray}
In the limit of very large $n$ we expand the hyperbolic functions in a Taylor
series and retain terms up to second order in $\delta t$. Thus,  
\begin{eqnarray} && P^n(|q_B,p_B\!>|q_A,p_A\!>)=\exp(\frac{T}{\delta t}
((q_Aq_B+p_Ap_B)\delta t+
(\frac{1}{2}(1+\frac{\delta t^2}{2}+\nonumber \\ && + \delta t)
-\frac{1}{4}(2+2\delta t^2+2\delta t))((q_A +p_A)^2
 +(q_B+p_B)^2)))=\label{e3.13} \\ && =
\exp(T((q_Aq_B+p_Ap_B)- 
\frac{\delta t}{4}((q_A+p_A)^2+(q_B+p_B)^2))) \nonumber 
\end{eqnarray}
Taking the limit of $n\to \infty$ ($\delta t \to 0$) we obtain 
\begin{equation} \label{e3.14}  \lim_{n\to \infty}P^n(|q_B,p_B\!>|q_A,p_A\!>)= 
\exp(T(q_Aq_B+p_Ap_B)) \end{equation}
The last  probability tends to unity when the $c$-numbers of either
$A$ or $B$ (or both) are zeroes, that is, when $A$ or $B$ (or $A$ and $B$) 
  are in their ground states (in which case they are    
  represented by only the first term of the sums in Eqs
 (\ref{e3.1})).  
Needless to say  that all the former
 discussion  remains valid also for  the opposite direction 
 $A+C \to A+B$.  Thus,  we conclude that when either direction of the 
 reversible reaction $A+B
 \leftrightarrow C+D$ is   repeated
a large number  of times $n$  in a finite total  time and when at least 
one of the 
reacting particles 
was in the  ground state so that its  $c$-numbers 
 are zeroes  one obtains, in the limit of very large $n$,   a result as if 
 the repeated reaction  did  not occur at all.    \par 
 It can be
shown that the general reversible reaction 
 $A_1+A_2+\cdots A_r \leftrightarrow B_1+B_2+\cdots +B_s$, 
where $r$, $s$ are any two 
positive natural numbers,  also
results in a similar outcome if at least one of
the reacting particles has zero $c$ numbers. We note that  the last  
condition  is not necessary when we begin with only one reacting particle
as we see from the reaction $A \leftrightarrow B$. 
\protect \section{ The probability to find given final configuration  
 different from the initial one \label{sec2}} \noindent 
 We now discuss the more general and natural case in which we 
have an ensemble of particles and we 
calculate the
probability to find at the time $t$ a subsystem of this ensemble 
at some given
configuration if at the initial time $t_0$ it was at another given configuration.  We 
assume that the corresponding time difference  
$(t-t_0)$ is not
infinitesimal and that during this time  the subsystem has undergone a series of 
reactions.  The passage from some reaction at some intermediate 
time $t_i$ to the neighbouring one at the time $(t_i+\delta t)$ 
is governed by the correlation  between the corresponding resulting configurations 
of the subsystem at these times. 
Thus, restricting, for the moment, our
attention to the case in which a particle  in the subsystem that was  
 at the time $t_0$ in  the state $A$ and  at the time $(t_0+\delta t)$ 
 in    $B$   we can write    the relevant correlation \cite{Klauder} 
 between these two states as 
\begin{equation} \label{e3.15} \tau(A,B;t_{0_A},(t_0+\delta t)_B)=
<\!q_{A_{t_0}},p_{A_{t_0}}|q_{B_{t_0+\delta t}},p_{B_{t_0+\delta t}}\!>,
  \end{equation}  where  $|q_{A_{t_0}},p_{A_{t_0}}\!>$  and 
  $|q_{B_{t_0+\delta t}},p_{B_{t_0+\delta t}}\!>$
  are the coherent states that represent the particles A 
and B 
at the times $t_0$ and  $t_0+\delta t$ respectively 
(see Eqs (\ref{e3.3})-(\ref{e3.4})) 
 and the angular brackets denote 
an   ensemble average over all the particles of it. We note
that if $A=B$, $\tau$ measures \cite{Klauder} the autocorrelation of
either the particle A or B, and when $A \ne B$  $\tau$ is the
crosscorelation \cite{Klauder} of the two particles. It can be shown,
using Eqs (\ref{e3.1}) and (\ref{e3.15}) that the following relation  
\begin{equation} |\tau(A,B;t_{0_A},(t_0+\delta t)_B)|^2=\tau(A,A;t_{0_A},t_{0_A})
\tau(B,B;(t_0+\delta t)_B,(t_0+\delta t)_B),  \label{e3.16} \end{equation} 
is valid. That is, the modulus of the crosscorrelation of the particles A and B 
 at the times $t_0$ and $(t_0+\delta t)$  equals the product of the
 autocorrelation of the particle $A$ at the time $t_0$ by that  
 of  $B$ at the time $(t_0+\delta t)$. The last relation is interpreted
 \cite{Klauder} as the probability density for the occurence of 
 the reaction  $A \to B$ at the time
 $(t_0+\delta t)$. That is, given that the system was in 
 ``state'' $A$ at the time $t_0$, the probability to find it at the time 
 $(t_0+\delta t)$ 
 in  ``state'' $B$ is given by Eq (\ref{e3.16}). We can generalize to the joint
 probability density for the occurence of $n$ different reactions  between the 
 initial  and  final times $t_0$ and   $t$, where each   involves  
  two different particles and is of the kind $A_i \to  A_{i+1}$. That is, each
  reaction is composed of two parts; the first one is that in which a particle of the
  subsystem  is observed at the time $t_i$ to be in state $A_i$, and the second
  that in
  which it is observed at the time $t_i+\delta t$ to be  in  the state $A_{i+1}$.
  Thus, the total time interval $(t-t_0)$ is partitioned into $2n$ subintervals
  during  which the $n$ reactions occur.    The total correlation  is  
  \begin{eqnarray} && |\tau(A_1,A_2,\cdots,A_{2n};t_0,t_0+\delta t,\cdots,t)|^2=
 \tau(A_1,A_1;t_0,t_0)\cdot \nonumber \\ && \cdot 
 \tau(A_2,A_2;t_0+\delta t,t_0+\delta t)\cdots 
   \tau(A_{2n},A_{2n};t,t)= \label{e3.17} \\&& =
 \prod_{k=0}^{k=2n-1}\tau(A_{k+1},A_{k+1};t_0+k\delta t,t_0+k\delta t)
 = \prod_{k=0}^{k=n-1}\tau(A_{2k+1},A_{2k+1};t_0+\nonumber \\ && +
 2k\delta t,t_0+2k\delta t)
 \cdot \tau(A_{2k+2},A_{2k+2};t_0+(2k+1)\delta t,t_0+(2k+1)\delta t)= \nonumber \\ 
&& =\prod_{k=0}^{k=n-1}
|\tau(A_{2k+1},A_{2k+2};t_0+2k\delta t,t_0+(2k+1)\delta t)|^2 \nonumber 
 \end{eqnarray}
The last result was obtained by using 
Eq (\ref{e3.16}).  By the
notation $A_{2n}$ we mean, as remarked, 
that there are $n$ separate reactions that each involves two states 
(and not $2n$ different particles).   Now, it has
been established in  the previous section,  for either direction of the
reversible reaction
$A \leftrightarrow B$,  that the probability  
to remain in the initial  state $A$  (or $B$) 
tends to unity in the limit of a very large number of repetitions, 
in a finite total time, 
of $A \to B$ (or $B \to A$) 
which amounts to performing each such reaction in an infinitesimal
time $\delta t$.  That is, in this limit of vanishing $\delta t$  
each factor of the last product in 
Eq (\ref{e3.17}),  
which is  the probability for the reaction  $A_i \to  A_{i+1}$, 
tends 
to unity and with it the joint probability for the occurence of the $n$
reactions. Thus,     
     the specific
prescribed path of reactions  is followed through all of them  with a probability
of unity. 
\par 
From the last discussion we  may obtain the joint probability density for the
case in which some of the $n$ intermediate reactions may be of the more general
kind $A_1+A_2+\cdots + A_r \to B_1+B_2+ \cdots +B_r$, where $r$ is an arbitrary
natural positive number. That is, at some of the $2n$  times  there may occur, 
in a simultaneous manner, $r$ different reactions at
$r$ different places each of the kind $A \to B$. Thus, we assume 
 that
$r$ particles in the subsystem  that were  at the time $t_0+i\delta t$ in  the
states $A_j$ ($j=1, 3, 5,\cdots,2r-1$) were  observed at 
the time $(t_0+(i+1)\delta t)$ to
be in  the states $A_{j+1}$ ($j+1=2, 4, 6,\cdots,2r$).  
 We assume that at each of 
the other
intermediate times there happen only one 
 single reaction $A_i \to A_{i+1}$. 
Thus, there
are $(n+r-1)$ reactions each of them occurs  between two particles. In this
case the corresponding total coherence among all these reactions is \begin{eqnarray*}
&&\tau_{total}=\tau(A_1,A_2,\cdots,A_{i+1},A_{i+2},\cdots,A_{i+2r},\cdots,A_{(2n+2r-2)};t_0,t_0+\delta
t,\cdots \nonumber \\&& \cdots,\underbrace{t_0+i\delta t,t_0+(i+1)\delta t,\cdots,
t_0+i\delta t,t_0+(i+1)\delta t}_r,\cdots,t),  \nonumber \end{eqnarray*} 
where the underbrace denotes that the $r$ particles observed at the time
$(t_0+i\delta t)$ as 
$A_j$ ($j=1, 3,\cdots,2r-1$) were seen to be at the time $t_0+(i+1)\delta t$ as 
$A_k$ ($k=2, 4,\cdots,2r$) (in the former equation we use the index $i$ for $A$).  
Again the notation $A_{(2n+2r-2)}$ means that we have
$(n+r-1)$ reactions each involving, as remarked, two states. 
Using Eqs (\ref{e3.16})-(\ref{e3.17}) and the former equation for $\tau_{total}$ we 
find that the joint probability 
to find   at the time $t$  
the relevant subsystem at the given final configuration after the occurence of 
 these  $(n+r-1)$
reactions 
   is given by  
  \begin{eqnarray} && |\tau_{total}|^2=\tau(A_1,A_1;t_0,t_0)\tau(A_2,A_2;t_0+  
\delta t,t_0+\delta
t) \cdots 
\tau(A_{i+1},A_{i+1};t_0+\nonumber \\ && +i\delta t,t_0+i\delta t) 
 \cdot \tau(A_{i+2},A_{i+2};
t_0+(i+1)\delta t,t_0+(i+1)\delta t)\cdots  \nonumber \\ && \cdots 
 \tau(A_{i+2r-1},A_{i+2r-1};t_0+i\delta t,t_0+i\delta t)  
 \cdot \tau(A_{i+2r},A_{i+2r};
t_0+(i+1)\delta t,t_0+\nonumber \\ && + (i+1)\delta t) \cdots
 \tau(A_{(2n+2r-2)},A_{(2n+2r-2)},t_0+
 2n\delta t,t_0+2n\delta t)= \nonumber \\ && =  
\prod_{k=0}^{k=n-1}\tau(A_{2k+1},A_{2k+1};t_0+2k\delta t,t_0+2k\delta t)
\tau(A_{2k+2},A_{2k+2};t_0+\nonumber \\ && +(2k+1)\delta t,t_0+
(2k+1)\delta t) \cdot 
\prod_{j=1}^{r-1}\tau(A_{i+j},A_{i+j};t_0+i\delta t,t_0+i\delta t) 
\cdot  \label{e3.18} \\ && \cdot \tau(A_{i+j+1},A_{i+j+1};t_0+(i+1)\delta t,t_0+
(i+1)\delta t)= 
\prod_{k=0}^{k=n-1}
|\tau(A_{2k+1},A_{2k+2};t_0+\nonumber \\ && + 2k\delta t,t_0+(2k+1)\delta t)|^2  
 \prod_{j=1}^{j=r-1}|\tau(A_{i+j},A_{i+j+1};t_0 +i\delta t,t_0+(i+1)\delta t)|^2 
  \nonumber  \end{eqnarray}
   The first product of the last result is the same as that of Eq (\ref{e3.17}) and the second takes
 account of  $r-1$ simultaneous reactions at the time $(t_0+(i+1)\delta t)$ 
 (the first product involves also one of the $r$ simultaneous reactions at the  
 time $(t_0+(i+1)\delta t)$). Each of the reactions in both products is of the kind
 $A \to B$, and it has been shown in the former section (see also the discussion
 after Eq (\ref{e3.17})) that the probability to remain with the initial particle
 $A$ tends to unity in the limit in which the time duration of it becomes
 infinitesimal. That
 is, in this limit in which the time alloted for each reaction $A \to B$ 
 becomes very small  each  factor of each product of Eq (\ref{e3.18}), and
 with it the whole expression, tends to unity. If any particle $A$ of the 
 subsystem 
 does not react with any other particle at some of the $2n$ intermediate times
 then we may denote its no-reaction at these times as $A \to A$ and the
 probability for it to remain in the initial state (which is the same as the final
 one) is obviously unity.
   \par 
  Thus, we see that the probability to find at the time $t$ the given ensemble 
of particles at some prescribed configuration obtained after following 
a given path of reactions (from a large number of possible paths) 
tends to unity if the constituent reactions 
are performed in a dense manner. 
 \protect \section{  Billiard simulation of the reversible reaction $A+B
 \leftrightarrow A+C$ \label{sec3}} 
 \noindent
 We, now, numerically show 
  that this kind of dense reactions,  described  
 in the former sections,  may yield 
  similar results as the former  analytical ones.   
  This is shown for  the reversible reaction $A+B 
 \leftrightarrow A+C$. 
 We  simulate  this reaction by using 
  the two-dimensional circular billiard \cite{Bar10} which is
 composed of two concentric circles. We assume that  
  initialy we have a large ensemble of
 identical point particles each of them is the component $A$ of the  
 reaction. All of   these particles are  entered, one at a time, 
   into the billiard  in which they are elastically reflected by the two
   concentric circles. That is, the angles before and after each reflection are
   equal. We assume that on the outer circle there is a narrow hole through
   which the particles $A$ leave the billiard. Once a particle is ejected
   out a new one is entered and reflected inside the billiard
   untill it leaves and so  for all the particles of the ensemble. There are
   two different possible kinds of motion for each point particle $A$ before 
   leaving the
   billiard; either it is
   reflected between the two cocentric circles or, when the angle of reflection
   is large, reflected by the outer circle only without touching the inner one.
   Now, since both motions  are elastic each particle $A$,  once it 
   begins its reflections  in
   either kind of motion,    continues to move   only in this kind     
    untill it leaves through the narrow opening. The
   component $B$ of the reaction denotes the outer larger circle, and the
   component $C$ denotes both circles. That is, the left  hand side  $A+B$ of the 
   reaction  
   signifies that the point particle $A$ moves inside the billiard and is 
   reflected by the outer circle only, whereas,  the right  hand side  of it 
   $A+C$ denotes the second kind of motion in which the point balls $A$ are reflected
   between the two circles.  We call these two kinds of paths ``states''
   \cite{Bar10}, so that the path that
   touchs both circles is ``state'' 1 and the one that thouchs the outer circle only
   is ``state'' 2. This billiard model was studied in \cite{Bar10} as an
   example of a classical system that behaves  the same way   
    quantum systems  do  when exposed to a large number of repetitions, in a
   finite total time, of the same experiment 
   \cite{Zeno,Itano,Aharonov,Facchi,Simonius}. The concentric billiard is,
   schematically, represented in Figure 3.1. 
   
   \begin{figure}
\begin{minipage}{.48\linewidth}
\centering\epsfig{figure=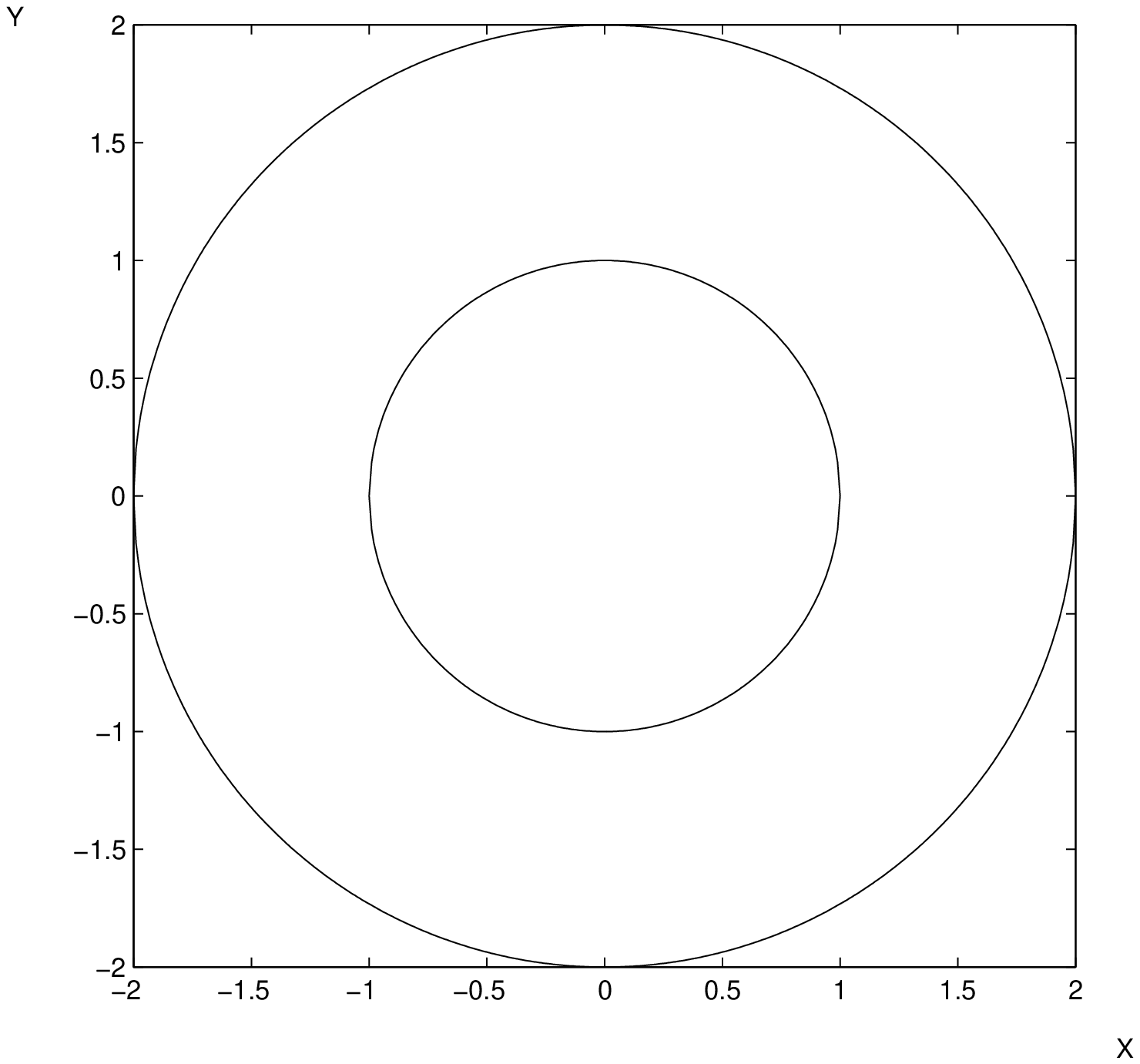,width=\linewidth}
\caption[fegf114]{A schematic representation of the concentric circular 
billiard that simulates the reversible reaction $A+B \leftrightarrow C+D$. A
narrow opening is assumed to exist on the outside larger circle through which
the particles leave the billiard.  }
\end{minipage} \hfill
\begin{minipage}{.48\linewidth}
\centering\epsfig{figure=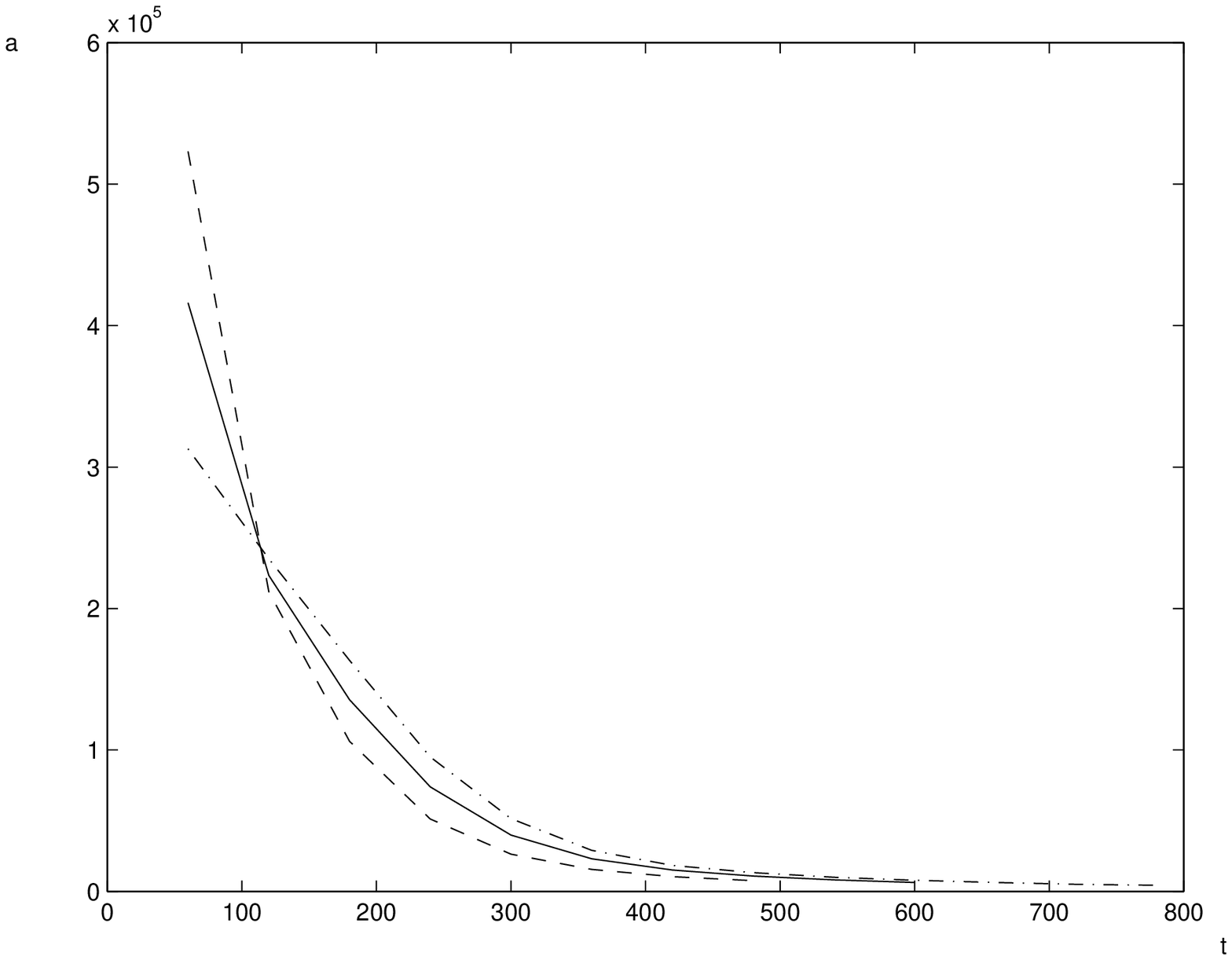,width=\linewidth}
\caption[fegf004]{The   dashed   curve shows the activity obtained when all
the $10^6$  particles of the ensemble are allowed to be only in ``state'' 2,  
in which  they
are reflected  between points on the outer circle only. The dashdot curve is the
activity    when all these  particles  are allowed to move only in
``state''  1 (between the two circles). For the values
assigned here to the outer and inner circles (6 and 3) the dashed (dashdot) curve
is the maximum (minimum) activity. 
The solid curve shows the activity obtained when the particles in either state
pass to the other after every 1100 reflections. The $x$ axis specifies time
binned in units of 60. }
\end{minipage}
\end{figure}
       Now, since in such a system we 
   can not
   follow the path of each particle and can not differentiate  between the two kinds 
   of motion  we have to consider, as done for the nuclear and radioactive 
   processes \cite{Bauer},
    the activities of these particles in either path.  
     That is, the rate at which
    the entire ensemble of particles, being at either  state,   
     leaves the
    billiard. We assume for the activity discussed here, as is assumed
    \cite{Bauer} for the nuclear and radioactive's activities, that each particle $A$
    enjoys arbitrary initial conditions, so in the following numerical
    simulations we assume that it may begin its journey inside the billiard at
    either ``state'' which is determined  randomly using a random number 
    generator. 
    As remarked, we want to show, 
    numerically,  that if either side of this reversible  reaction $A+B
    \leftrightarrow A+C$ is repeated a large number
     of times $N$ in a finite total time $T$, then,  in the limit of very large
    $N$,  the activity obtained is the same as the ``natural activity'' 
    \cite{Bar10} that  results
    when no such  repetitions are done \cite{Bar10}. 
       For that matter, we 
    take into account that the reversible reactions that  occur 
    in nature have either equal or different rates for  the two
    directions of the reactions and that the total activity of such ensemble 
     in which these reactions happen depend critically upon these
    rates \cite{Bauer}. 
     If, for example, we consider the equal rate case 
     then we have to discuss   the rate of evacuation
    of the billiard when each particle  
    is allowed,  after a  prefixed number of reflections in either state,  
     to pass, if it is still in the billiard,
    to  the other one.  This  activity is shown by the solid curve in 
      Figure 3.2 in which the
    ordinate axis denotes the number of particles $A$ that leave the billiard in
    prescribed time intervals binned in units of 60 \cite{Bauer}. We assume 
    \cite{Bauer} that
    each point particle $A$ in either state moves with the same speed of 3, and
    the hole through which they leave has a width of 0.15. We denote the outer
    and inner radii of the billiard by $r_1$ and $r_2$ respectively, and 
    assign them the values of $r_1=6$ and $r_2=3$. The initial number of the
    particles $A$ was $10^6$, and each one of them passes from one ``state'' to the
    other, if it did not leave the billiard through the hole, after every 1100 
    consecutive reflections. We note that this rate of one passage for every
    1100 reflections is typical and common for these kinds of billiard
    simulations \cite{Bauer,Gutkin}.  
    The  natural activity is obtained,  as remarked,   when the entire ensemble 
    of $10^6$ particles
    $A$ enter, one at a time, the billiard at the same definite ``state'' and
    remain all the time in this ``state'' without passing to the other  until
    they leave the  billiard.  The dashed curve in Figure 3.2 shows this natural 
    activity  when  all the particles $A$ are in ``state''  2 in which they are 
    reflected only between points of the outer circle untill they leave the
    billiard. The dash-dot curve shows  the activity when all the particles $A$ 
    are in ``state'' 1  in
    which  they  are reflected only between the two circles. 
      It has been found 
    that for the values assigned here to the radii of the outer and inner
    circles (6 and 3) the activity of ``state'' 2 shown by the dashed curve is 
    the maximum
    avilable and that of state 1 shown by the dashdot curve is the minimum. The
    large difference between the two activities has its source in the range of
    the allowed angles of reflections which is much larger in state 2 than in 
    state 1. This is because the minimum trajectory between two neighbouring
    reflections in state 2, where the particles $A$ are reflected between points
    of the outer circle only, may be infinitesimal  compared to the
    corresponding trajectory in state 1 which is (we denote the trajectories 
    between
    neighbouring reflections in states 1 and 2 by $d_1$ and $d_2$ respectively)  
    $d_{1_{min}}=r_1-r_2$.    For
    the values assigned here to the radii $r_1$ and $r_2$ of the two concentric
    circles ($r_1=6$ and $r_2=3$) $d_{1_{min}}=3$. We note that the 
    maximum trajectory between two
    neighbouring reflections in state 2 is equal to the corresponding one in
    state 1, that is $$d_{1_{max}}=d_{2_{max}}=\sqrt{r_1^2+r_2^2}$$ 
    Thus, the particles in $A$ have  many  more possibilities to be reflected 
    to the
    hole and leave the billiard in state 2 than in state 1 
    and,  accordingly,   their activity is much 
    larger. The solid curve in Figure 3.2 is, as remarked, the activity
    obtained when the particles $A$ are transferred between the two states at
    the rate of one passage for every 1100 reflections and
    so, as expected, its activity  is between the two other activities 
    shown in Figure 3.2.  
    \par 
     We numerically  interfere  with the rate of the systematic passage of the 
     point
    particles $A$ between the two states such that this rate is accelerated. 
    It is found that the activity of the entire ensemble is directly (inversely)
    proportional to    the rate of the
    passage from state 1 (2) to state 2 (1) when the
    opposite passage from state 2 (1) to state 1 (2) remains at the  rate 
     of one  for every 1100
    reflections.  Thus, we have 
    found that when the particles in  state 1 (2) are transferred to  
    state 2 (1) 
     at the maximum rate of one passage after  each single reflection and the 
     particles in
      state 2 (1) are passed to the state 1 (2)   at the  
      rate  of one for every 1100
    reflections then the activity of the
     particles $A$ is maximal (minimal).  But as we have  remarked the
     maximal (minimal) activity is obtained only when each particle of 
     the entire ensemble is always in state 2 (1).  
     In other words, as we have  remarked,  a very large number of 
     repetitions of the left (right)  direction $A+B \to A+C$ ($A+C \to A+B$)  
     of the   reaction    where the   right (left) direction $A+C \to A+B$ 
     ($A+B \to A+C$)   occurs   every 1100 reflections,  yields a   result as  
     if the densely repeated  reaction never happened and the 
     activity obtained is  the natural  one in which no repetition is present. 
        The dashed curve in Figure 3.3, which is the same as the dashed one of
      Figure 3.2,  shows the activity obtained when all the $10^6$ particles $A$
      of the ensemble are allowed to move only in state 2 until they leave the
      billiard. The solid curve is the activity obtained when the reaction 
      $A+C \to A+B$ is repeated after each single reflection and the opposite
      one   $A+B \to A+C$ after every 1100 reflections. 
      
      \begin{figure}
\begin{minipage}{.48\linewidth}
\centering\epsfig{figure=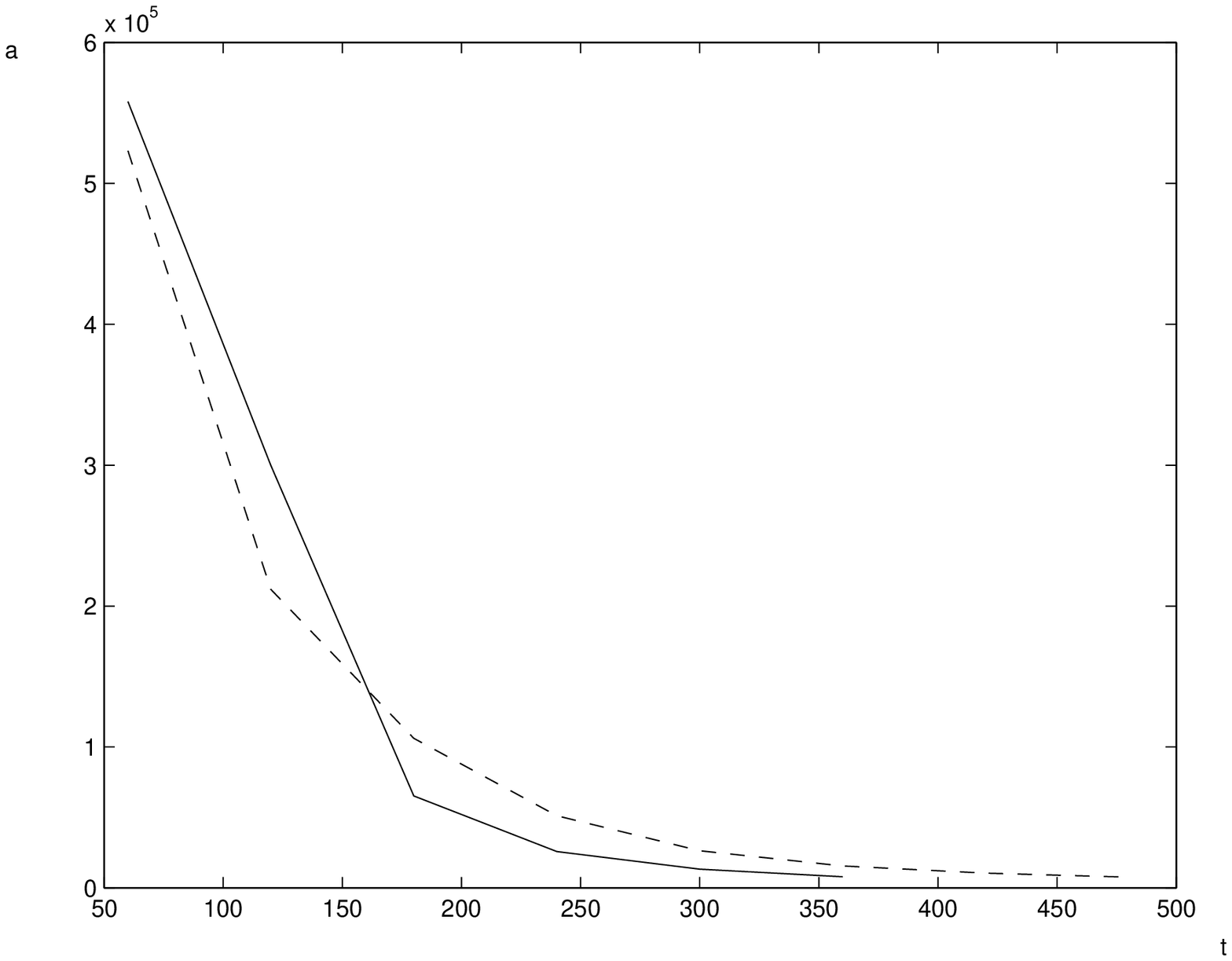,width=\linewidth}
\caption[fegf14]{The  dashed  curve, which is the same as the dashed curve of
Figure 3.2 (they look slightly different since the abcissa axes of these figures
are different),  shows the activity obtained when all the
$10^6$ particles
$A$ of the ensemble are numerically constrained to be only  in ``state'' 2 
until they evacuate the billiard. State 1 is not allowed for them. The solid
curve is the  activity obtained when each particle in state 1 is passed 
 to  ``state'' 2  after each single reflection, whereas 
those in ``state'' 2  pass to the opposite one only after every 1100
reflections.  As for Figure 3.2 the abcissa axis denotes time binned in units of
60.  Note the similarity between the two curves.   }
\end{minipage} \hfill
\begin{minipage}{.48\linewidth}
\centering\epsfig{figure=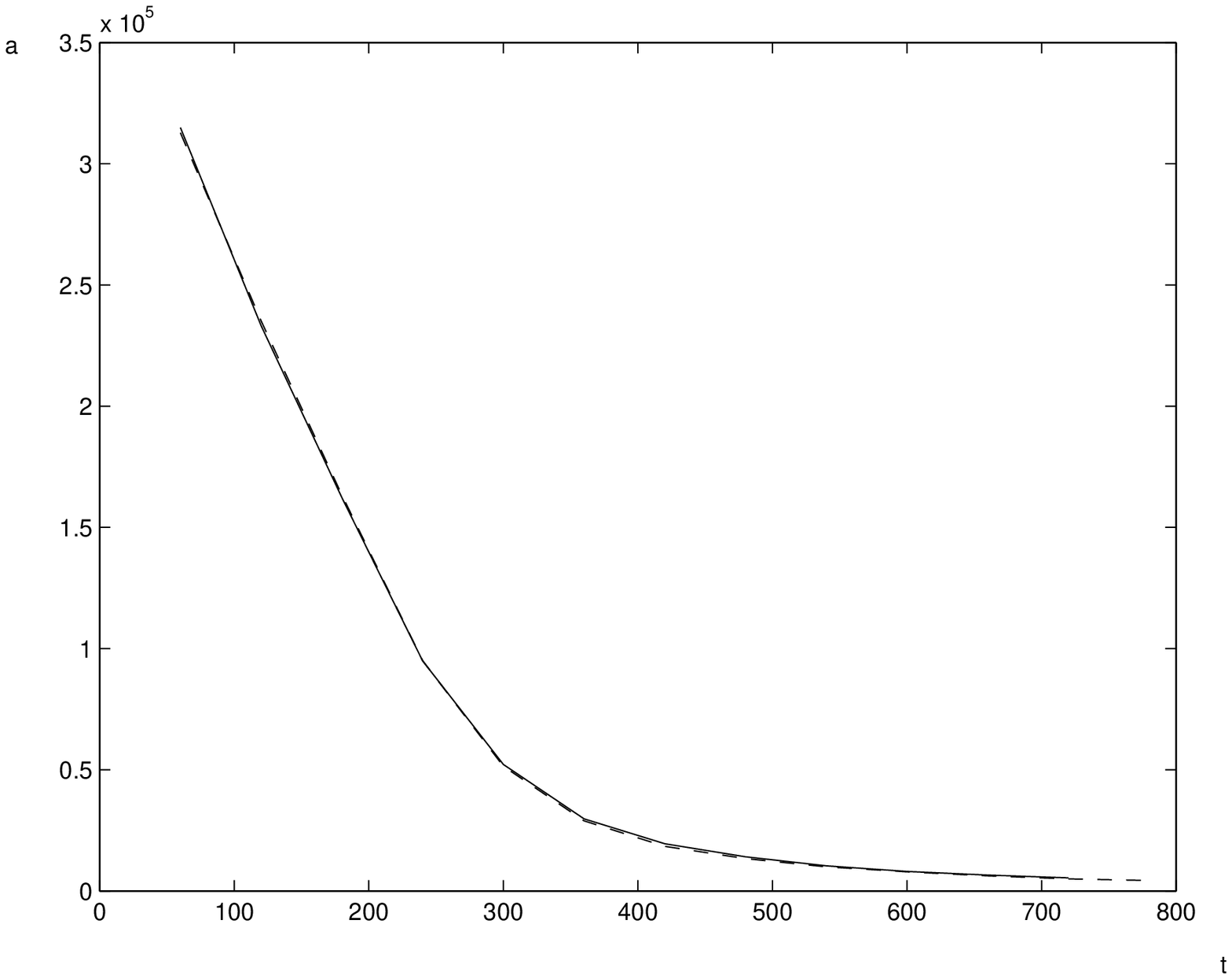,width=\linewidth}
\caption[fegf04]{The apparently one curve shown in the figure is actually two
curves one dashed and the other solid. The  dashed  curve, which is the same 
as the dashdot curve of
Figure 3.2,  shows the activity obtained when all the
$10^6$ particles
$A$ of the ensemble are numerically constrained to be only  in ``state'' 1 
until they evacuate the billiard. State 2 is not allowed for them. The solid
curve is the  activity obtained when each particle in state 2 is passed 
 to  ``state'' 1  after each single reflection, whereas 
those in ``state'' 1  pass to the opposite one only after every 1100
reflections.  As for Figure 3.2 the abcissa axis denotes time binned in units of
60.  Note that the two curves are almost identical (the dashed curve has a
longer tail (for large $t$) than the solid one).}
\end{minipage}
\end{figure}

       It is seen that the  curves of Figure 3.3  
      are  similar to each other.  That is,  the results obtained are 
      in accordance
     with the former sections where  a large number of repetitions of the 
     reaction 
     yields a result that characterizes the activity obtained in the absence of
     such repetitions. This is seen, in a much more clear way, in 
     Figure 3.4 for 
     the
     other direction $A+B \to A+C$  of the reaction. The apparent  single graph
     of the figure is actually composed of two curves; one solid and the other
     dashed. The solid curve  shows the activity obtained
     when the reaction $A+B \to A+C$ is repeated after each single 
     reflection and  the
     opposite one $A+C\to A+B$ after 
     every 1100 reflections. The dashed curve,  which is identical to the 
     dash-dot
     one from Figure 3.2,  is the activity  obtained when all the particles $A$ of
     the ensemble are constrained to move only in state 1 until they leave the
     billiard. Note that the two curves are almost the same except for the
     longer tail of the dashed curve. \par
        From both Figures 3.3 and 3.4 we realize that 
       the large number of repetitions of either direction  of the reversible 
      reaction $A+B \leftrightarrow A+C$ has the effect as if it has not 
      been performed at all and
     the actual activity obtained is  that of the natural one that does not
     involve any repetitions. \par
      We note that as the analytical results are obtained in the limit 
      of the  {\it largest}  number (actually infinite) of repetitions so the 
      similar 
      numerical results are obtained in 
       the limit of the largest  number of repetitions of
      the reaction.  That is, of numerically repeating it after each single
      reflection. In other words, a  mere high rate (which is not the maximal)
       of one side of the reaction compared to the slow one  is not enough to
      produce the results shown in Figures 3.3-3.4. This is clearly shown by the
      solid curve in 
      Figure
      3.5 which shows the activity obtained 
       when each particle in ``state'' 1 is passed to ``state'' 2 after every two
      consecutive reflections (the high frequency reaction)  
      whereas those of ``state'' 2 are passed 
      (one at a
      time) after every 1100 reflections (the low frequency reaction).
      
     \begin{figure}  
    \begin{minipage}{.60\linewidth}
\centering\epsfig{figure=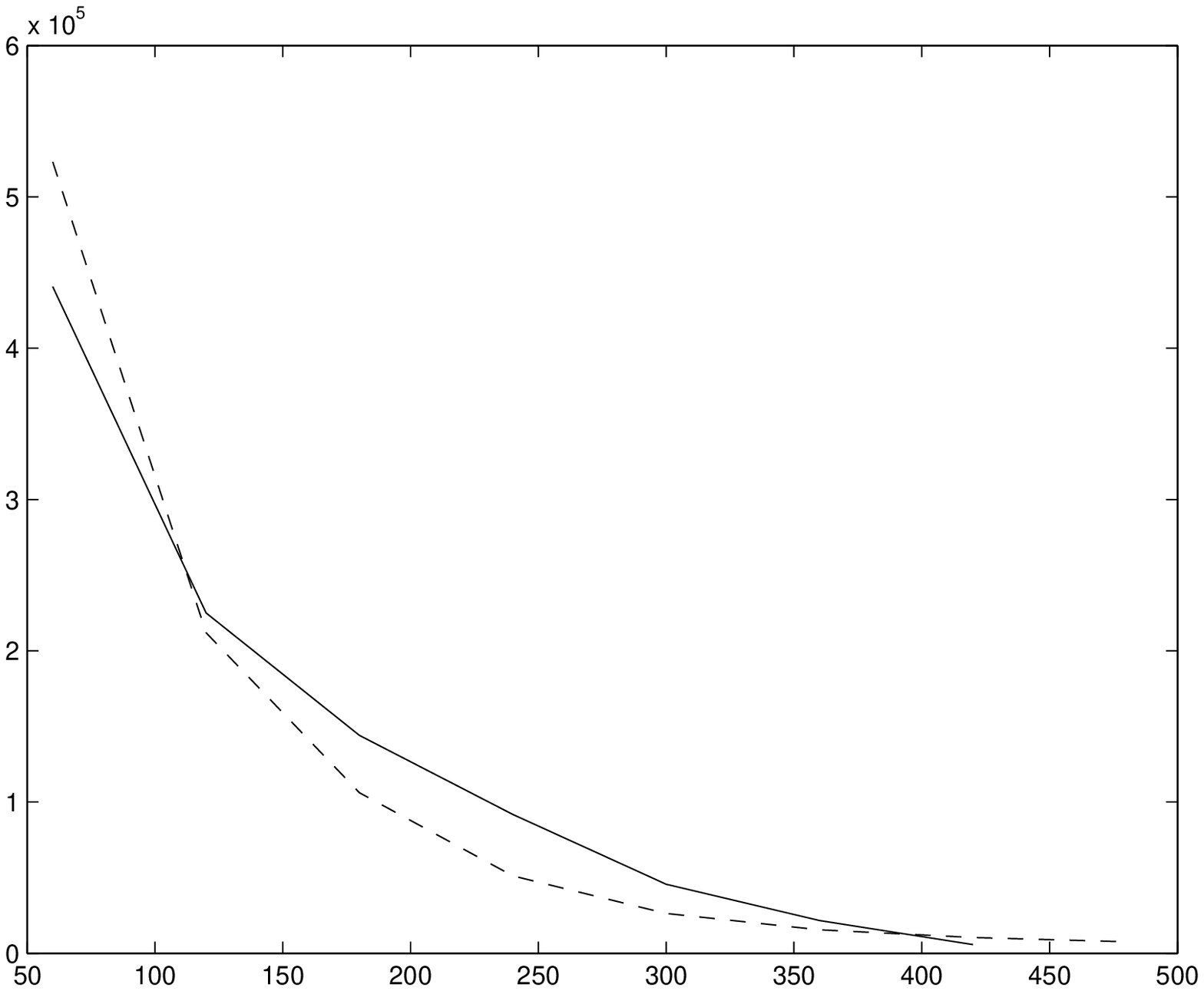,width=\linewidth}
\caption[fegf05]{The  dashed  curve, which is the same as the dashed curve of
Figure 3.3,    shows the activity obtained when all the
$10^6$ particles
$A$ of the ensemble are moving inside the billiard  only  in ``state'' 2 
until they are evacuated outside of it. State 1 is not allowed for them. 
The solid
curve is the  activity obtained when each particle in state 1 is passed 
 to  ``state'' 2  after every  two reflection, whereas 
those in ``state'' 2  pass to the opposite one only after every 1100
reflections.  Note that although the two solid curves of Figures 3.3 and 3.5 
are
obtained under almost the same conditions the activities are very different (see
text). 
As for all the former figures  the abcissa axis denotes time binned in units of
60.}
\end{minipage}
\end{figure}

      Note that the solid curve in 
      Figure 3.3
      shows the activity obtained when the particles in ``state'' 1 are passed
      to ``state'' 2 after each reflection and those of 2 passed to 1 after
      every 1100 reflections. That is, although the two high rates represented
      by  the two solid curves in 
      Figures 3.3 and 3.5 are almost the same nevertheless the resulting activities,
      contrary to what one may expect, are very different. That is, that of
      Figure 3.3 is much higher than that of Figure 3.5 as may be seen from the
      solid curve   that begins at $t=60$ (note that our  abcissa axis
      is binned in units of 60) from the high 
      value of $5.65 \cdot 10^5$ and ends at $t=360$. The corresponding solid
      curve of Figure 3.5 begins at $t=60$ at the much smaller value of $4.45
      \cdot 10^5$ and ends at the later time of $t=420$. That is, by only
      increasing the rate of repeating the same reaction from one for every two 
      reflections to one for each reflection results in an additional 120000
      particles that leave the billiard already at the first binned time unit. 
        The two dashed curves
       of Figures 3.3 and 3.5 are identical  and denote the same activity obtained
        when all the $10^6$ particles
$A$ of the ensemble are numerically constrained to be only  in ``state'' 2 
until they evacuate the billiard.  
 Thus, as remarked,  the important factor
that causes a result of maximum activity is the highest possible 
rate and not merely a large ratio between the higher and slower frequencies.       
      This is in accord with the analytical results obtained in Sections 3.2 
      and 3.3 
      in which the largest rate (actually infinite) 
       of repeating the same direction of the
      general reversible reaction $A_1+A_2+\cdots A_r \leftrightarrow 
      B_1+B_2+\cdots +B_s$, 
      where $r$, $s$ are any two arbitrary natural positive numbers,  yields the 
      results of remaining with a unity probability with 
      the initial reacting particles as if the  repeated reaction did not occur
      at all. \par  
      All the former simulations were done when the outer and inner 
     circles radii were 6 and 3 respectively. We note that we obtain similar
     numerical results for all other assigned values of  $r_1$ and $r_2$ up to 
     the extreme limits of $r_1>\!>r_2$ and $r_1 \approx r_2$ provided we always
     have $r_1>r_2$.  \par
      These results may be explained along the same line  used to
   interpret the similar results obtained analytically
   \cite{Zeno,Aharonov,Facchi,Harris,Simonius} and experimentally \cite{Itano} 
    in the quantum regime.  That is, a  quantum  
      system,  which may reduce  through
    experiment 
    to any of its relevant  
   eigenstates,  
     is preserved in its initial state by  repeating  the experiment of checking
     its state    
     a large number  of times in a finite total time which is  
    the static Zeno effect
   \cite{Zeno,Itano,Aharonov,Facchi,Harris,Simonius}. The similar results obtained
   theoretically in Section 3.2 suggest that this
    effect may be effective also in the classical reactions. That is, repeating
    them a   large number of times, in a finite total time,  may result  in 
    remaining with
    the initial reacting particles  as if the repeated 
    reaction did not happen at all.
    Moreover, the dynamic Zeno  effect \cite{Aharonov,Facchi} may also be
    obtained as 
    in Section 3.3, in which we show that the joint probability density for the 
    occurence of $n$ special different  reactions between the initial 
    and final
    times $t_0$ and $t$  tends to unity in the limit of $n \to \infty$. 
      The Zeno effect has been  shown also in the numerical
    simulations from which we realize that repeating a large number of times any
    direction of the reversible reaction $A+B \leftrightarrow A+C$ has, in the
    limit of numerically repeating it   after each single reflection,   
    the effect as if it has never happened and the activity obtained is the 
    natural one in which no repetitions occur.  That is, the very large
    number of repetitions, in a finite total time,  causes the resulting activity to be the same as if
    these repetitions never happened as obtained in the Zeno effect in which 
    the
    system is preserved in the initial state due    to the very
    large number of measurements.  \par 
We see, therefore,   that the large number of
repetitions of either the same reaction (corresponds to the static Zeno effect),
or along a consecutive sequence of different ones (corresponds to the dynamic 
 Zeno effect) causes the relevant system, as seen   in  the former chapters,  
 to respond differently compared to its 
 response in the absence of these repetitions. That is, referring to the first
 case we see that although the related reaction is done a very large number of
 times one remains with  the initial reacting particles  only as if no 
  reaction has ever been done. For the second case the new response, as a 
  result
  of these dense reactions along the specific path of reactions, is to
  ``realize'' this path so that the probability to proceed along these reactions
  is unity.    That is, we see for both
  cases that the mere act of repetitions of the kind involved changes the
  behaviour of the system to an entirely new and unexpected one. Moreover, we
  can at any time reconstruct this unexpected response of the system by
  going once more through the dense repetitions process. That is, under the
  condition of dense measurement we may establish and validate these  
   new responses  of the system. Note that exactly the same results were 
   obtained
   in Chapter 2 with respect to the quantum field examples discussed there. 
   We will
   also obtain  the same results in the following Chapters  of this work. 
These  unique effects of the dense repetitions  have 
 also been, numerically, shown using the
circular billiard model \cite{Bar10}.



 \pagestyle{myheadings}
\markright{CHAPTER 4. \ \ \ SPACE ZENO EFFECT}






 \newpage

 \noindent  
\chapter{\label{chap4}Space Zeno effect}

\protect \section{Introduction  \label{intr}}
\smallskip \noindent  The effect  of performing the same 
experiment simultaneously in a very large 
number of   regions of space  all 
occupying a finite space is similar to that of performing an experiment
repetitively a large number of times in a finite interval of time.  The 
difference is that the repetition in the second case is over independent units of 
equal steps in  time, while in the first case it is over independent 
units of equal shifts in space.  We have shown \cite{Bar3} that as the Zeno 
effect \cite{Zeno,Itano,Aharonov,Facchi,Harris,Simonius} is 
 obtained in the second case when these 
equal intervals of time tend 
to infinitesimal values, so this effect occurs also in the first case when 
the equal shifts in space tend to be infinitesimal.  \par  
Piron  has discussed in \cite{Piron} a physical example of how this procedure 
can be seen as an actual evolution. He considers an array of Geiger counters 
 at each of a closely spaced set of points 
along the $x$ axis. This type of apparatus treats the value of $x$ at which an 
event occurs as a classical parameter, since the $x$ value of each counter is 
known in advance. What is unknown is the time $t$ at which the counter will 
trigger, and this $t$ then becomes a quantum observable. Passing from a counter 
 at $x$ to a counter at $(x+\rho)$ corresponds to 
a Hamiltonian type evolution $e^{-\frac{ip_1\rho}{\hbar}}$, generated by the evolution 
operator $p_1$, now a function on the phase space $(H,t,y,z,p_y,p_z)$. The 
survival amplitude is $(\phi_{x_0},e^{-ip_1\rho}\phi_{x_0})$, 
with $\phi_{x_0}(t,y,z)$ some ``initial'' state at position $x_0$. The successive 
performances of such experiments along the $x$ axis at small intervals 
$\frac{X}{n}$, as $n\to \infty$ corresponds precisely to the analogous process
of the time Zeno effect.  Since, in this limit, 
the state $\phi_{x_0}$ is stabilized (as for the time Zeno effect), the 
distribution in $t$, $y$, and $z$ becomes stationary, and we see that the 
effect is that of  essentially ``simultaneous'' (at the peak $t$) measurements 
over an interval of $x$. \par 
As pointed by Piron \cite{Piron} the two formulations,  1) according to the 
parameter 
$t$ (as in a bubble chamber type of experiment where $t$ is known, but the 
locations $x$, $y$, and $z$ are subject to measurement), and  2) according to 
the parameter $x$ (as in the set of Geiger counters described above, where the 
$x$ is determined, but the times $t$ at which the counters are triggering are 
the results of measurement) are classically completely equivalent, 
as can be seen 
by a change of variables. In the quantum case, the difference is profound, i.e, 
in the first case $t$ is an evolution parameter and $x$, $y$, $z$ are physical 
observables, while in the second case, $x$ is the parameter of evolution, and 
$t$, $y$, $z$ are the observables. Our comparison 
here of the two interpretations  
corresponds 
to a qualitative equivalence which carries over,  under suitable conditions, to 
the quantum theory.  \par       
In  Section 4.2 we show, using Appendix $C_1$,  
that  the suitable expression for  
 the space Zeno effect may be  obtained by exchanging  the time and space 
 variables and making the necessary changes.
      In  Section 4.3  we demonstrate,  by referring to 
  different 
articles \cite{Bar7,Bar8,Bar11,Bar111,Bar15},   the existence of the spatial  Zeno  
effect in
both quantum and classical systems and by this we  show that it is a general
effect and not only a quantum one.  We have just 
shown in Chapter 3 that it may be found also in chemical classical reactions.    

\bigskip  \noindent \protect \section{ A coherent and general state examples 
 of the space Zeno effect 
 \label{sec1}}
  \smallskip \noindent          
As seen from \cite{Bar3} one may obtain the relevant analytical expressions
for the space  Zeno effect by substituting the time variable $t$ by the space variable
$r$ and doing the resulting necessary changes. This may be seen from Eqs 
(1)-(9)   of \cite{Bar3}  ( Sections 1-3 of which  are   
shown  in Appendix $C_1$). These equations discuss the specific case of the
ground state of the harmonic oscillator 
$\phi (x)=({w\over \hbar \pi })^{1\over 4}e^{-{1\over 2\hbar }wx^2}$ and 
demonstrate the kind of transfer one do from the time variable to the 
spatial one.   The more general case is also discussed 
(see  Eqs (9)-(11)  in  \cite{Bar3}  and in  Appendix $C_1$)   
 and one may see
the appearance of the space Zeno effect in the limit of $n \to \infty$ 
where $n$
is the number of systems (observers) in the finite spatial section 
that perform these experiments. 
 \bigskip 
\section{\label{sec6} The one-dimensional arrays of the multibarrier potential
and multitrap systems of finite range}
\bigskip
We refer in the following to the articles \cite{Bar7,Bar8,Bar11,Bar111,Bar15}
which discuss the quantum and classical systems of the one dimensional arrays of
 multibarrier potential and multitraps systems of finite ranges. 
In  \cite{Bar8}   
 the quantum one-dimensional multibarrier potential of finite range is 
 discussed,  and in 
 \cite{Bar11}     the corresponding classical  multitrap system. 
  We may see that although these two systems are entirely
physically different and necessitate different methods of analysis as should 
be
for quantum and classical systems, nevertheless, they clearly show the same 
somewhat unexpected behaviour.  That is,  as the number of either the 
quantum potential
barriers or the classical traps increases the value of the respective 
 probability amplitude  
or  the classical density after crossing these systems  tend to remain in 
their former values
before  passing  them. This behaviour is   contrary to what one may expect 
that the values of the former variables must   
 decrease  with increasing the  number of barriers or traps. This is shown
 explicitly in Appendices  $C_2$ and $C_3$ in which we represent  respectively 
  the relevant 
 paper \cite{Bar8} and Sections 1-2 of \cite{Bar7} for the
 one-dimensional multibarrier potential of finite range and in Appendix $C_4$ 
  which
 shows the paper  \cite{Bar11} for the  corresponding    
 one-dimensional multitrap  system. \par 
 In the two  systems one sees the
 same correspondence not only  for a variable total length of each  
 \cite{Bar7,Bar11} but also for a constant  length.  
 Moreover, the same parameter
 $c=\frac{b}{a}$,  which denotes in both systems 
 \cite{Bar7,Bar8,Bar11,Bar111,Bar15}  
 the ratio of the total interval
 between the barriers or traps to their total length, has turns out to have the
 same influence in these two different systems.  Both papers \cite{Bar8,Bar11}
 use the transfer matrix  method \cite{Merzbacher,Yu} to obtain analytical expressions 
 for the relevant
 transmission probability or the classical density  in  the respective limits 
  of
 infinite number of barriers or
 traps which are the limits for which the space Zeno effect appears. 
  In Section 4.4 we  show that both systems behave  the same also for the case of
  finite number $N$ of potential barriers or traps and, especially, in the
  transfer to large $N$. For this we  use, for both
  systems, 
  numerical simulations in contrast to the explicit expression we have obtained
  fot the $N \to \infty$ case. In the following section we use the 
 numerical  
 $4n \times 4n$ matrix  approach which is applicable for both systems.  
   
 \section{The $4n \times 4n$ matrix approach}
We, now,  show that both one-dimensional systems  of quantum potential 
barriers and classical traps
demonstrate the same behaviour also for finite number of barriers and traps and
especially in the transfer to the  large  numbers of them.  
The quantum case has already been discussed  
  in \cite{Bar7} and the relevant Sections  
  are  appended to Appendix $C_3$. We, now,  show the same behaviour  also for the multitrap system by
 starting from Eqs (7)-(10) in \cite{Bar11} (appended to Appendix $C_4$) 
 which are valid
 for the first trap.  Continuing in the same manner 
for all the other $n-1$ traps we obtain $4n$ simultaneous equations with $4n+2$ 
 unknowns $A, B, C, \ldots $ etc.  Now,  since these $4n$  simultaneous 
 equations are 
 not sufficient to determine the $4n+2$ unknowns   we must reduce the number 
 of the unknown variables from $4n+2$ to $4n$ by using the following two
 assumptions: (1) Noting that  the traps act as imperfect ones 
 (see the original problem
 from the set (1) in \cite{Bar11}) we divide all the $4n$ equations by  
  the value of the  coefficient of the density function of the imperfect 
  problem  at the left hand side of the first trap.   
 That is, this is the value of the coefficient of the density of the diffusing
 particles before their interaction with the traps.  
  (2) The second assumption is that at the 
 point $x=a+b$ at the right hand side of all the $n$ traps we ignore the ideal trap 
 component of the density function since it is clear that the density of the 
 particles that survive the $n$ traps can not be represented by any ideal trap
 function. We remark that these two assumptions  of dividing by some nonzero unknown 
 and ignoring another  in order  to equate the number of 
 unknown variables to the number of equations are common  
 in  the literature (see, for example, the potential barrier problem in 
 \cite{Schiff} and \cite{Merzbacher}). These two assumptions reduce the number 
 of unknown variables to $4n$, and so they can be determined by the $4n$
 equations.    We  denote the ideal and imperfect trap components by
 $\rho_1(D,x,t)$, and  $\rho_2(D,x,t)$ respectively  
 where $D$ is either  the external  diffusion constant outside the traps 
 in which case it is
 denoted by  
 $D_e$ or inside them  whereby its notation is  $D_i$. The 
 notation $\grave \rho$ denotes the first derivative of the 
 density function $\rho$ with respect to the $x$  variable. 
  The $4n$ simultaneous equations system is, therefore,   given by  
 \begin{eqnarray}
 &&\rho_2(D_e,\frac{b}{n},t)= C\rho_2(D_i,\frac{b}{n},t) \nonumber \\
&&\grave \rho_2(D_e,\frac{b}{n},t)+B\grave \rho_1(D_e,\frac{b}{n},t)=
C\grave \rho_2(D_i,\frac{b}{n},t)+D\grave \rho_1(D_i,\frac{b}{n},t)
\nonumber \\
&&C\rho_2(D_i,\frac{a+b}{n},t)= E\rho_2(D_e,\frac{a+b}{n},t) \nonumber \\
&&C\grave \rho_2(D_i,\frac{a+b}{n},t)+D\grave \rho_1(D_i,\frac{a+b}{n},t)=
E\grave \rho_2(D_e,\frac{a+b}{n},t) + F\grave \rho_1(D_e,\frac{a+b}{n},t)
\nonumber \\
 &&...................................................................... \nonumber \\
 &&
.........................................................................\label{4.1}  \\
 && ......................................................................... \nonumber  \\
&&R\rho_2(D_e,\frac{(n-1)a+nb}{n},t)=T\rho_2(D_i,\frac{(n-1)a+nb}{n},t) \nonumber \\
&&R\grave \rho_2(D_e,\frac{(n-1)a+nb}{n},t)+S\grave
\rho_1(D_e,\frac{(n-1)a+nb}{n},t)=  \nonumber \\ && = 
T\grave \rho_2(D_i,\frac{(n-1)a+nb}{n},t) +U\grave \rho_1(D_i,\frac{(n-1)a+nb}{n},t)
\nonumber \\
&&T\rho_2(D_i,a+b,t)=V\rho_2(D_e,a+b,t) \nonumber \\
&&T\grave \rho_2(D_i,a+b,t)+U\grave \rho_1(D_i,a+b,t)=V\grave \rho_2(D_e,a+b,t)
\nonumber 
 \end{eqnarray}
 The last set (4.1) can be written as the matrix equation $$Nx=c,$$ 
 where $N$ is the square matrix with $4n$ rows and $4n$ columns,  $x$ is the 
 unknown vector with the $4n$ unknowns $(B, C, \cdots T, U, V)$ and  $c$ is the 
 constant vector whose first two elements are $-\rho_2(D_e,\frac{b}{n},t)$ and 
 $-\grave \rho_2(D_e,\frac{b}{n},t)$, and all its other $(4n-2)$ elements are
 zeroes. $a$ is the total width of the traps and $b$ is the total interval
 among them \cite{Bar11}. Note that the last set is very similar in form to the corresponding
 sets of the quantum case (see the sets  (1) and (3) in \cite{Bar7}) 
  and like them  can be managed, especially for large $n$, only by 
 numerical simulations.  We can compute, numerically,  each one of the $4n$ 
 variables not only for constant $a$, $b$, and $n$, but also can determine 
  how each of
 these variables behaves as a function of $a$ for constant $n$, or as a function
 of $n$ for constant $a$ (usualy $b=a$, or $b=\frac{a}{2}$). Typical values of 
 the diffusion constant are conventionaly found in the literature in the cgs 
 units and range from $0.1\frac{(cm)^2}{sec}$  to $0.8 \frac{(cm)^2}{sec}$ 
  \cite{Reif}.  
   At $t=0$  we find that $V$ (see the set (4.1)), which is 
  the  ratio of the imperfect trap density function coefficient at the point 
  $x=a+b$ 
  (after passing  the $n$ trap system) to that at the point $x=\frac{b}{n}$ 
(before passing  it) is  unity for all values of $n$ and
$a$.  We have obtained this result for $k=1$, $D_e=0.5$, $D_i=0.1$ 
and $b=\frac{a}{2}$.  
 That is, at the initial 
time the densities at the two extreme sides of the trap system are equal. When 
$t$ departs from zero  the coefficient  $V$ 
becomes smaller for the same values of $n$
and $a$. That is, as time progresses the density after the $n$ trap system
becomes smaller than that before it  as expected from 
a physical point
of view.  \par 
 The interesting and unexpected findings, obtained analytically and
numerically, are found when the number $n$ of either the potential barriers 
in the quantum
system  or traps in the classical one increases.  In this  case the respective 
probability amplitude or the 
 particle density {\it grows}  so that in the limit 
 $n \to \infty$ both variables tend to the initial value they have  before
 passing through the systems. That is, contrary to what one may expect the larger
 is the number of barriers or traps the easier will be for the entire 
 ensemble of particles (quantum or classical) 
 to pass through them.  That is, we see, as remarked, that experimenting
 uninterruptedly 
 on either a quantum or a classical system totally changes the known response of
 these systems to that of a new, even unexpected,  
  one. Moreover, as remarked in the former parts of this work with
 respect to the bubble and open-oyster examples (as well as the 
  programming  examples of \cite{Bar2,Bar001})  we may
 cause these systems to repeat, as many times as we wish,  their new and
 unexpected behaviours by experimenting again with  them 
 in a  dense  manner.   In other words, by this kind of dense 
 experimenting
 we may establish and validate these new and unexpected responses of these 
 systems.  Note that, as we have remarked in Section 1.2, we do not have to
 reach  the limit of $n \to \infty$ or even large $n$ in order to obtain again
 these new responses (new phenomena) of these systems. As seen in
 \cite{Bar7,Bar11,Bar111} (see Appendices $C_2$, $C_4$, $C_5$  and 
 the figures there) one does
 not have to take more  than 30 barriers (even less for the case  of traps 
 as seen in \cite{Bar111} (see Appendix $C_5$)) for obtaining a significant
 transmission of the respective quantum or classical particles. These large
 transmissions constitute, as remarked, the new phenomena that may be repeated
 by preparing again these 30 barriers. \par 
  The establishing effect that results from the space Zeno effect
 where a large number of similar systems, confined in a finite region of space, 
   perform  similar experiments turns our attention to directly consider the 
  effect of a
  large ensemble of {\it related} observers.   We do this in the following chapter.

\pagestyle{myheadings}
\markright{CHAPTER 5. \ \ \ THE EFFECT OF THE ENSEMBLE OF .....}

 \chapter{\label{Chap5}The effects of the  ensemble of 
 related  observers  in quantum  and classical phenomena}       
 \protect \section{Introduction }

  \footnote{This chapter
 was later published with some changes, after the thesis's submission, under 
 the title "The effect of
 related experiments" in IJTP, {\bf 44}, 1095-1116 (2005)}
 
In this chapter we discuss the influence of observation and observers upon the 
obtained experimental results and although the effect of these have been 
 discussed in the literature \cite{Adriana} (see also \cite{Wheeler} 
 and references therein)  a little was said regarding the influence 
upon the obtained results of the presence or absence of mutual relationship 
and cooperation among them. 
We concentrate our discussion here to   
  the effects that result
 when a large ensemble
of {\it related}  observers  perform 
experiments. By the term "related observers" we mean that the members of the
ensemble are related to each other by some kind of connection such as, for
example, that each one of them do his experiment on  the same kind of system as
all the others. The last condition ensures that any result obtained by any one
of them are valid for all the  others in the sense that if the relevant 
experiment 
is repeated, under the same conditions,  by any other member of the ensemble  
 the results obtained will
be the same. This will not be the case for the unrelated observers that do not
have the same kind of systems so they can not perform the same experiment 
under
the same conditions.  Thus,  the results obtained by any one of them  are 
valid only for the specific involved system and not for the other different 
ones.  \par 
We note that when the observers of the ensemble participate in some 
collective experiment then the relation among them is established not only 
by having the same kind of systems but also by the degree of their mutual 
cooperation in the experiment. Thus, they will be maximally correlated if 
they perform the collective experiment in such a way that not two of them 
prepare and do their specific parts independently of each other. This kind of 
connection among the members of the ensemble will be analytically  
 discussed in Sections 5.2 and 5.4.    
We show that  the  probability for the physical validation  of   real
phenomena   increases  for  large number    
 of related  observers  and   this is obtained for the whole 
  of
them  without having  each one  doing dense
measurement along the specific Feynman paths that represent these phenomena.
\par 
In Section 5.2 we use the Feynman path
integral method \cite{Feynman2,Schulman,Gert} to show this behaviour and 
in Section 5.3 use is made, for 
this purpose, of Everett's    
  relative state theory \cite{Everett,Graham}. We show, using the unique  
  character  of the last theory, that  the number of
observers increases in an asymptotic manner as   the number of experiments grows 
as  seen from the following Eq (\ref{e10}) and Table 1. For example, for  100
experiments we have over $10^{306}$ different observers.
Both theories,
together with Figures 5.1-5.3 and Table 1,  clearly demonstrate   that the
presence of a large ensemble of {\it related} observers  not only makes possible
the resulting physical validation  of  real phenomena but it is more 
enabled   the larger is the ensemble. 
  This
 does not hold for the ensemble of unrelated observers or for the single
one as will be shown.  Note that the appearance of physical reality for 
the large number of {\it identical}  systems all prepared in the same initial
states    have 
already been discussed in the literature \cite{Hartle,Finkelstein,Smolin} 
from different points of view. As remarked, we discuss also in this work 
 the case where the large ensemble of systems  was aligned in such a way  
  that  no two of them are prepared in  
the same initial state (see, especially, Sections 5.2 and 5.4). 
   \par  
In Section 5.4 we show,  using the classical thermodynamical
system of cylinder and  pistons \cite{Szilard,Reif},   
that the influence of the ensemble of related observers  
is effective  also in 
classical systems.   Moreover, since
this  system   is simple the effect of the
ensemble is shown in a more direct  manner. We follow the discussion in 
\cite{Szilard} 
and generalize it to include the large ensemble of related observers. That is, we test  
some assumed relation between the variables of the thermodynamical system 
and find  the conditions
under which this relation becomes established or refuted. We follow the whole
procedure in \cite{Szilard} of first attempting some initial relation so that if it turns out
through experiments 
  that it does not cover all the possible motions
of the particles in the cylinder a new and better relation is proposed. We show
that the new theory   becomes established only for a {\it related} large  
 ensemble of
observers  that  perform   similar experiments.  
It is shown   that the
  effect of such a {\it related} ensemble of observers is to physically
  establish  the new 
  assumed relation  compared to that of 
  an
  unrelated ensemble that have no effect  in this respect.  \par
  We note that this obtained effect of the ensemble of observers,   compared to
that of the single one, is in accordance with the space Zeno effect
discussed in the former chapters. That is, when the large ensemble of related
observers operate cooperatively any result  obtained  from any
experiment done by any observer of the related ensemble becomes valid 
 for the whole of them. This corresponds to the
preserving  of the {\it given} initial probability amplitude  or the
 particle density in passing respectively through either a large number
of a one-dimensional array of potential barriers or  traps.  This
 does not hold for the ensemble of unrelated observers or for the single
one which corresponds, as remarked, to the time Zeno effect and not to the spatial
one. Note that exactly the same results were obtained in \cite{Bar001}  
with respect to
the Internet websites where we see that if  the clusters of doubly
linked websites are large enough then adding only a small amount of connecting links results in, 
actually,  a phase transition-type  strengthening of the overall connectivity
among them. This  unusual   increase of the overall connectivity 
due to a small addition of it corresponds to the former overall physical
realization for the whole ensemble of the result obtained by only one of them.
\bigskip 

\protect \section{The Feynman path integrals of the ensemble of observers} 
The collective  measurement is 
performed by first preparing $N$ similar systems at   
$N$ arbitrarily selected states,  from actually the very large number 
 which constitute
the specific Feynman path of states which we want to ``realize'' 
\cite{Aharonov,Facchi} by  
 densely experimenting  
  along it.  These systems are then delivered  to the $N$ observers of the ensemble so 
 that  the  system  $i$  $(i=1, 2, \ldots N)$,  prepared at the state $\phi_i$,  
   is assigned to the observer $O_i$.   
   Thus,  we may write for the probability amplitude that the
first observer $O_1$ finds his system,   after doing the experiment of 
checking its present state, at 
the state $\phi_2$ of the second 
observer $O_2$ \begin{equation} \label{e1} \Phi_{12}=\sum_i\phi_{1i}\phi_{i2} 
\end{equation} 
The summation is over all the possible secondary  paths \cite{Bar0}  
(as those
shown along the middle   path of Figure 5.1)  
between $\phi_1$ and $\phi_2$ and the quantities $\phi_{1i}$ and $\phi_{i2}$
denote \cite{Feynman2} the probability amplitudes to proceed from the state $\phi_1$ 
to $\phi_i$ and from $\phi_i$ to $\phi_2$ respectively. In the same manner one may
write for the conditional probability amplitude that the second observer $O_2$
finds his system     at the
state $\phi_3$ (of the  observer $O_3$),   provided that
the observer $O_1$ finds  his system   at the state $\phi_2$  
\begin{equation} \label{e2} \Phi_{23|12}=\sum_{ij}\phi_{1i}\phi_{i2}
\phi_{2j}\phi_{j3} 
\end{equation} 
Where $\Phi_{23|12}$ is the remarked conditional probability amplitude and 
$\sum_{ij}$ is the summation over all the secondary  paths  that lead 
from the state $\phi_1$ to $\phi_2$ and over those  from $\phi_2$
to  $\phi_3$. Correspondingly, the conditional probability amplitude that the
$(N-1)$-th observer finds  his system  at the state $\phi_N$ of the observer 
$O_N$  
provided that all the former
$(N-2)$ observers find their respective systems,  that were initially
prepared at the states $\phi_i \ \ (i=1, 2, \ldots N-2)$,  to be at the   
states $\phi_{i} \ \ (i=2, 3, \ldots N-1)$ 
\begin{equation} \label{e3} \Phi_{N-1N|12,23,\ldots,N-2N-1}=
\sum_{ij\ldots rs}\phi_{1i}\phi_{i2}\phi_{2j}\phi_{j3}\ldots 
\phi_{N-2r}\phi_{rN-1}\phi_{N-1s}\phi_{sN}
\end{equation}
Figure 5.1 shows 7  Feynman paths, from actually a large number of paths that all
begin at $\phi_1$ and end at $\phi_8$  (only 8 states are shown in the figure for
clarity).  The middle path is the specific one
along which the described collective dense measurement is performed. 

 \begin{figure}
 \begin{minipage}{.78\linewidth} 
\centering\epsfig{figure=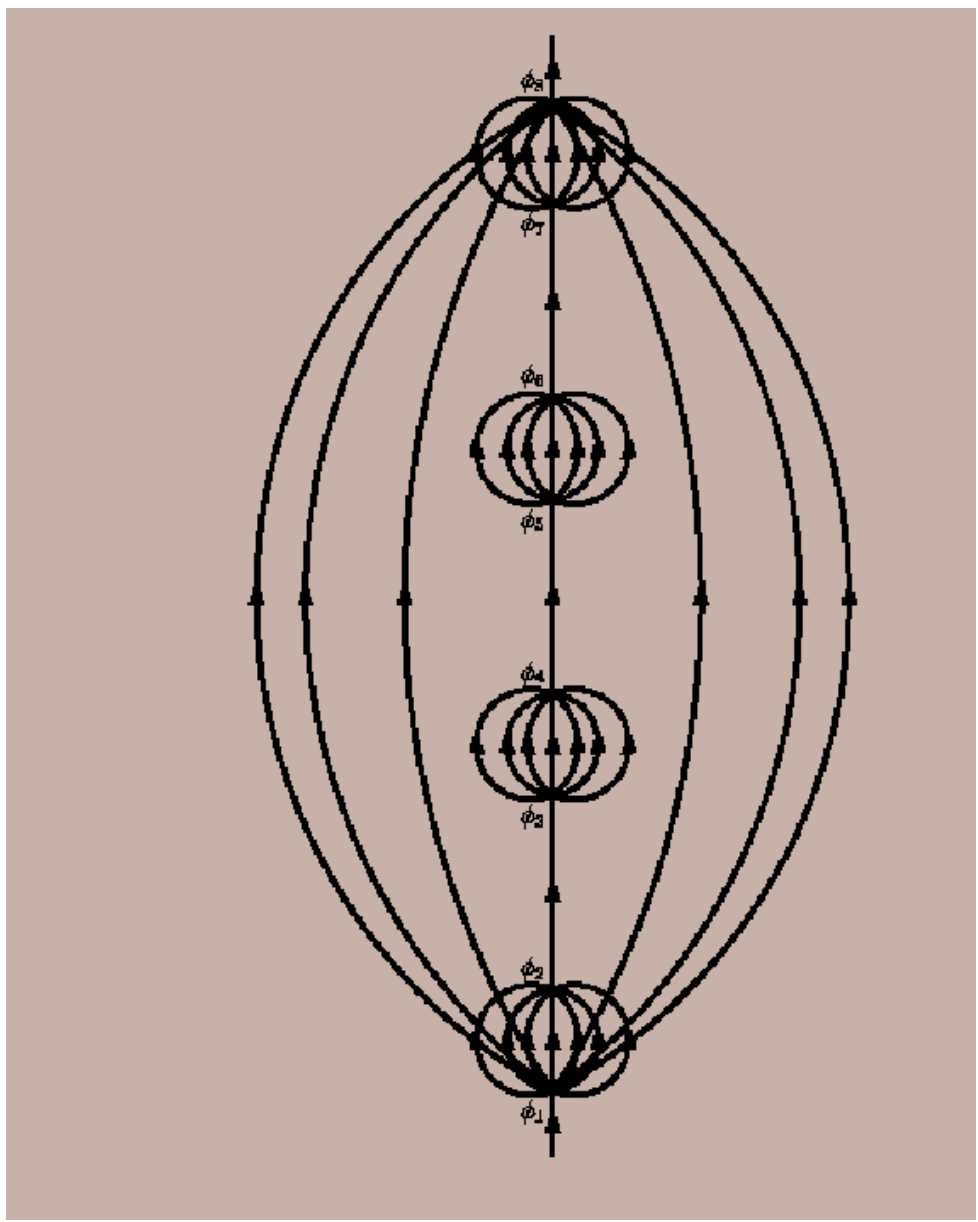,width=\linewidth}
\caption[fegf4]{seven  Feynman paths of states 
that all begin at the state $\phi_1$ and end
 at $\phi_8$ are shown in the figure. The middle path is the one along 
which   the collective dense measurement is performed by the ensemble
members $O_i\ \ \  i=1, 2, \ldots N$.   The $N$  separate systems of these 
observers have been initially prepared in the
states $\phi_i \ \ \ i=1, 2, \ldots N$.  Only eight  states are shown in the figure for
clarity. Note the secondary Feynman paths between neighbouring states in the
middle path.}
\end{minipage}
 
\end{figure}

 Along this line
we have the $N$ ($N=8$  in the figure)  initially prepared 
states $\phi_1, \phi_2, \ldots \phi_N$ as well
as the secondary Feynman paths that lead from each $\phi_i$ to $\phi_{i+1}$
where $i=1, 2, \ldots,  (N-1)$  . The relevant conditional probability 
is found by
multiplying the last probability amplitude from Eq (\ref{e3})  by its 
conjugate to obtain, omitting
the subscripts of the $\Phi$'s for clarity
\begin{eqnarray} && \Phi^{\dagger}\Phi=\sum_{\grave i \grave j\ldots \grave r 
\grave s}\sum_{ij \ldots rs} \phi_{\grave i1}\phi_{1i}\phi_{2\grave i}\phi_{i2}
\phi_{\grave j2} \phi_{2j}\phi_{3\grave j}\phi_{j3} \ldots 
\phi_{\grave rN-2}\phi_{N-2r}\phi_{N-1\grave r}\phi_{rN-1} \cdot 
\nonumber \\ && \cdot \phi_{\grave sN-1} \phi_{N-1s}\phi_{N\grave s}\phi_{sN} 
=(\sum_{\grave i i}\phi_{\grave i1}\phi_{1i}\phi_{2\grave i}\phi_{i2})
(\sum_{\grave j j}\phi_{\grave j2} \phi_{2j}\phi_{3\grave j}\phi_{j3}) \ldots
 \label{e4} \\ && \ldots (\sum_{\grave r r}\phi_{\grave rN-2}\phi_{N-2r}\phi_{N-1\grave r}\phi_{rN-1})
(\sum_{\grave s s}\phi_{\grave sN-1} \phi_{N-1s}\phi_{N\grave s}\phi_{sN}),  
\nonumber \end{eqnarray}  
where the number of all the double sums $\sum_{\grave i i} \sum_{\grave j j} \ldots
\sum_{\grave r r} \sum_{\grave s s}$ is $N$. \par 
We note  that we regard the traversed Feynman path as actually composed of a 
very large number of states. Thus, no matter how large is the number $N$ of observers 
there will always be a sufficient number of states to assign to all of them so that no 
two observers have the same state. \par 
As remarked, we are interested in
the limit of dense measurement along the relevant Feynman path in order to
realize it so we take  $N\to \infty$.  In this limit the length of the secondary
Feynman paths among the initially prepared $N$ states (where  now
 $N \to \infty$) tends to zero \cite{Bar0} so that  the former 
 probabilities  to proceed along  the secondary paths between the given states 
 become the   
 probabilities  
for these  states.   Thus, 
 we may write for   Eq  (\ref{e4}) in the limit of $N \to \infty$
\begin{eqnarray} && \lim_{N \to \infty}<\!\Phi^{\dagger}|\Phi \!>=
\lim_{N \to \infty}<\! \phi_{\grave i1}| \phi_{2\grave i}\!>
<\!\phi_{i2}|\phi_{1i}\!><\! \phi_{\grave j2}| 
\phi_{3\grave j}\!><\!\phi_{j3}| 
\phi_{2j}\!> \ldots
\nonumber  \\ && \ldots <\! \phi_{\grave r(N-2)}|\phi_{(N-1)\grave r}\!><\!
\phi_{r(N-1)}|\phi_{(N-2)r}\!> <\! \phi_{\grave s(N-1)}| \phi_{N\grave
s}\!>\cdot \label{e5} \\ && \cdot <\!
\phi_{sN}|\phi_{(N-1)s}\!> = 
\delta_{ \phi_{\grave i1}  \phi_{2\grave i}}\delta_{\phi_{1i} \phi_{i2}}
\delta_{ \phi_{\grave j2}  \phi_{3\grave j}}\delta_{\phi_{2j} \phi_{j3}}
\ldots 
\delta_{ \phi_{\grave r(N-2)} \phi_{(N-1)\grave r}} \cdot 
\nonumber \\ && \cdot \delta_{\phi_{r(N-1)} \phi_{(N-2)r}}\delta_{ \phi_{\grave s(N-1)}
 \phi_{N\grave s}} 
\delta_{\phi_{sN} \phi_{(N-1)s}} =1   \nonumber 
\end{eqnarray}
The last result of unity follows because in the limit of $N \to \infty$
successive states differ infinitesimally from each other so we may write  
as in \cite{Facchi} 
$<\! \phi_{\grave k-1}| \phi_{\grave k}\!>=<\!\phi_{k-1}|\phi_{k}\!> 
=\delta_{ \phi_{\grave k-1} \phi_{\grave k}}=
\delta_{\phi_{k-1}\phi_k}\approx 1$.  \par
Thus, we see that performing dense measurement along any Feynman path of states 
results
in its realization in the sense that the probability to proceed through 
 all of its  states
 tends to unity. Moreover, as described, the dense measurement is
performed through the joint action of all  the members of the ensemble 
without having to perform it  separately by each one of them.  Thus,  even 
when each 
observer performs his experiment only once,
nevertheless, when  $N \to \infty$  the obtained
realized path is now {\it for all them}.    Figure 5.2 shows a schematic representation of the state of the 
ensemble  after the remarked collective dense measurement. Each separate
batch of 4 similar curves  denotes  a member of the 
ensemble  that  has, as known,  a large number of 
different possible Feynman paths (only 4 are shown for clarity).  In the 
middle 
part of the figure 
we have a large number of different batches of paths all mixed  among them 
so 
it becomes  difficult to discern which curve belongs to which batch. The emphasized
path in Figure 5.2 is the definite Feynman path along which the described
collective dense measurement has been done. Note that this path, actually,
belongs to all the different interwoven batches which means that although each
one of the observers performs his experimental part only once,  nevertheless,  
after completing  the described collective 
measurement each one of those that participates in it has now the same realized
Feynman path.  The reason is that although each  observer $O_i$ of the ensemble 
performs his  experiment on  only his
specifically prepared state $\phi_i$, nevertheless, the results he obtains are
valid for all the others  since any other observer 
that
acts on the same state $\phi_i$ under the same conditions obtains 
 the same result.   
  In other words,   
the realized Feynman path has been made  tangible and real for all of them in 
the
sense that the probability for each to move  along its 
 constituent states tends to unity as seen from Eq (\ref{e5}).  \par
 \begin{figure}
\begin{minipage}{.68\linewidth} 
\centering\epsfig{figure=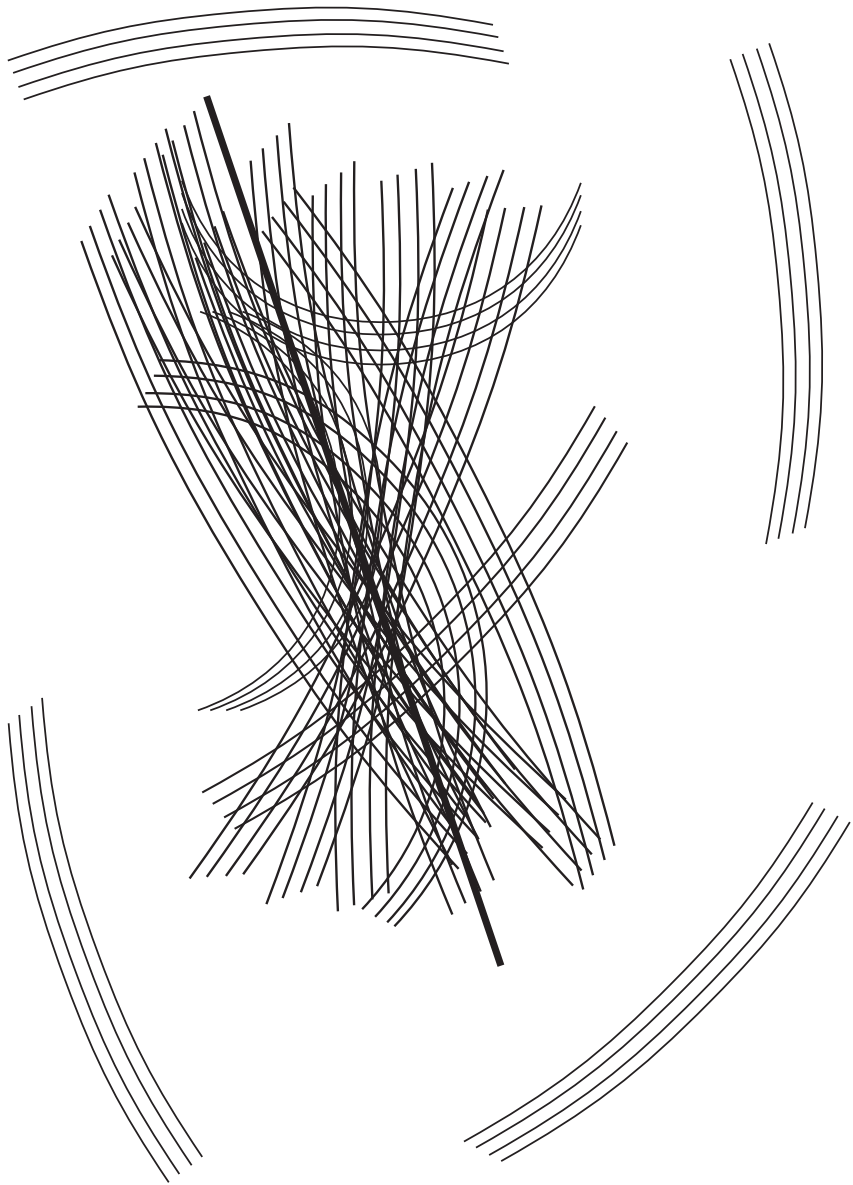,width=\linewidth}
\caption[fegf5]{A schematic representation of the physical situation after
performing  the collective 
dense measurement symbolized by   Figure 5.1. Note that  although no member of 
the 
ensemble has done dense
measurement by himself, nevertheless, the joint action of all or most of the
observers has resulted in ``realizing'' the specific Feynman path from Figure
5.1
for {\it all the participating observers}. This "realized"  path is shown emphasized
in the figure.}
\end{minipage}
\end{figure}

We note that using the large ensemble of similar systems for analysing experimental results 
has been fruitfully done in the 
literature \cite{Hartle,Graham,Finkelstein,Smolin}. 
It has been shown, for example, that considering an $N$ {\it identical} systems all prepared 
in the same initial state one may derive the probability interpretation of quantum 
mechanics in the limit of $N \to \infty$. That is, this probability is not imposed 
upon the theory as an external assumption as done in the conventional Copenhagen 
interpretation of quantum mechanics  but is derived from other principles of quantum 
mechanics \cite{Smolin}. This is done using Finkelstein 
theorem \cite{Finkelstein,Smolin}.     

\protect \section{\label{sec1}  The relative state theory of Everett}
  The last results may be demonstrated in a more natural and
appealing manner by using the relative state theory of Everett 
 \cite{Everett,Graham} that has been
formulated, especially, for taking observers into account. We use, in the
following, the special notation and terminology of this theory. Thus, if the
initial state was some eigenstate of an operator $A$ the total initial state of
the (system $S$ $+$ observer $O$) is denoted by $\Psi^{S+O}=\phi_i\Psi^O[...]$,
where $\phi_i$ is the initial eigenstate of the system $S$ and $\Psi^O[...]$ 
denotes the observer's state before the measurement. After the experiment the
observer's state is denoted by $\Psi^O[...\alpha_i]$, where $\alpha_i$ stands
for recording of the eigenvalue $\phi_i$ by the observer so that the total final
state of  the (system $S$ $+$ observer $O$) is $\Psi^{S+\grave
O}=\phi_i\Psi^O[...\alpha_i]$. Now, if the initial state of the system is not an
eigenstate but a superposition of them $\sum_ia_i\phi_i$ then the total states
before and after the measurement are \cite{Everett,Graham}   
$\Psi^{S+O}=\sum_ia_i\phi_i\Psi^O[...]$,  and 
$\Psi^{S+\grave O}=\sum_ia_i\phi_i\Psi^O[...\alpha_i]$ 
 respectively where $a_i=<\!\phi_i|\Psi^{S+O}\!>$. Suppose  we 
 continue our experiments and measure  some other physical observable $B$
 beginning from the state $\sum_ia_i\phi_i\Psi^O[...\alpha_i]$  as the initial one. In such a case one 
 may 
 expand the eigenfunction $\phi_i$ of the observable $A$  in
 terms of the eigenfunctions of $B$ $\phi_i=\sum_jb_{ij}\phi_j$ so that the 
  state $\sum_ia_i\phi_i\Psi^O[...\alpha_i]$   before the new experiment may be written as
 \cite{Everett}
 \begin{equation} \label{e8} 
 \Psi^{S+O}=\sum_j\sum_ib_{ij}a_i\phi_j\Psi^O[...\alpha_i],  \end{equation}
 where $\phi_j$ are the eigenfunctions of the operator $B$. After measuring $B$
 one  obtains \begin{equation} \label{e9} 
 \Psi^{S+\grave O}=\sum_j\sum_ib_{ij}a_i\phi_j\Psi^O[...\alpha_i,\beta_j],  
 \end{equation} where $b_{ij}=<\!\phi_j|\phi_i\!>$ and
 $\Psi^O[...\alpha_i,\beta_j]$
 denotes that now the observer records the eigenfunctions $\alpha_i$ and
 $\beta_j$ after the two experiments. Continuing  and
 measuring, for example, $n$ observables  one
 obtains the following wave function \begin{equation} \label{e10} 
 \Psi^{S+\grave
 O}=\sum_k\sum_l\sum_h\ldots\sum_j\sum_ic_{hl}d_{lk}\ldots a_i\phi_k\Psi^O
 [\alpha_i,\beta_j,\ldots,\lambda_l,\xi_k], \end{equation} 
 where $c_{hl}=<\!\phi_l|\phi_h\!>$,  $d_{lk}=<\!\phi_k|\phi_l\!>$, and
 $\phi_i$,  $\phi_j$ $\ldots$ are eigenfunctions of the $A$, $B$, $\ldots$
 operators. Note that each term in Eq (\ref{e10})   denotes an observer
 with his specific  sequence  $[\alpha_i,\beta_j,\ldots,\lambda_l,\xi_k]$ 
 that results 
 from the $n$ experiments. Thus, Eq (\ref{e10}), termed the Everett's universal 
 wave function \cite{Everett,Graham}, yields all the possible results that may be obtained from
 measuring the $n$ observables. We, now,  count the number of observers
 that have the same or similar  sequences  
 $[\alpha_i,\beta_j,\ldots,\lambda_l,\xi_k]$ which  record, as remarked, the $n$
 measured eigenvalues. For this we assume that each measurement of any  of the
 $n$ observables may yields $K$ possible different results  where 
 the $n$ observables do not have to be all different and so some eigenvalues in
 the sequence $[\alpha_i,\beta_j,\ldots,\lambda_l,\xi_k]$ may be identical. Thus,
 denoting by  $R_1,R_2,\ldots,R_r$ the numbers of times 
 the $r$ particular eigenvalues
 $l_1,l_2,\ldots,l_r$ appear respectively in some specified sequence 
 $[\alpha_i,\beta_j,\ldots,\lambda_l,\xi_k]$ we may see from Eq (\ref{e10}) that
 each possible value of $R_i$ in the range $0 \le R_i \le n$, and for each 
 $i
 \ \ \ 
 (1 \le  i \le r)$, may be realized in some observer. Now, the number of
 sequences in which $l_1,l_2,\ldots,l_r$ occur, respectively, at 
 $R_1,R_2,\ldots,R_r$ predetermined positions is 
 $(K-1)^{(n-\sum_{i=1}^{i=r}R_i)}$ since for each position in the sequence 
  $[\alpha_i,\beta_j,\ldots,\lambda_l,\xi_k]$ in which $l_1,l_2,\ldots,l_r$ are
  absent there are $(K-1)$ possibilities (note that each position is related
  only to
  its specific observable and so  to, at most,  only one of the $l$'s). Thus,  the total number of sequences 
  in which 
$l_1,l_2,\ldots,l_r$ occur respectively in  $R_1,R_2,\ldots,R_r$ positions (we
denote this number by $N_{l_1,l_2,\ldots,l_r}$) is 
\begin{eqnarray} &&  N_{l_1,l_2,\ldots,l_r}=\left( \begin{array}{c} n \\ R_1 
\end{array} \right)\left( \begin{array}{c} (n-R_1) \\ R_2 
\end{array} \right)\left( \begin{array}{c}(n-(R_1+R_2)) \\ R_3 
\end{array} \right)\ldots \label{e11} \\ && \ldots \left( \begin{array}{c} (n-\sum_{i=1}^{i=r-1}R_i) \\ 
R_r \end{array} \right)(K-1)^{(n-\sum_{i=1}^{i=r}R_i)},  \nonumber  
\end{eqnarray} where $\left( \begin{array}{c} n \\ R_1 
\end{array} \right)$ is the number of possible ways to choose in the $n$ member
sequence $[\alpha_i,\beta_j,\ldots,\lambda_l,\xi_k]$ $R_1$ places for $l_1$, 
$\left( \begin{array}{c} (n-R_1) \\ R_2 
\end{array} \right)$ is the number of possible ways to choose $R_2$ places from
the remaining $(n-R_1)$ etc. The  calculation in Eq (\ref{e11}) was done for the more simple case in which
all the $l_1,l_2,\ldots,l_r$ are different.  The
relevant measure may be found \cite{Graham} by taking into consideration  
 the expected
relative frequency of the eigenvalues $l_1,l_2,\ldots,l_r$ which is 
$P_{l_1,l_2,\ldots,l_r}=|\!\!<\!\Psi_{l_1,l_2,\ldots,l_r}|\Psi\!>\!\!|^2$, where
 $|\Psi_{l_1,l_2,\ldots,l_r}\!>$ is the state in which the eigenvalues $l_1,
 l_2, \ldots, l_r$ occur among those of the sequence  
 $[\alpha_i,\beta_j,\ldots,\lambda_l,\xi_k]$, 
 and also the
corresponding relative frequency of any other eigenvalue $m$ different from 
$l_1,l_2,\ldots,l_r$, which is 
$Q_m=\sum_{m\ne l_1,l_2,\ldots,l_r}|\!\!<\!\Psi_m|\Psi\!>\!\!|^2=1-
P_{l_1,l_2,\ldots,l_r}$. That is,  the measure of all the sequences that have 
the eigenvalues $l_1,l_2,\ldots,l_r$  at $R_1,R_2,\ldots,R_r$ predetermined
positions      
 respectively 
 is $P_{l_1,l_2,\ldots,l_r}^{\sum_{i=1}^{i=r}R_i}Q_m^{(n-\sum_{i=1}^{i=r}R_i)}$. 
  The last expression 
 must be multiplied by the number of  possible ways to choose first $R_1$
 places for $l_1$ from the  $n$ positions of the sequence  
 $[\alpha_i,\beta_j,\ldots,\lambda_l,\xi_k]$, then to choose $R_2$ places for $l_2$
 from the
 remaining $n-R_1$ etc,  
  until the last step of choosing $R_r$ places from   
 $(n-\sum_{i=1}^{i=r-1}R_i)$  (see  Eq (\ref{e11})).   
 That is, the sought-for measure $M_e$ is  
\begin{eqnarray} && M_e=\left( \begin{array}{c} n \\ R_1 
\end{array} \right)\left( \begin{array}{c} (n-R_1) \\ R_2 
\end{array} \right)\left( \begin{array}{c} (n-(R_1+R_2)) \\ R_3 
\end{array} \right)\ldots \label{e12} \\ && \ldots \left( \begin{array}{c} (n-\sum_{i=1}^{i=r-1}R_i) \\ 
R_r \end{array} \right)
P_{l_1,l_2,\ldots,l_r}^{\sum_{i=1}^{i=r}R_i}Q_m^{(n-\sum_{i=1}^{i=r}R_i)}
\nonumber , 
\end{eqnarray} which is, up to a constant coefficient,  the Bernoulli 
distribution \cite{Spiegel}. Note that for
large $n$ the measure from the last equation may be approximated by a Gaussian
distribution with mean $\mu=nP_{l_1,l_2,\ldots,l_r}$ and standard deviation 
$\sigma=\sqrt{nP_{l_1,l_2,\ldots,l_r}Q_m}$. For large $n$ this Gaussian
distribution has a sharp peak \cite{Spiegel} around $nP_{l_1,l_2,\ldots,l_r}$ 
since $nP_{l_1,l_2,\ldots,l_r}>\!>\sqrt{nP_{l_1,l_2,\ldots,l_r}Q_m}$ and the
measure of all the sequences that lie between $(nP_{l_1,l_2,\ldots,l_r}-
3\sqrt{nP_{l_1,l_2,\ldots,l_r}Q_m})$ and $(nP_{l_1,l_2,\ldots,l_r}+
3\sqrt{nP_{l_1,l_2,\ldots,l_r}Q_m})$ is greater than 0.99.  We calculate now 
an explicit expression  for $P_{l_1,l_2,\ldots,l_r}(r)$ and $Q_m(r)$ as functions of
$r$, for  
$n=100$.   The probability $P_{l_1,l_2,\ldots,l_r}(r)$ to find
the values   
$l_1,l_2,\ldots,l_r$ among the eigenvalues of the sequence 
$[\alpha_i,\beta_j,\ldots,\lambda_l,\xi_k]$ is 
$P_{l_1,l_2,\ldots,l_r}(r)=|<\!\Psi_{l_1,l_2,\ldots,l_r}|\Psi\!>|^2=\frac{r}{n}=
\frac{r}{100}$,  and the probability to find any other eigenvalue 
$m \ne l_1,l_2,\ldots,l_r$  is 
$Q_m(r)=\sum_{m\ne l_1,l_2,\ldots,l_r}|<\!\Psi_m|\Psi\!>|^2=1-
P_{l_1,l_2,\ldots,l_r}=1-\frac{r}{n}=\frac{(100-r)}{100}$.  
 Now, in order to 
simplify
the following calculations we assign to all the different values of 
$R_i \  \ i=1,2,\ldots r$ the unity value, in which case each of the given
eigenvalues $l_1,l_2,\ldots,l_r$ may occur only once in the sequence 
$[\alpha_i,\beta_j,\ldots,\lambda_l,\xi_k]$,  
  so that the relevant total number 
 $N_{l_1,l_2,\ldots,l_r}(K,r)$  and the corresponding measure $M_e(r)$ from 
 Eqs (\ref{e11})-(\ref{e12}) 
are given by  
 \begin{eqnarray} &&  N_{l_1,l_2,\ldots,l_r}(K,r)=\left(
 \begin{array}{c} 100 
 \\ 1 
\end{array} \right)\left( \begin{array}{c} 99 \\ 1 
\end{array} \right)\ldots \left( \begin{array}{c} (100-r) \\ 1 
\end{array} \right) \cdot \label{e13} \\ && \cdot (K-1)^{(100-r)}= 
\prod_{i=0}^{i=r}(100-i)(K-1)^{(100-r)}
\nonumber \end{eqnarray}   and 
  \begin{eqnarray} && 
M_e(r)=\left( \begin{array}{c} 100 
 \\ 1 
\end{array} \right)\left( \begin{array}{c} 99 \\ 1 
\end{array} \right)\ldots \left( \begin{array}{c} (100-r) \\ 1 
\end{array} \right)(\frac{r}{100})^r  \cdot \label{e14}  \\ && \cdot (\frac{(100-r)}{100})^{(100-r)}
= \prod_{i=0}^{i=r}(100-i)
(\frac{r}{100})^r(\frac{(100-r)}{100})^{(100-r)} \nonumber 
\end{eqnarray} respectively. In Table 1 we show the number of observers that have $r$
predetermined different eigenvalues in their respective $n$-place  sequences 
 for $n=100$,     five  different values 
of $K$:  $1100, 100, 10, 5, 2$,    and   even
values of $r$ between $r=0$ and $r=98$.    Note that for the large values of 
$K$, which
signifies  a large  number of  possible results for any experiment done  
 by any  observer, the sequences most frequently encountered are the 
 ones that have 
 small $r$  as should be and as we have seen by other methods 
in  the former section. This is so because a large 
$K$ signifies  a large number of possible results for each experiment 
 which entails  a
comparatively large number of observers with small $r$ 
so   that the probability to find in their   sequences   
a large  
 number of the $r$ 
predetermined eigenvalues is small.  That is, the larger is $K$ the smaller the ensemble's 
members are related among them. For example, for $K=1100$ and  $K=100$ the
number of observers with $r=0$,  that have not even one of the preassigned 
 eigenvalues, are $1.258257 \cdot 10^{306}$ and  
$3.660323 \cdot 10^{201}$  respectively  compared  to 
$1.025655 \cdot 10^{161}$ and  
$9.23929 \cdot 10^{159}$ that have in their sequences 98 places  occupied by
such eigenvalues.  That is, for $K=1100$ and  $K=100$ the number of observers 
with $r=0$ 
are large by the  factors  of $1.2268 \cdot 10^{145}$ and 
$3.9617 \cdot 10^{41}$ respectively compared to  those with $r=98$.  \par 
For smaller $K$, which signifies a small number of possible different  results 
for any  observer  one finds a large number of observers that
have among their  sequences,  after the $n$ experiments,  a comparatively
large number of the $r$ predetermined eigenvalues. That is, 
the smaller is $K$ the larger is the relationship  among the ensemble's members.  
Moreover, as seen from Table 1, the number of observers increases proportionally
to $r$ for small values of $K$,    compared to the large values of $K$ for 
which the
number of observers decreases as $r$ increases.  The results of  Table 1 are
corroborated also from Figure 5.3 which shows a three-dimensional surface of the
relative rate $R(K,r)$ of the number of observers which is given by 
\begin{equation} \label{e15} R(K,r)= \frac{N_{l_1,l_2,\ldots,l_r}(K,r)-
N_{l_1,l_2,\ldots,l_r}(K,r-1)}{N_{l_1,l_2,\ldots,l_r}(K,r)},  \end{equation} 
where $N_{l_1,l_2,\ldots,l_r}(K,r)$ is given by Eq (\ref{e13})  and
the ranges of $K$ and $r$ are $0 \le K \le 250$ and $0 \le r \le 100$
respectively. 

\begin{figure}
\begin{minipage}{.68\linewidth} 
\centering\epsfig{figure=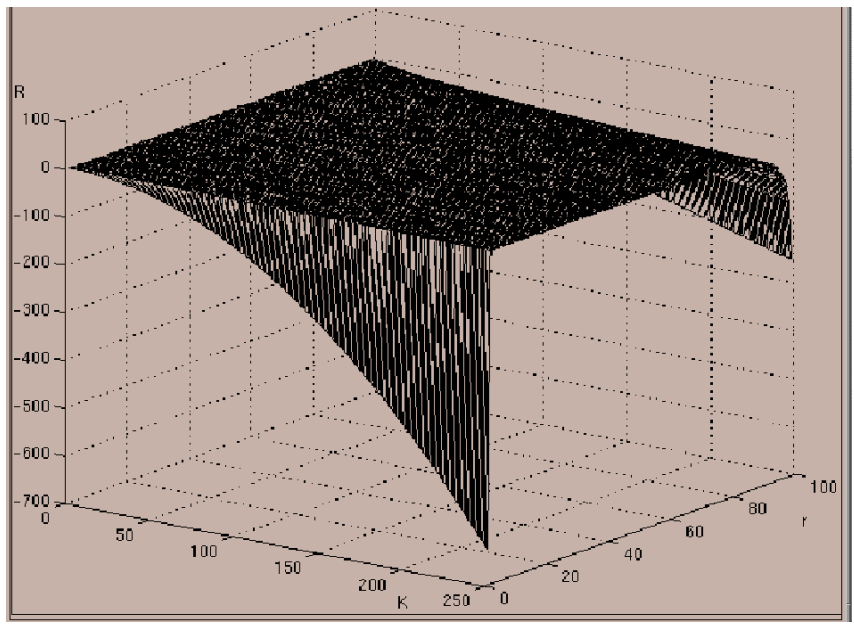,width=\linewidth}
\caption[fegf6]{The figure shows a three-dimensional surface of the relative
rate of the number of observers as function of the number of possible results 
$K$ for each experiment and the number $r$ of places occupied by 
 preassigned eigenvalues (see text). 
The ranges of $K$ and $r$ are  $2 \le K \le 250$    
and  $0 \le r \le 100$  respectively. Note the large jump
towards zero for
large $K$ when $r$ increases from zero. }
\end{minipage}

\end{figure}

We see from the figure that the surface, including the planar form on top of it,
is inclined  from positive values of  $R(K,r)$,  for small $K$,  towards 
 negative values
for large $K$ which means that the large numbers of observers are found at large
$K$ and small $r$ as we have found from Table 1. Also, as seen from the figure, the rate 
$R(K,r)$ decreases sharply,  for small $r$,  as $K$ increases, whereas this
decrease is less pronounced for intermediate values of $r$  and then it
 strengthens again 
 for large $r$ but less  than for the small values of it as seen from the
 figure. 
  When $K=1$,  which means that
there is only one result for any experiment   we must have, for each
observer,   $r=n$ since there is no eigenvalue in any place of any $n$-sequence 
that is
different from  the specified ones.  In this case all the  
 sequences of all the observers are identical to each other and the probability
 to find in them all the $r$ specified eigenvalues, where $r=n$, is unity.    
In other words,   
the more known are  the obtained results from
any experiment,    
 the  larger is the probability to
find in  the sequences of most of them a large   number of 
the specified
eigenvalues.
\par 
We, thus, see that an important necessary aspect of the physical validation of
real phenomena 
is that a large ensemble of {\it related} observers must be involved in the relevant
experimentation. This is affected, as remarked, through the magnitude of $K$ 
so that the smaller is 
$K$  
the
larger is the number of observers that obtain the same results in their
sequences and consequently the more valid is the tested theory.  
When $K=1$ which means that there is only one possible result for
any experiment  then the whole ensemble of
observers obtains exactly the same results in their sequences.  
In other words, 
 the  smaller is $K$, in which case the obtained result from any experiment 
 is less unknown,  the
more related are  the observers of the ensemble where this relationship  is
demonstrated   by obtaining 
the same  results.  
 When $K=1$   then this relationship is maximal and the whole ensemble obtains
exactly the same results. Note that if they do not  measure the same observables
then they are totally unrelated and our former results would not be obtained 
  even for $K=1$.   As remarked,  the number of possible
observers  increases in an asymptotic manner with the number of experiments (see
Eq (\ref{e10}) and Table 1) and so  all  the  results in this section are 
 for large ensemble of observers. \par 
$K=1$
corresponds, for example, to the {\it given}  initial probability amplitude 
or particle
density in the one-dimensional arrays of potential barriers or  traps
respectively discussed in the former chapter. It has been shown there that as
the number of barriers or traps increases the corresponding initial values of
the probability amplitude or particles density is preserved in passing through
them. This corresponds to the results obtained here and strengthened by Table 1
that the more small is $K$, in which case the initial state is less unknown then
a larger number of observers obtain similar results. 
 When the initial state is known ($K=1$) then the whole ensemble obtains
exactly the same results. Thus, we see that the new response of the system,
identified here by the large probability to find in the sequences of most 
observers the same
eigenvalues for small $K$, is possible only for large ensemble of {\it related} 
observers. This relationship is manifested by the fact that all the observers 
 measure the same observables, so that  the new response is obtained only 
 as a result
 of it.  
 Otherwise, this  new phenomenon 
would not be obtained even for $K=1$. Also, as for the other examples discussed
in the former chapters, this new response may be physically established by
merely 
repeating again this collective experiment for small $K$. As remarked,  
the number of possible
observers  increases in an asymptotic manner with the number of experiments 
(see
Eq (\ref{e10}) and Table 1). Thus, all  the  results in this chapter are 
 for large ensemble of observers.

 \begin{table}
\caption{\label{table1} The table shows the  number of observers that have $r$
    positions in their 100 places sequences occupied by the preassigned eigenvalues, where
    the numbers $K$ of possible values for each experiment are 1100, 100, 10, 5
    and 2}
     \begin{center}
      \begin{tabular}{cccccc} 
        \  r&Number \ \ of  &Number \ \ of  &Number \ \ of &
       Number \ \ of  &Number \ \ of  \\ &observers for&
      observers for& 
      observers for & observers for&observers for \\&\ \ K=1100&\ \ K=100&\ \ 
      K=10&\ \ K=5&\ \ K=2\\
   \hline \hline
  $  0. $&$ 1.2582567\cdot 10^{306} $&$ 3.6603234\cdot 10^{201} $&$  2.6561399\cdot 10^{97}$&$   1.6069380\cdot 10^{62}  $&$ 1.0000000\cdot 10^{02}$ \\
  $  2. $&$ 9.1968147\cdot 10^{300} $&$ 3.6599499\cdot 10^{199}  $&$ 3.5349615\cdot 10^{98} $&$  2.4360176\cdot 10^{64} $&$  9.7020000\cdot 10^{05}$ \\
 $  4. $&$ 7.0906332\cdot 10^{298} $&$ 3.4773445\cdot 10^{199} $&$ 4.0638965\cdot 10^{100}$&$  1.4177623\cdot 10^{67}  $&$ 9.0345024\cdot 10^{09}$ \\
  $  6. $&$ 5.2425320\cdot 10^{296} $&$ 3.1683182\cdot 10^{199} $&$ 4.4803204\cdot 10^{102}  $&$ 7.9128857\cdot 10^{69}  $&$ 8.0678106\cdot 10^{13}$ \\
   $  8. $&$ 3.7137826\cdot 10^{294} $&$ 2.7658535\cdot 10^{199} $&$ 4.7325459\cdot 10^{104} $&$   4.2314156\cdot 10^{72}  $&$ 6.9028188\cdot 10^{17}$ \\
  $ 10.  $&$2.5182857\cdot 10^{292} $&$ 2.3112275\cdot 10^{199} $&$ 4.7851297\cdot 10^{106}  $&$ 2.1659559\cdot 10^{75}  $&$ 5.6534086\cdot 10^{21}$ \\
$ 12. $&$ 1.6329854\cdot 10^{290} $&$ 1.8469068\cdot 10^{199} $&$ 4.6268069\cdot 10^{108}  $&$1.0602354\cdot 10^{78}  $&$ 4.4277496\cdot 10^{25}$ \\

   $ 14. $&$ 1.0115902\cdot 10^{288} $&$ 1.4099129\cdot 10^{199} $&$ 4.2737987\cdot 10^{110}  $&$  4.9579258\cdot 10^{80}  $&$ 3.3128423\cdot 10^{29}$ \\

   $ 16. $&$ 5.9800861\cdot 10^{285} $&$ 1.0271175\cdot 10^{199}  $&$3.7672744\cdot 10^{112}  $&$ 2.2124744\cdot 10^{83}  $&$ 2.3653694\cdot 10^{33}$\\
   
   $ 18. $&$ 3.3697990\cdot 10^{283} $&$ 7.1324983\cdot 10^{198} $&$ 3.1654407\cdot 10^{114}  $&$ 9.4113129\cdot 10^{85}  $&$ 1.6098704\cdot 10^{37}$\\

   $ 20. $&$ 1.8079384\cdot 10^{281} $&$ 4.7157013\cdot 10^{198} $&$ 2.5323525\cdot 10^{116}  $&$ 3.8115817\cdot 10^{88} $&$  1.0431960\cdot 10^{41}$\\

   $ 22. $&$ 9.2238011\cdot 10^{278} $&$ 2.9648150\cdot 10^{198} $&$ 1.9264637\cdot 10^{118}  $&$ 1.4679354\cdot 10^{91} $&$  6.4281738\cdot 10^{44}$\\
  
   $ 24. $&$ 4.4690875\cdot 10^{276} $&$ 1.7702375\cdot 10^{198} $&$ 1.3918106\cdot 10^{120} $&$  5.3689737\cdot 10^{93}  $&$ 3.7617673\cdot 10^{48}$\\

   $ 26. $&$ 2.0536029\cdot 10^{274} $&$ 1.0024301\cdot 10^{198} $&$ 9.5364800\cdot 10^{121}  $&$ 1.8623628\cdot 10^{96} $&$  2.0877809\cdot 10^{52}$\\

   $ 28.$&$  8.9366846\cdot 10^{271} $&$ 5.3757503\cdot 10^{197} $&$ 6.1881159\cdot 10^{123}  $&$ 6.1178617\cdot 10^{98}  $&$ 1.0973376\cdot 10^{56}$\\

   $ 30. $&$ 3.6773709\cdot 10^{269} $&$ 2.7259952\cdot 10^{197}  $&$3.7969057\cdot 10^{125} $&$ 1.9003608\cdot 10^{101} $&$  5.4537680\cdot 10^{59}$\\

   $ 32. $&$ 1.4285652\cdot 10^{267}  $&$1.3050066\cdot 10^{197}$&$ 2.1993928\cdot 10^{127} $&$ 5.5728080\cdot 10^{103}  $&$ 2.5589079\cdot 10^{63}$\\

   $ 34. $&$ 5.2302616\cdot 10^{264} $&$ 5.8879085\cdot 10^{196} $&$ 1.2007055\cdot 10^{129} $&$ 1.5401848\cdot 10^{106} $&$  1.1315491\cdot 10^{67}$\\

   $ 36. $&$ 1.8014464\cdot 10^{262} $&$ 2.4991021\cdot 10^{196} $&$ 6.1665865\cdot 10^{130} $&$ 4.0044805\cdot 10^{108}  $&$ 4.7072442\cdot 10^{70}$\\

   $ 38. $&$ 5.8258354\cdot 10^{259} $&$ 9.9596905\cdot 10^{195} $&$ 2.9736650\cdot 10^{132} $&$ 9.7759381\cdot 10^{110}  $&$ 1.8386496\cdot 10^{74}$\\

   $ 40. $&$ 1.7654032\cdot 10^{257} $&$ 3.7192600\cdot 10^{195} $&$ 1.3436561\cdot 10^{134} $&$ 2.2362458\cdot 10^{113}  $&$ 6.7294575\cdot 10^{77}$\\
 
   $ 42. $&$ 5.0018254\cdot 10^{254} $&$ 1.2985724\cdot 10^{195} $&$ 5.6765321\cdot 10^{135} $&$ 4.7827708\cdot 10^{115}  $&$ 2.3028204\cdot 10^{81}$\\

   $ 44. $&$ 1.3218922\cdot 10^{252} $&$ 4.2292041\cdot 10^{194} $&$ 2.2369741\cdot 10^{137} $&$ 9.5416277\cdot 10^{117} $&$ 7.3506026\cdot 10^{84}$\\

   $ 46. $&$ 3.2505518\cdot 10^{249} $&$ 1.2815770\cdot 10^{194} $&$ 8.2022385\cdot 10^{138} $&$ 1.7711646\cdot 10^{120}  $&$ 2.1831290\cdot 10^{88}$\\

   $ 48. $&$ 7.4172159\cdot 10^{246} $&$ 3.6037407\cdot 10^{193} $&$ 2.7907863\cdot 10^{140} $&$ 3.0508311\cdot 10^{122}  $&$ 6.0167034\cdot 10^{91}$\\

   $ 50. $&$ 1.5659782\cdot 10^{244} $&$ 9.3761236\cdot 10^{192} $&$ 8.7858088\cdot 10^{141} $&$ 4.8622621\cdot 10^{124}  $&$ 1.5342594\cdot 10^{95}$\\

   $ 52. $&$ 3.0494930\cdot 10^{241} $&$ 2.2500401\cdot 10^{192} $&$ 2.5511386\cdot 10^{143} $&$ 7.1475252\cdot 10^{126} $&$  3.6085781\cdot 10^{98}$\\
 
   $ 54. $&$ 5.4586840\cdot 10^{238} $&$ 4.9633574\cdot 10^{191} $&$ 6.8093353\cdot 10^{144} $&$ 9.6580935\cdot 10^{128} $&$ 7.8017458\cdot 10^{101}$\\

   $ 56. $&$ 8.9486549\cdot 10^{235} $&$ 1.0026985\cdot 10^{191} $&$ 1.6645042\cdot 10^{146} $&$ 1.1951891\cdot 10^{131} $&$ 1.5447457\cdot 10^{105}$\\

   $ 58. $&$ 1.3380740\cdot 10^{233} $&$ 1.8476415\cdot 10^{190} $&$ 3.7112278\cdot 10^{147} $&$ 1.3490697\cdot 10^{133} $&$ 2.7898107\cdot 10^{108}$\\

    $60. $&$ 1.8168898\cdot 10^{230} $&$ 3.0916560\cdot 10^{189} $&$ 7.5140909\cdot 10^{148} $&$ 1.3827964\cdot 10^{135} $&$ 4.5752895\cdot 10^{111}$\\
   
\end{tabular} 
\end{center}
\end{table}
 \newpage
 
 \begin{table}

     \begin{center}
      \begin{tabular}{cccccc} 
       \  r&Number \ \ of  &Number \ \ of  &Number \ \ of &
       Number \ \ of  &Number \ \ of  \\ &observers for&
      observers for& 
      observers for & observers for&observers for \\&\ \ K=1100&\ \ K=100&\ \ 
      K=10&\ \ K=5&\ \ K=2\\
   \hline \hline  
      $ 62. $&$ 2.2293661\cdot 10^{227} $&$ 4.6748640\cdot 10^{188} $&$ 1.3748003\cdot 10^{150} $&$ 1.2808152\cdot 10^{137} $&$ 6.7805790\cdot 10^{114}$\\

   $ 64 $&$ 2.4586134\cdot 10^{224} $&$ 6.3533505\cdot 10^{187} $&$ 2.2607828\cdot 10^{151} $&$ 1.0662786\cdot 10^{139} $&$ 9.0317313\cdot 10^{117}$\\

   $ 66.$&$  2.4223775\cdot 10^{221} $&$ 7.7139956\cdot 10^{186} $&$ 3.3213969\cdot 10^{152} $&$ 7.9304473\cdot 10^{140} $&$ 1.0747760\cdot 10^{121}$\\
 
  $  68. $&$ 2.1179239\cdot 10^{218} $&$ 8.3113758\cdot 10^{185} $&$ 4.3301175\cdot 10^{153} $&$ 5.2340952\cdot 10^{142} $&$ 1.1349635\cdot 10^{124}$\\

  $  70. $&$ 1.6307895\cdot 10^{215} $&$ 7.8865212\cdot 10^{184} $&$ 4.9716164\cdot 10^{154}$&$ 3.0423178\cdot 10^{144} $&$ 1.0555160\cdot 10^{127}$\\

  $  72. $&$ 1.0963736\cdot 10^{212} $&$ 6.5338795\cdot 10^{183} $&$ 4.9838920\cdot 10^{155} $&$ 1.5439763\cdot 10^{146} $&$ 8.5707902\cdot 10^{129}$\\

  $  74. $&$ 6.3723598\cdot 10^{208} $&$ 4.6799137\cdot 10^{182} $&$ 4.3193730\cdot 10^{156} $&$ 6.7741960\cdot 10^{147} $&$ 6.0166947\cdot 10^{132}$\\

   $ 76. $&$ 3.1656008\cdot 10^{205} $&$ 2.8649609\cdot 10^{181} $&$ 3.1995356\cdot 10^{157} $&$ 2.5403235\cdot 10^{149} $&$ 3.6100168\cdot 10^{135}$\\

   $ 78. $&$ 1.3262069\cdot 10^{202} $&$ 1.4791044\cdot 10^{180} $&$ 1.9987222\cdot 10^{158} $&$ 8.0337731\cdot 10^{150} $&$ 1.8266685\cdot 10^{138}$\\

   $ 80. $&$ 4.6117440\cdot 10^{198} $&$ 6.3383721\cdot 10^{178} $&$ 1.0363745\cdot 10^{159} $&$ 2.1088654\cdot 10^{152} $&$ 7.6720078\cdot 10^{140}$\\

   $ 82. $&$ 1.3058579\cdot 10^{195} $&$ 2.2117368\cdot 10^{177} $&$ 4.3758034\cdot 10^{159} $&$ 4.5076999\cdot 10^{153} $&$ 2.6238267\cdot 10^{143}$\\

  $  84. $&$ 2.9408267\cdot 10^{191} $&$ 6.1380718\cdot 10^{175} $&$ 1.4694056\cdot 10^{160} $&$ 7.6630898\cdot 10^{154} $&$ 7.1368085\cdot 10^{145}$\\

  $  86. $&$ 5.1132066\cdot 10^{187} $&$ 1.3151669\cdot 10^{174} $&$ 3.8095700\cdot 10^{160} $&$ 1.0057805\cdot 10^{156} $&$ 1.4987298\cdot 10^{148}$\\
  
   $ 88. $&$ 6.6042355\cdot 10^{183} $&$ 2.0933174\cdot 10^{172} $&$ 7.3369497\cdot 10^{160} $&$ 9.8063602\cdot 10^{156} $&$ 2.3380185\cdot 10^{150}$\\

  $  90. $&$ 6.0147815\cdot 10^{179} $&$ 2.3494022\cdot 10^{170} $&$ 9.9637588\cdot 10^{160} $&$ 6.7418726\cdot 10^{157} $&$ 2.5718203\cdot 10^{152}$\\

  $  92. $&$ 3.5855598\cdot 10^{175} $&$ 1.7259153\cdot 10^{168} $&$ 8.8566745\cdot 10^{160} $&$ 3.0338427\cdot 10^{158} $&$ 1.8517106\cdot 10^{154}$\\

  $  94.$&$  1.2468404\cdot 10^{171} $&$ 7.3960252\cdot 10^{165} $&$ 4.5923497\cdot 10^{160} $&$ 7.9638371\cdot 10^{158} $&$ 7.7771846\cdot 10^{155}$\\

  $  96. $&$ 2.0646454\cdot 10^{166} $&$ 1.5092389\cdot 10^{163} $&$ 1.1339135\cdot 10^{160} $&$ 9.9547963\cdot 10^{158} $&$ 1.5554369\cdot 10^{157}$\\

  $  98. $&$ 1.0256551\cdot 10^{161} $&$ 9.2392953\cdot 10^{159} $&$ 8.3993594\cdot 10^{158} $&$ 3.7330486\cdot 10^{158} $&$ 9.3326215\cdot 10^{157}$\\
 \hline
\end{tabular} 
\end{center}
\end{table}

\newpage

 \section{\label{}The classical effect of an ensemble of 
  observers} 
    
\bigskip  \bigskip

 We discuss now the same system used in \cite{Szilard} for demonstrating the effect 
of observers upon the experimental results.  The discussion 
in \cite{Szilard} is generalized to include 
the large ensemble of related observers so  we assume that we 
have $N$ thermodynamical systems,  of the
 kind discussed in \cite{Szilard}, 
  that is,    a hollow cylinder that contains  $n$
  particles, not all of the same species,  among    
     four pistons as
 shown in Figure 5.4.  The pistons $A$ and $\grave A$ are fixed while $B$ and
 $\grave B$ may move along the cylinder. Also  the pistons $\grave A$ and 
 $B$  do not allow passage of particles through them, whereas $A$ and $\grave
 B$ are permeable so that each permits some kind of particles to move
 through it where those that are allowed to pass through $A$ are not allowed
 through $\grave B$ and vice versa. The pistons $B$ and $\grave B$ move in such
 a way that the distances $B\grave B$ and $A\grave A$ are always equal as seen
 in Figure 5.4. These
 distances are measured using the $x$ axis which is assumed to be upward along
 the cylinder. 
 
 \begin{figure}
\begin{minipage}{.58\linewidth} 
\centering\epsfig{figure=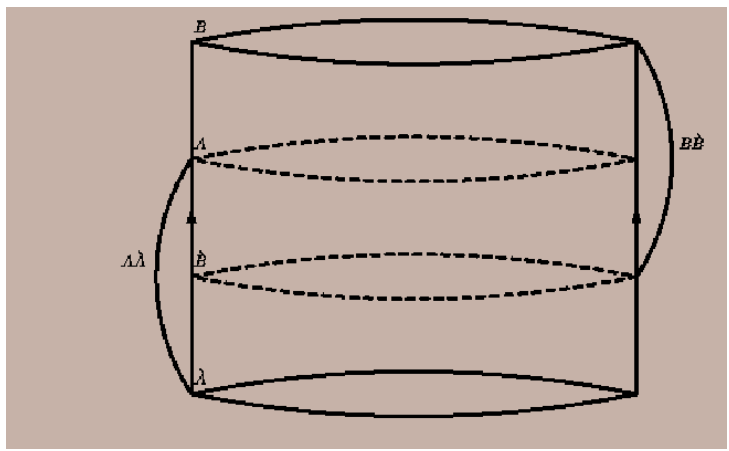,width=\linewidth}
 \caption[fegf7]{The cylinder with the four pistons. The pistons $A$ and 
$\grave A$ are
fixed  and $B$ and $\grave B$ may move along the cylinder.  Also the piston $A$
is permeable to the molecule inside the interval $(x_1,x_2)$ and $\grave B$ to
those outside it.  } 
\end{minipage}

\end{figure}

 We wish to test  the validity of the  connection   
 (suggested in \cite{Szilard}) 
 between the two variables $x$ and $f$, 
 where the
 latter  denotes  the  property  that if
 any of the $n$ particles is found in some preassigned interval $(x_1,x_2)$ 
 then  we    
 assign to $f$ the value of $+1$   otherwise   $f$  assumes the value of $-1$. 
 That is, the relevant proposed connection between $x$ and $f$ 
 is  \cite{Szilard}  
 \begin{equation} \label{e16}   f(x)= 
 \left\{ \begin{array}{ll}  +1& {\rm for~} 
 x_2 \ge x  \ge x_1 \\
 -1 & {\rm for \    x   \ outside~}  (x_1,x_2)  \end{array} \right. \end{equation} 
 We assume that the piston $A$ is permeable only to  the particles inside the
 interval $(x_1,x_2)$ and $\grave B$  only to those outside it.  We
 denote by $w_1$ the initial probability  
  that any randomly selected particle is found
 to be in the interval $(x_1,x_2)$ and by $w_2$ that it is outside it. 
At  first the pistons $B$ and $\grave B$ were at the positions of 
$A$ and $\grave A$ respectively
 and all the $n$ particles were in the one space between. Now, we  wish to test
 the assumed relation from Eq (\ref{e16}) by performing, reversibly and with no
 external force, a complete cycle of first moving up the pistons $B \grave B$
 and then retracing them back to their initial places so that the only assumed
 relation between the  particles and their positions along the axis $x$ 
 is that from Eq (\ref{e16}). Thus,  we
 first  move, without
 doing work, the pistons $B$ and $\grave B$ so that, as remarked, the volume
 enclosed between them equals that between $A\grave A$ 
and  we obtain two separate equal volumes, each of which equals to
 the initial one.     Now,  since $A$ is
 permeable to the particles in the interval $(x_1,x_2)$ and $\grave B$ to the
 rest the result is that the upper volume  $B\grave B$ contains only the particles from
 the predetermined interval $(x_1,x_2)$ and the lower $A\grave A$ only the
 others.  \par  We want now to retrace our former steps and   move,  again without doing
 work,  the pistons $B$ and $\grave B$ to the places of  $A$ and $\grave A$
 so as to have, as before, the same initial volume and thus to complete one
 cycle.  We must  take into account, however,   that  
  during the upward motion  some particles that were 
  inside (outside) 
 the interval $(x_1,x_2)$ may come out of (into) it 
  due to thermal or other kind of
 fluctuation so that  these particles change 
  from the kind that may pass through the piston $A$  ($\grave B$)  
  into the kind
 that is not allowed to do that. Thus, the last step of retracing the pistons
 $B$, $\grave B$ into their former initial positions at the pistons $A$, $\grave
 A$ respectively can not be performed without doing work since the molecules
 that have come out of (into) the  interval $(x_1,x_2)$ are not permitted now
 to pass through $A$ ($\grave B$). That is,  the former process of expanding 
 the volume is not reversible as
 described because we have to exert force on these molecules to move them back
 into (out of) the interval $(x_1,x_2)$ so that they can pass through 
 $A$ ($\grave B$).  Thus, the relation (\ref{e16}) is
 not valid any more since it does not take into account the external force just
 described. \par We may express this in a quantitative manner by noting that 
 there
is  now \cite{Szilard}  a decrease of entropy per molecule 
 after the first step of moving up the pistons.  This  may be calculated by taking
 account of the fact that now  the  probabilities   to find
 any randomly selected molecule out of (in) the preassigned interval 
 $(x_1,x_2)$   are different from the 
  initial values $w_2$ and $w_1$ before moving up the pistons.  Thus, suppose that during
 the first stage of expanding  the initial volume of the cylinder $n_o$ molecules,
 from the total number $n$,  have come out of  the remarked interval and $n_i$
 from outside have entered   so that
 the probability to find now any randomly selected molecule out of it is
 $(w_2+\frac{(n_o-n_i)}{n})$ and that  to find it in  is  
 $(w_1+\frac{(n_i-n_o)}{n})$.  Thus, the initial entropy per molecule, 
 denoted by $s_i$, before moving up the pistons is \cite{Szilard} 
 \begin{equation} \label{e17} s_i=-k(w_1\ln w_1+w_2\ln w_2), 
 \end{equation}
 and after moving-up the pistons the corresponding entropy per molecule, denoted
 by $s_m$, is \begin{eqnarray}  && s_m=
- k((w_1+\frac{(n_i-n_o)}{n})\ln (w_1+\frac{(n_i-n_o)}{n})+\label{e18} 
\\ &&+(w_2+\frac{(n_o-n_i)}
{n})\ln
 (w_2+\frac{(n_o-n_i)}{n})) \nonumber  
 \end{eqnarray}
 The difference in the entropy per molecule  between the two situations from Eqs
 (\ref{e17})- (\ref{e18}) is \begin{eqnarray} && s=-(s_m-s_i)=-(kw_1(\ln
 (w_1+\frac{(n_i-n_o)}{n})-\ln w_1)+kw_2(\ln
 (w_2+\nonumber \\ && +\frac{(n_o-n_i)}{n})- \ln w_2)
 + k\frac{(n_o-n_i)}{n}(\ln (w_2+\frac{(n_o-n_i)}{n})-\label{e19} \\ && - 
   \ln (w_1+\frac{(n_i-n_o)}{n}))) =
   -(kw_1\ln (1+\frac{(n_i-n_o)}{w_1n})+ 
   kw_2\ln (1+\frac{(n_o-n_i)}{w_2n})+ \nonumber \\ &&+ k\frac{(n_o-n_i)}{n}
 \ln (\frac{w_2+\frac{(n_o-n_i)}{n}}
 {w_1+\frac{(n_i-n_o)}{n}}))= -(k(w_1(1+\frac{(n_i-n_o)}{w_1n})
 \ln (1
 +
 \frac{(n_i-n_o)}{w_1n})+ \nonumber \\ && +
 w_2(1+\frac{(n_o-n_i)}{w_2n})\ln (1+\frac{(n_o-n_i)}{w_2n})+ 
  \frac{(n_o-n_i)}{n}\ln
 (\frac{w_2}{w_1}))) \nonumber \end{eqnarray}
  Eliminating $w_2$ through use of 
  the relation $w_1+w_2=1$ one may write the last equation as 
 \begin{eqnarray} &&   s=-(s_m-s_i)= 
 -k(w_1(1-\frac{(n_o-n_i)}{nw_1})\ln (1-\frac{(n_o-n_i)}{nw_1})+ \label{e20} \\ 
&& + (1-w_1)(1+\frac{(n_o-n_i)}{n(1-w_1)})\ln (1+\frac{(n_o-n_i)}{n(1-w_1)})+ 
  \frac{(n_o-n_i)}{n}\ln
 (\frac{(1-w_1)}{w_1})) \nonumber \end{eqnarray} 
 If $n_o=n_i$,  the entropy difference from Eqs (\ref{e20})  
  is  obviously zero which results in the validation of the relation 
 (\ref{e16}) after returning the pistons back to their initial places as
 remarked. When $n_o \ne n_i$ 
 the expression (\ref{e16}) can not be
 validated by retracing, without doing work,     the volume back to its 
 initial value 
  since now the  molecules that come out of the interval $(x_1,x_2)$ and those
  that have entered it  prevent 
  this reversible motion which is necessary for its validation. Thus, a new
  expression, instead of the invalid one from Eq (\ref{e16}), that takes account
  of these molecules must be adopted as in \cite{Szilard}.  
    But  before writing this expression we 
  remark   that the
  probability $w_1$ must be  proportional to the length of the remarked interval $x_2-x_1$, 
  so that a small or large value for one indicates a corresponding value for the
  other. Thus, we may define a probability distribution for $w_1$ in terms of
  the variable $x$ and assume a normal distribution \cite{Spiegel} 
   so that we may write for the density function of $w_1(x)$  
  $f_{w_1}(x)=\frac{\exp(-\frac{(x-\mu)^2}{2\sigma^2})}{\sqrt{2\pi}\sigma}$,  
  where $\mu$ is the mean value of $x$ and $\sigma$ is the standard deviation.
  To simplify the following calculation we assume a standard normal distribution
  \cite{Spiegel} $z=\frac{(x-\mu)}{\sigma}$ for which  $\mu=0$ and  $\sigma=1$. Thus,  the density function
   $f_{w_1}(x)$ may be written as $ f_{w_1}(z)=
  \frac{\exp(-\frac{(z)^2}{2})}{\sqrt{2\pi}}$ and the probability $w_1(x)$ to
  find any randomly selected molecule in the interval $(-x,x)$, where now this
  interval is symmetrically located around the origin $x=0$, is 
  \cite{Spiegel}
  \begin{equation} \label{e22} w_1(x)=\int_{-x}^xf_{w_1}(z)dz
  =\frac{1}{\sqrt{2\pi}}\int_{-x}^xdze^{-\frac{z^2}{2}}=
  erf(\frac{x}{\sqrt{2}}) \end{equation} 
 $erf(x)$ is the error function defined as $erf(x)=\frac{2}{\sqrt{\pi}}\int_0^x
 e^{-u^2}du$. Note that $erf(0)=0$, $erf(\infty)=1$, and $erf(-x)=-erf(x)$ so
 that this function is appropriate for a representation of the probability
 $w_1(x)$. Substituting from Eq (\ref{e22}) into Eq (\ref{e20}) we obtain 
\begin{eqnarray} &&   s=-(s_m-s_i)= 
 -k(erf(\frac{x}{\sqrt{2}})
 (1-\frac{(n_o-n_i)}{n\cdot erf(\frac{x}{\sqrt{2}})})
 \ln (1-\frac{(n_o-n_i)}{n\cdot erf(\frac{x}{\sqrt{2}})})+ \nonumber \\ 
 &&+ (1-erf(\frac{x}{\sqrt{2}}))(1+\frac{(n_o-n_i)}
 {n(1-erf(\frac{x}{\sqrt{2}}))})\ln (1+\frac{(n_o-n_i)}
 {n(1-erf(\frac{x}{\sqrt{2}}))})+ \label{e23}  \\ 
 &&+ \frac{(n_o-n_i)}{n}\ln
 (\frac{(1- erf(\frac{x}{\sqrt{2}}))}
 {erf(\frac{x}{\sqrt{2}}) })) \nonumber 
  \end{eqnarray}  
 Note that in order  to have no negative
  expression under the $ln$ sign, especially for the following numerical
  simulations, we take   the absolute values 
  of these  expressions  which  does not change the real 
  calculated
  results.  The right hand side   of Eq (\ref{e23}),  which yields the entropy decrease
  per molecule,  must be multiplied by the number $n$ of molecules in the cylinder
  in order to obtain the total decrease of entropy after moving up the pistons. 
  Figure 5.5 shows a three-dimensional representation 
   of the entropy $s$  
  per molecule from the
 last equation as function of $\frac{n_i}{n}$ and $\frac{n_o}{n}$.

\begin{figure}
\begin{minipage}{.58\linewidth} 
\centering\epsfig{figure=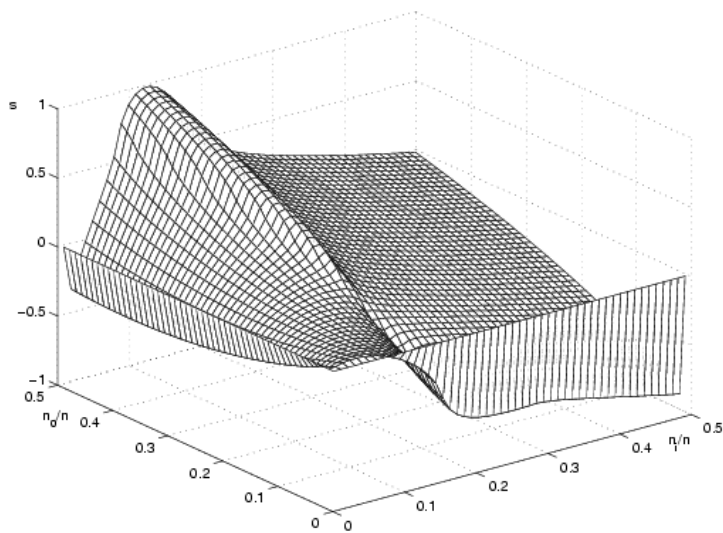,width=\linewidth}
\caption[fegf8]{The figure shows a three-dimensional surface  of the entropy per 
molecule $s$ 
as function of $\frac{n_o}{n}$ and $\frac{n_i}{n}$.  Both ranges of 
$\frac{n_o}{n}$ and $\frac{n_i}{n}$ are $(0.005, 0.5)$ since it is unexpected
that in a reversible motion more than half of the total molecules will leave or
enter the given interval $(x_1,x_2)$. Note that for large $\frac{n_o}{n}$ 
($\frac{n_i}{n}$) and small $\frac{n_i}{n}$  ($\frac{n_o}{n}$) $s$ tends to $+1$
$(-1)$. }
\end{minipage}
\end{figure}

     The
 relevant range of $w_1=erf(\frac{x}{\sqrt{2}})$  
must  begin from the  minimum value
 of $\frac{n_0}{n}$  since $w_1$ can 
 not be smaller than $\frac{n_0}{n}$. The ranges of both $\frac{n_i}{n}$ 
 and $\frac{n_o}{n}$ are $0.005 \le \frac{n_i}{n}, \frac{n_o}{n} \le 0.5$
 because in the reversible motion discussed here it is unexpected that more 
 than
 half of the total particles will enter or leave the interval $(x_1,x_2)$. 
 For large  values of 
  $\frac{n_0}{n}$ ($\frac{n_i}{n}$)  and comparatively small values of  
  $\frac{n_i}{n}$ ($\frac{n_o}{n}$) the entopy
  differences tend to $+1$ ($-1$) and when  both $\frac{n_0}{n}$ 
  and $\frac{n_i}{n}$ are large $s$ tends to zero from negative values. 
        \par 
 The remarked problem of moving back the pistons, without
 doing work, to their original volume 
     has been solved in \cite{Szilard} by taking
 into account  Eq (\ref{e16}) which assign to $f(x)$
 the value of $+1$ when the relevant molecule was in the interval $(x_1,x_2)$ 
 and $-1$ otherwise. That is,  
 after the first step of doubling the initial  volume the cylinder 
  includes now, except for the
 particles that remain inside (outside)  the noted interval and characterized by
 $f(x)=1$ ($f(x)=-1$), also those that {\it were} 
  in (outside) it and {\it were}  denoted by these values. All these particles 
  that were in (out of) the interval  $(x_1,x_2)$ continue,  for the short time
  interval between moving the pistons up and down,  to be denoted by  
  $f(x)=1$ ($f(x)=-1$). 
  Thus, as noted in 
 \cite{Szilard},  for  the last step of retracing to  the original volume, 
 without doing work,  one has only to replace the pistons $A$ by $A^*$ that is
 permeable not with respect to the molecules  in the interval
 $(x_1,x_2)$  but to those that their $f(x)$ is $+1$. Correspondingly,  the piston
 $\grave B$ is replaced by $\grave B^*$ that is permeable to those  
 that their
 $f(x)$ has the value of $-1$. Thus,  the external  intervention in this case is 
 changed in  \cite{Szilard} from Eq (\ref{e16}) to 
 \begin{equation} \label{e24}   f(x)= 
 \left\{ \begin{array}{ll}  +1& {\rm for \ x \ that \ is \ or \ was \ in~} 
 x_2 \ge x  \ge x_1 \\
 -1 & {\rm for \    x   \ that \ is \ or \ was  \  outside~}  (x_1,x_2)  \end{array} \right. \end{equation} 
 In such a way one is able to perform a complete
 cycle of first expanding the volume with the original permeable pistons $A$ and
 $\grave B$ and then retrace this step reversibly with the pistons $A^*$ and     
$\grave B^*$ instead of $A$ and  $\grave B$ as remarked. Thus, all the possible
motions of the molecules, including their coming out of or into the given
interval $(x_1,x_2)$ are accounted for by Eq (\ref{e24}) which
results in its validation as remarked (see the discussion after Eq
(\ref{e16})). \par 
Now, if we take into account  the  possible Feynman paths
  \cite{Feynman2} through which the system may evolute during 
 the remarked
complete cycle then such paths may be characterized also by those that conform either 
to Eq (\ref{e16})  or to (\ref{e24}). That is, the Feynman paths that 
may result in a decrease of  the
entropy are those  in accordance  with (\ref{e16}) and those that do not
result with such a change  correspond  to (\ref{e24}). Thus, using (\ref{e24}) is 
the same as
passing along the specific  path that preserve the entropy and rejecting
those that change it (that conform to (\ref{e16})). All one have to do is to
``realize'' the correct path,  in the sense of \cite{Aharonov}, which is done 
in accordance with the former section 
by having  
 a large number of observers  each   moving his respective
 piston up and down  in the described manner.   We calculate 
  the correlation among
 the $N$ separate systems by assuming that all begin with the original pistons $A$, 
 $\grave A$,  $B$ and $\grave B$ and finding the number of them that after
 completing one cycle  are found with the pistons $A$,  
 $ A^*$,  $B$ and $\grave B^*$ which denote that the  expression
 (\ref{e16}) is not valid for them.  When, after expanding 
  the initial volumes of the $N$ cylinders we find,  for some 
 of them,    that no molecule 
 from   
 the interval $(x_1,x_2)$  has come out of 
  it and no one from outside has entered then
 they end, after  returning the volume to its
 initial state, with the same pistons they begin with and in such a case the
 expression (\ref{e16}) is obviously valid for  them.  
 But suppose that 
 for other   observers 
 $n_{o_j}$ molecules come out of the interval $(x_1,x_2)$ and $n_{i_j}$ 
 have entered where $n_{o_j} \ne n_{i_j}$.   In such case 
  the total decrease of entropy, after moving-up the pistons,  
    using  Eq (\ref{e23}) and  assuming  that the total number of 
    molecules $n$  are
 the same for all the ensemble members,   is \begin{eqnarray}
 && s_{total}=-k\sum_{j=1}^{j=N}n(w_{1_j}\ln (1+\frac{(n_{i_j}-n_{o_j})}{w_{1_j}n})+
 w_{2_j}\ln (1+\frac{(n_{o_j}-n_{i_j})}{w_{2_j}n})+ \nonumber \\ && + \frac{(n_{o_j}-n_{i_j})}{n}
 \ln (\frac{w_{2_j}+\frac{(n_{o_j}-n_{i_j})}{n}}
 {w_{1_j}+\frac{(n_{i_j}-n_{o_j})}{n}})= 
 -k\sum_{j=1}^{j=N}n(erf(\frac{x_j}{\sqrt{2}})
 (1-\label{e25} \\ && -\frac{(n_{o_j}-n_{i_j})}{n\cdot erf(\frac{x_j}{\sqrt{2}})})
 \ln (1- 
 \frac{(n_{o_j}-n_{i_j})}{n\cdot erf(\frac{x_j}{\sqrt{2}})})+  
 (1-erf(\frac{x_j}{\sqrt{2}}))(1+\frac{(n_{o_j}-n_{i_j})}
 {n(1-erf(\frac{x_j}{\sqrt{2}}))})\cdot \nonumber \\ && \cdot 
 \ln (1+\frac{(n_{o_j}-n_{i_j})}{n(1-erf(\frac{x_j}{\sqrt{2}}))})
 + \frac{(n_{o_j}-n_{i_j})}{n}\ln
 (\frac{(1-erf(\frac{x_j}{\sqrt{2}}))}
 {erf(\frac{x_j}{\sqrt{2}})})) \nonumber \end{eqnarray} 
  We, now,  show that when the $N$ observers are related to each other in the
  sense that all the $N$ experiments of moving the pistons up and down are
  prepared in such a way that no two observers share the same values of 
 $\frac{n_{o_j}}{n}$,  $\frac{n_{i_j}}{n}$ and $x_j$ then  the  
  larger  is  
    $N$ the more probable is that the majority of them obtain negative entropy
    difference.  If, on the other hand, they are not related in this manner   
     so that some observers  may have the same $\frac{n_{o_j}}{n}$,  
     $\frac{n_{i_j}}{n}$ and $x_j$ 
     then the mentioned probability will not be obtained even for large
    $N$. We first note  that     
   since for  $x \ge 3$ 
  $erf(x)$  is approximately the same as for $x=\infty$, we may
  assume a range of $(0,3)$ from which we take the values for the preassigned
  interval $(-x,x)$. Thus,  we
  subdivide the interval $(-3,3)$ into $N$ subintervals, where $N$ is   the
  number of  observers, so that each  has his 
  respective
  interval $(-x_j,x_j)$  where $x_j$ is the corresponding  real number from 
  the range $0 \le
  x_j \le 3$. We note, as remarked after Eq (\ref{e23}),   that each 
  probability $w_{i_j}$ for any 
  observer $O_j \ \ (j=1, 2,\ldots, N)$ must begin from the  minimum value 
  of $\frac{n_{o_j}}{n}$ and assume (see the discussion after Eq (\ref{e23})) 
   that both  $\frac{n_{i_j}}{n}$ and $\frac{n_{o_j}}{n}$ are 
   in the range $0.005 \le \frac{n_{o_j}}{n}, \frac{n_{i_j}}{n} \le 0.5$.  
   We assign to each observer that obtains negative entropy difference after
   moving-up the pistons the value of $+1$ (this has nothing to do with the $+1$
   or $-1$  in Eqs (\ref{e16}) and (\ref{e24})) and 0 otherwise.  \par  We 
   assume
   that $N_c$ observers, of the total $N$,   lift up their respective
   pistons  and  we calculate the fraction \begin{equation} \label{e333} 
   f_{N_c}(N)=\frac{1}{N}\sum_{i=1}^{i=N_c}g_i(N),  \end{equation}  
  where $g_i(N)=1$ if the result of any of these experiments yields a 
  negative
   value for the entropy difference 
 and $g_i(N)=0$ otherwise.  It is shown, as 
   remarked, that as $N$ grows this fraction increases and with it 
   the probability that most of
  them obtain negative entropy differences  in which case they end, according to
  the rules of the experiment (see the discussion after Eq (\ref{e24})),  with 
  the replaced pistons $A^*$ and $\grave B^*$ in place of the original ones 
   $A$ and
  $\grave B$ so that the relation (\ref{e24}) is established for most of them. 
  Figure 5.6  shows the remarked fraction $f_{N_c}(N)$  
   as a
  function of the $N$ observers and we see that $f_{N_c}(N)$ grows as $N$
  increases. That is, the presence of a large number of observers 
  resulted with the outcome that  a large number of
  experiments  end with a negative entropy difference. The same result has
  been obtained in the former section for the effect of a large ensemble of
  related observers that perform  experiments where any  result obtained by any one  
  is valid for all
  the others (see the discussion after Eq 
  (\ref{e5})
  and also Figure 5.2).  
  \begin{figure}
\begin{minipage}{.48\linewidth} 
\centering\epsfig{figure=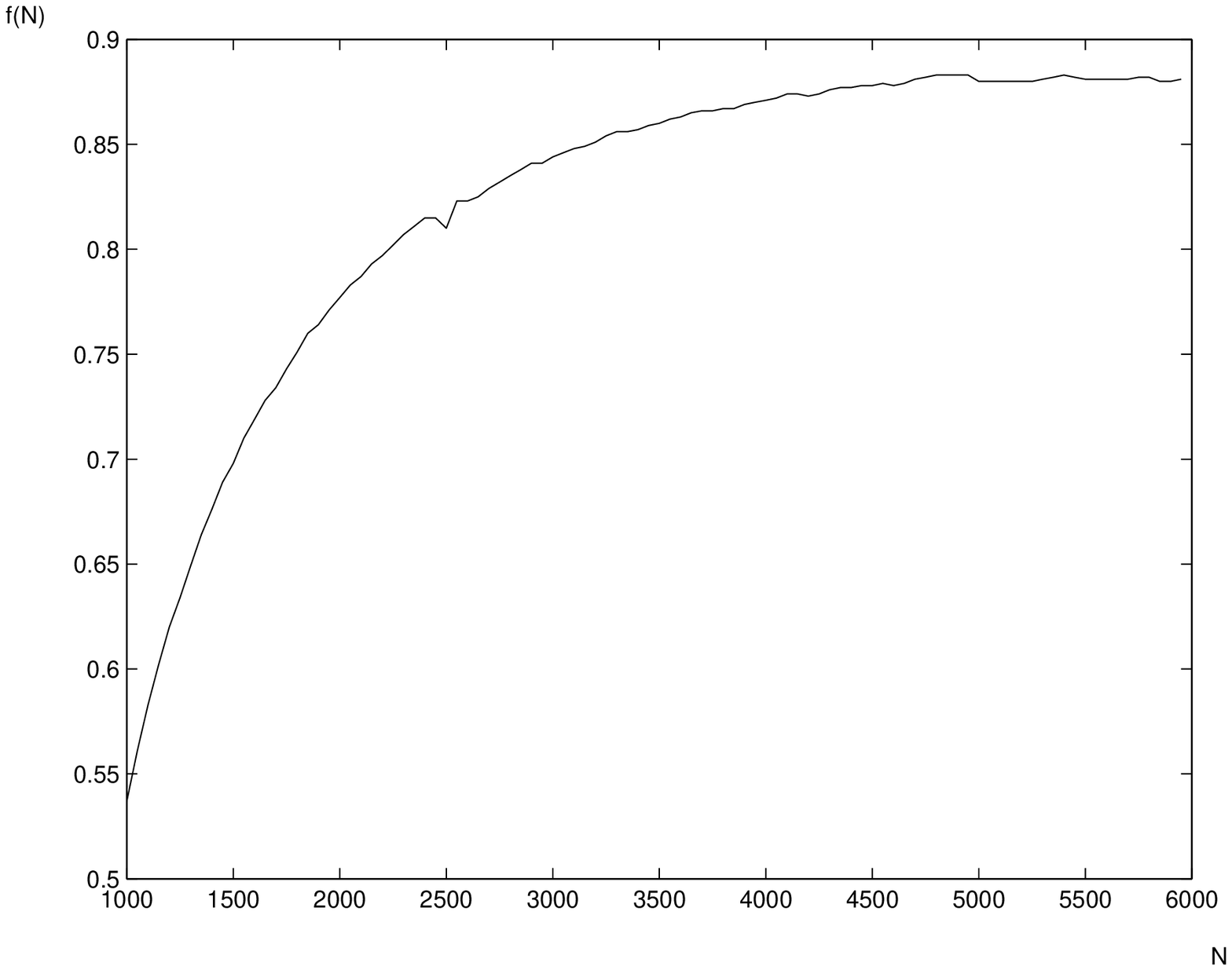,width=\linewidth}
\caption[fegf9]{The curve shows the result of performing 1000 different
experiments of lifting up the pistons as a function of the number of observers
$N$ (that only 1000 of them perform the experiments). Note that no two of the
1000 experiments are identical and that each is deliberately performed 
 for different  values of the
intervals $(-x_j,x_j)$, $\frac{n_{o_j}}{n}$ and $\frac{n_{i_j}}{n}$ where 
$x_j=6\cdot \frac{n_{o_j}}{n}$.  Note 
that as  $N$ grows the number of experiments that end in negative entropy
decrease increases.}
\end{minipage} \hfill
\begin{minipage}{.48\linewidth} 
\centering\epsfig{figure=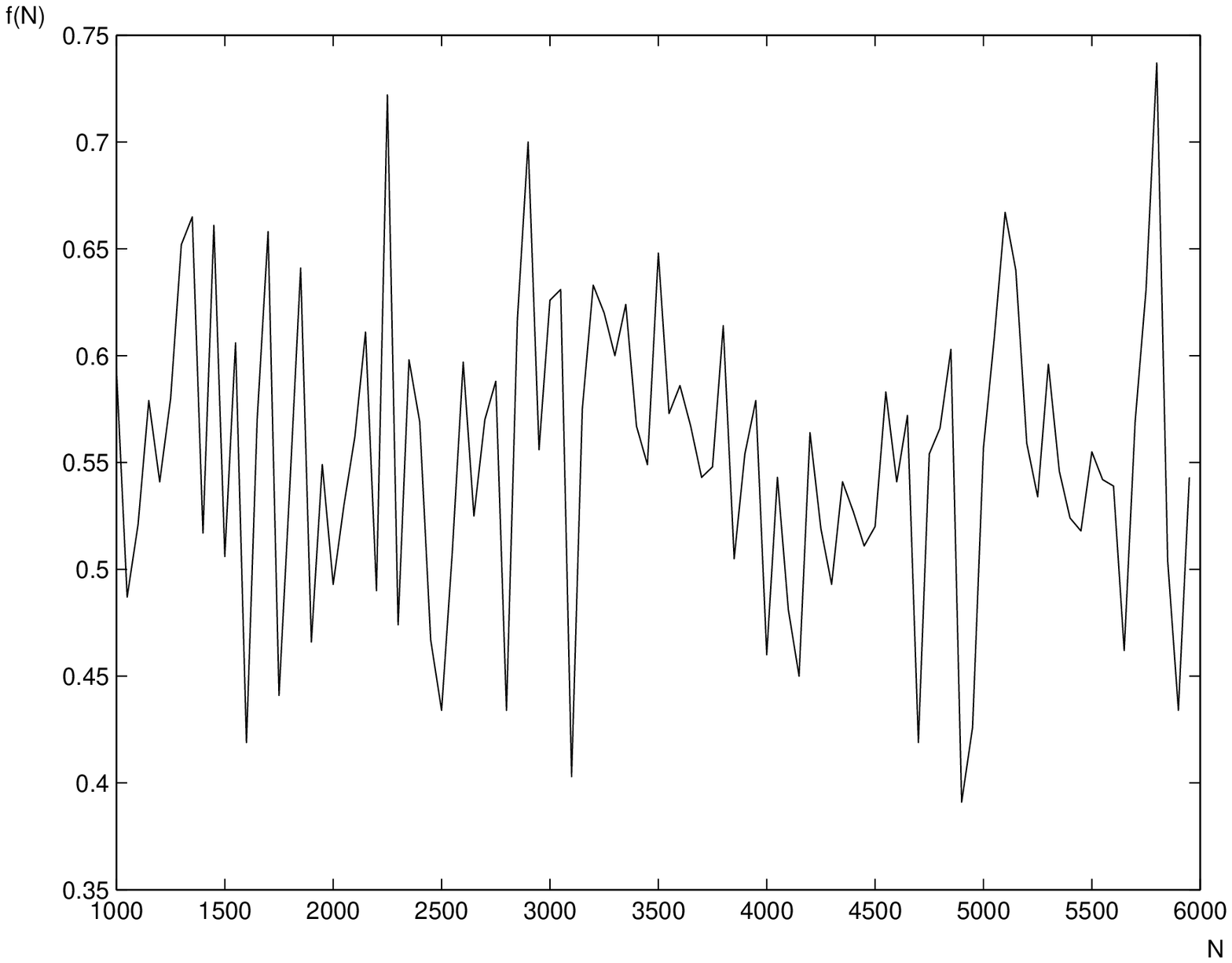,width=\linewidth}
\caption[fegf10]{The stochastic curve shown is drawn for exactly the same
conditions as those of Figure 5.6, except that the values of  $(-x_j,x_j)$, 
$\frac{n_{o_j}}{n}$   and $\frac{n_{i_j}}{n}$ are chosen randomly.  That is, the
curve shows the results of 1000 experiments as a function of the number of
observers $N$.  Note  that some of these experiments  may be identical
due to the random conditions under which they are performed. Also, as seen from
the graph there is no discernable increase or decrease  of $f(N)$ with $N$ 
(compare  with Figure 5.6). }
\end{minipage}
\end{figure}
    The results of Figure 5.6 are obtained by choosing for 
  $\frac{n_{o}}{n}$ and   $\frac{n_{i}}{n}$, that are both confined 
   in the range $(0.005, 05)$, the  values of 
  $\frac{n_{o}(j)}{n} =0.005+\frac{(0.5-0.005)j}{49}$, 
    and $\frac{n_{i}(l,k)}{n} = 
  0.005+\frac{(0.5-0.005)l}{(20+k)}$, 
  where  $0 \le j \le 49$,   $0 \le l \le 120$  and 
  $0 \le k \le 100$. Note that the common total interval $(0.005, 05)$  of 
  $\frac{n_{o}}{n}$ and   $\frac{n_{i}}{n}$ has been subdivided differently for 
  them. That is, for  $\frac{n_{o}}{n}$ the division is to 50 equally spaced
  values and for $\frac{n_{i}}{n}$ it is to 121. That is, the maximum possible
  number of  observers  is $N_{max}=121 \cdot 50=6050$  and we assume, for the
  numerical analysis,  that 
    1000 of them
 participate in the   experiments so   that $N_{c}=1000$. 
 Thus, 
      all the observers  are related in the
  sense that each has its specific   
  $\frac{n_{o_j}}{n}$ and    $\frac{n_{i_j}}{n}$.  If, on the other hand, this kind of
  connection is absent as when   
   assigning randomly to any observer $O_j \ \ (j=1, 2, \ldots, N)$ an
  interval $(-x_j,x_j)$ and also $\frac{n_{o_j}}{n}$,   
  $\frac{n_{i_j}}{n}$ from $(0.005, 0.5)$ we obtain a stochastic result for 
  $f(N)$ that
  implies no increase (and no
  decrease)  of the
  number of observers that get negative entropy difference.  This is seen 
   clearly  from the sawtooth
  form of the curve of Figure 
  5.7 which is drawn under exactly the same conditions as those of  Figure 5.6
   except that the values of $(-x_j,x_j)$, $\frac{n_{o_j}}{n}$ and  
  $\frac{n_{i_j}}{n}$ are  randomly assigned to the observers.   
   \par We, thus,  see that when the 
  observers are related among them 
  the probability to find any one of them (that   
 begin  with the pistons $A$, $\grave A$, $B$ and $\grave B$)   ending   with 
 $A$, $ A^*$, $B$ and $\grave B^*$ is large so that, as remarked,  
 the relation (\ref{e24}) 
 is physically established the  larger is the number $N$ of related observers.   
  We note that 
for Figures 5.6 - 5.7 we have expressed the values of $x_j$ from the interval 
$(-x_j, x_j)$ in terms of $\frac{n_{o_j}}{n}$ as 
$x_j=6 \cdot \frac{n_{o_j}}{n}$. In such a way each observer $O_j (j=1, 2,
\ldots N)$ that has been assigned a specific value of $\frac{n_{o_j}}{n}$ from
the remarked range $0.005 \le \frac{n_{o_j}}{n} \le 0.5$ is automatically
assigned a corresponding value from the range $0.03  \le x_j \le 3$.  Thus, 
the maximum value of $x_j$ is in accordance with the discussion after Eq  
(\ref{e25}). 
    \par 
   We note that the same results may be obtained by using other methods 
  and terminology.
  Thus, 
   it is shown \cite{Gisin2} that the ``localization''  (in the
 sense of smaller dispersion) for the state $|\phi\!>$ is greater the  smaller 
  is
 the entropy which results when  the rate of ``effective interaction
 with the environment'' \cite{Gisin2} increases.  Localization is another 
 phrase  for 
 what we call here
 ``realizing or preserving a specific state''   and the interaction with the 
 environment is equivalent to
 performing  experiment \cite{Harris,Davies,Zeh,Amann}, so that as the rate of performing
 experiment grows the more realized and  
  localized is the state one  begins with or the path of states along which one
 proceeds. \par  
  We see, therefore, that the classical thermodynamical  system discussed here obeys
also the same development  we have encountered in the former chapters  
regarding the
evolution  of  new    phenomena (and their theories).  That is, the 
establishment 
of the
assumed physical connection from Eq (\ref{e24}) between the variables $f$ and
$x$ proceeds by first trying  the form (\ref{e16}) but it soon
becomes clear,  by testing it through experiments,  that there is a 
gap
between it  and the experimental results with regard
to the molecules that come out of (into) the preassigned interval $(x_1,x_2)$. 
Thus, in
order to conform to the experimental findings one have to replace  the weak
theory of (\ref{e16}) by  the new expression 
 from Eq (\ref{e24}) which takes into account these  molecules. The new 
 theory entails a
corresponding change in the experimental set-up that is supposed to validate it.
That is, the replacement of the pistons $A$ and $\grave B$ by $A^*$ and $\grave
B^*$ in the second stage of reversing back, without doing work, to the former
volume. The important step that assigns to the relation  (\ref{e24}) 
 a physical aspect is obtained, as remarked,  
   when  a large number of related 
observers perform the relevant experiments with their cylinders   
 and obtain similar results (see Figure 5.6). Unrelated ensemble of observers, 
 no matter how
 large it is, does not obtain the sought-for physical validation of the new
 relation from Eq (\ref{e24}) as seen clearly from Figure 5.7. \par The new
 response (new phenomenon) in this case is clearly illustrated in Figure 5.6
 which shows that the larger is the ensemble of related observers the larger is
 the number of them that obtains negative entropy difference. The last result
 indicates, as remarked, that the system responds in a new  and different 
  manner from the  expected response of  the old theory (\ref{e16}). Thus, the
  failure of the relation (\ref{e16}) to cope with the experimental findings
  necessitates a new relation (\ref{e24}) which has been shown to be more
  validated the larger is the number of related observers. In other words, as
  seen in all our work thus far, the large number of related observers that
  perform the experiments not only give rise to 
   this new response of finding most of them with a negative entropy
  difference but also physically establish this outcome.     
   \par


  
 \pagestyle{myheadings}
\markright{ \ \ \ EPILOG} 

\protect\chapter{\bf   EPILOG}

We have discussed, using examples that range from  quantum and classical field events,  
 through 
Internet webmastering to classical thermodynamics,   the possible 
evolutions of
 real phenomena especially at  the initial stage of which when 
 they are first encountered and their nature is not clear.  The physical 
 character of these new-encountered phenomena is established  only after initially
 trying to explain them   by  suitable theories  that must stand 
  the
 tests of  experiments which  are prepared  to reconstruct these phenomena
under various conditions.  We have shown  
 that the  larger is the number of repetitions of the relevant 
experiments the more valid and real these  new phenomena  will be 
assuming that most of these 
experiments result in establishing them.  We have also shown that the process 
of validating and establishing the physical aspect of  real phenomena
may be explained as if it results from   experimenting upon 
a physical system  in
an uninterrupted manner, that is, to the Zeno effect.  
Under this condition of dense measurement the known
response of the system, in the absence of this dense experimentation,  changes
entirely to a new,  and even unexpected,  one as shown with respect to the
one-dimensional systems discussed in Chapter 4.  This new behaviour of the system
may always be repeated and reconstructed any number of times by performing this
dense measurement again. That is, it may be physically established under  this 
kind of experimentation.  \par 
 We note that at the initial stage 
before the new response becomes physically
established by dense measurement the system's response and evolution under a
measurement (not dense)  may assumes any of a large number 
of possible different ones. Thus, the  corresponding initial theory, 
before
it becomes physically established through the remarked experiments,   may be
described as having a possible 
dependence   upon   an extra variable  that takes account of the 
large possible evolutions,  
allowed at this initial time,  for the relevant system.  
In Section 2.2 we have discussed (see Appendix $A_2$)  
the generalization of
quantum mechanics through the Flesia-Piron extension of the Lax-Phillips theory
in which the extra time variable  $t$ has been explicitly introduced besides the
laboratory time parameter $\tau$. We do so also in \cite{Bar2}  by using  
 the stochatic quantization  theory of Parisi-Wu \cite{Parisi1,Namiki1}
 where  the introduced  extra variable $s$ 
  takes account of an assumed stochastic process (in
this variable) that allows, as all stochastic processes do, a large different
possible behaviours of the  system. The equilibrium configuration is
obtained \cite{Parisi1,Namiki1} when this variable is eliminated through 
equating all its different
values to each other and taking to infinity.   
This equating of all the possible $s$ values 
to each other introduces an element of repetitions  
of the same process through     
 which the
system is stabilized and brought to its physical equilibrium configuration. We
have shown this in \cite{Bar2}  for the examples of the Internet websites of the 
harmonic oscillator and those 
 of the
energy shift (Lamb shift) \cite{Lamb,Hansch}. The same result was obtained 
in \cite{Bar001} for the
general Internet websites without specifying  any particular one. For this
 discussion of the Internet statistics we use the formalism  of
Ursell-Mayer \cite{Mayer,Ursell} in which an extra variable has been introduced.
We note that this additional variable is not of the same kind as that of the SQ
extra variable \cite{Namiki1,Parisi1}   discussed in \cite{Bar2}. This 
extra variable is
implicitly referred to in Section  2.1 (see Appendix $A_1$) when we discuss 
the bubble and open-oyster processes. This is  because we do not sum there the
Feynman diagrams of these processes to all orders, as usually done in quantum
field theory, but their $n$ time repetitions where $n \to \infty$. The resulted
double sum has the same effect as introducing an extra variable and summing over
all its possible values.  \par
The same effect of  establishing the physical character of the  new phenomena 
and their theories 
is obtained  through a collective
experiment performed by a large number of {\it related} observers.   
It    results, as remarked,  in ``realizing''  \cite{Aharonov,Facchi} 
 the relevant Feynman path of states, that represents the tested phenomenon, 
  {\it for all} the observers 
   as if each has performed dense measurement along it.  This is because,  although
 each one is restricted to do only  his specific experiment,  nevertheless,
 since all the observers have  similar systems the specific results
  obtained by any  one  may, under the same conditions, be obtained by any 
  other.  
 Thus,   the  experimental results obtained by any one  are valid  for  all 
 the others. \par  
  This outcome  of the ensemble of related systems was  corroborated  also upon 
 comparing the mechanism just described to
the procedure of numerical simulations.    This has been shown  in
\cite{Bar2} with regard to  a large ensemble of related computers and in  
\cite{Bar001} 
in connection to a corresponding large cluster of mutually linked websites  
and in both cases the same outcome was obtained. 
Thus, the relationship among the ensemble of related computers is strengthened
  by seeing to it  that  the actions $S$ of all or most of the path integrals 
  that
  represent them contain the same expression \cite{Bar2} that characterizes the involved
  interaction simulated by them such as that of the  
  harmonic oscillator or of  the Lamb shift \cite{Bar2}.
  In such a way all the related computers may be brought, as shown in
  \cite{Bar2},
  to the situation in which all of them have the same common sites and,
  therefore, their correlation is maximal.   
  These  effects of the  ensemble of related systems were demonstrated also in
  \cite{Bar001}   by   
  using the cluster formalism of
 Ursell \cite{Ursell} and Mayer \cite{Mayer} where   a large ensemble of 
 connected
  computers (users) may aquire a very large additional amount of
 connectivity \cite{Havlin} among them  by adding only a small number  of
 connecting website links.   The correspondence between computer simulations and
 real experimentation is sharply reflected   at the last  stage of running
 the written  code  on the computer screen, especially if it simulates
 some physical process which is governed, as most physical processes are,   by  
 a differential equation. 
 This is because the numerical solutions of these  equations are 
 obtained   by updating them a
large number of times. Moreover, 
the larger is the number of these  iterations 
 the larger is the   number of samples and the  better is the resulting 
 statistics. 
 All the remarked  effects of the ensemble of related
 observers have been demonstrated even more pronouncedly for the classical
 thermodynamical system of cylinder and pistons discussed in Chapter 5. 
 We have also shown that when 
 the observers are not related to
  each other then this ``realization'' of the specific evolution will not
  be obtained.  \par
 All these processes  correspond to  the Zeno effect
\cite{Zeno,Itano,Aharonov,Facchi,Harris,Simonius}  by which an  equilibrium 
physical configuration is
 obtained as a consequence of these repetitions.  We have shown that it   
 is effective not
 only for repeating the same experiment  on   the same system  a large 
 number of times  
  in a finite time but also when  a large number of  
  observers,  all confined in a finite region of space,  perform   similar  
  experiments  as shown by the
  various examples discussed here and in \cite{Bar7,Bar8,Bar11,Bar3}. 
  The last effect is the
  space Zeno effect, described in details in Chapters 4-5,  which has been 
  demonstrated 
  for both quantum and classical regimes and which, actually, stands in the
  basis of  the
   influence of the ensemble of related observers compared to that of 
  the 
  single observer   or of the unrelated ensemble of them. The discussion in 
  Chapter 4  of the
  quantum and classical systems of the respective one-dimensional arrays of potential
  barriers and traps confined to finite spatial section demonstrates in a 
  clear fashion 
  the unique nature of space  dense measurement. The responses of both systems to
  increasing the number of barriers or traps in the finite section, which is
  equivalent to increasing the rate of experiments, is not only new 
  but also contrary to what one may expect as explained in Chapter 4  (see
  Appendices $C_2$, $C_3$ and $C_4$). That is, the dense measurement process has
  produced, as remarked, a new phenomenon that has not been encountered before. 
  Moreover, this new phenomenon is not a temporary one but may be repeated and
  reconstructed any times we wish by performing again these dense measurements.
  In other words, this kind of experimentation not only gives rise to new
  phenomena but also may physically establish and validate them. 
     We remark that the appearance of physical phenomena due to only repeating 
 the same experiment  a
 large number of times  has been experimentally demonstrated
 \cite{Itano} using various methods and techniques.\par
 We note that this Zeno effect by which one Feynman path of states, from a 
 large number of
 possible ones, is ``realized'' whereas the probability for the other paths  
 tends to zero satisfies the consistency conditions
 of the histories  formalism of Gell-Mann-Hartle 
 and Griffiths  \cite{Hartle,Isham,Griffiths}. That is, only one evolution is
 effected in the final stage and not a superposition of them (see Appendix $A_2$).
 
\newpage 
 
 \pagestyle{myheadings}
\markright{APPENDICES} \vspace{40 pt}

\chapter{\bf  APPENDICES}

\newpage
 
 \section{APPENDIX $A$: }  \vspace{20 pt}
 
 \centerline{\bf \large ABSTRACT}
 
 \footnote{The full abstract, as represented in the thesis submitted to the
 Bar-Ilan University, is shown here.}

\bigskip \bigskip \bigskip \bigskip \bigskip \bigskip \bigskip 

{\large  It is accepted  that among the means through which a quantum 
phenomenon decoheres and becomes a classical one is   what is termed 
in the literature the Zeno effect. This effect is named in honor of the  
ancient Greek philosopher Zeno of Elea (born about 485 B.C) 
which is known for his logical paradoxes.  
A representative  example is   that in a race contest between the best runner
of these times Achilles and 
a tortoise  the latter must win if it was permitted to start a small distance
before Achilles.  This is so if we  analyze  the running  of Achilles in terms 
of how much time  it  takes him to cover the distance between them and take 
into account that in this time the tortoise proceeds ahead a small  additional
distance which must also  be covered by Achilles.  Thus,   one  may 
conclude that in the limit 
in which the  distance  unit becomes infinitesimal Achilles will  not proceed 
at all in any finite time. 
This effect has been used in 1977 to analytically predict that one may 
preserve an initial 
quantum state in time  
 by merely repeating a large number of times in a 
finite total time the experiment of checking its state. Since then this effect 
has been experimentally strengthened and has become an established physical fact. 
It has been argued by Simonius  
that the Zeno  effect   must be  related not only  to quantum phenomena but also to many macroscopic 
and classical effects.  Thus, since it operates in both quantum and classical
regimes it must cause to a more generalized kind of decoherence than the
restricted one that ``classicalizes'' a quantum phenomenon. We show that 
this generalized decoherence,  
{\it obtained as a result of  dense measurement},  
not only gives rise to new phenomena that are
demonstrated through new responses of the densely
interacted upon system  but also may physically establish these 
phenomena.    
   For that matter  
we have found and established the  analogous  {\it space Zeno effect} 
and demonstrate its existence 
using analytical and numerical methods. The last effect leads to the necessity  
of an ensemble of related observers (systems) for the remarked physical validation  of  new  
phenomena. As will be shown in Chapters 3-5  of this work  the new phenomena 
(new responses of the
system) that result from  the space Zeno effect may be of an 
unexpected nature. \par  An important process that corresponds 
to testing  physical theories through  experiments    
   is the numerical simulations that
run  and test  programs  on the computer screen. We use this correspondence 
for obtaining a better understanding of real phenomena.     \par  
We use quantum and classical field theories in addition to 
the more conventional methods of analysis.  We also corroborate our analytical 
findings by numerical simulations. 
 Some conclusions and results have already 
been published in articles, especially those dealing with the space Zeno effect 
and its possible realizations in quantum and classical systems.  We present these 
papers or part of them in the Appendices  of this work.}     
   
\newpage 

\bigskip \bigskip \bigskip \bigskip \bigskip \bigskip

\section{ \large  Appendix $A_1$: \\ Quantum field theory and dense measurement}

\bigskip \bigskip \bigskip \bigskip \bigskip \bigskip \bigskip \bigskip 

\footnote{Due to limitations imposed upon  submissions's size the former
paper, which fully appears  in the thesis submitted to the Bar-Ilan
University, is omitted here. This paper may be  downloaded at 
http://www.arxive.org/abs/quant-ph/0112070}

{\noindent \Large The paper \\ ``Quantum field theory and dense measurement'' \\
by D. Bar \\
 Published  in  \\ Int. Jour. Theor. Phys,  {\bf 42}, 443-463  (2003)}

\hbox{}

\bigskip \bigskip \bigskip \bigskip \bigskip \bigskip \bigskip \vspace{6cm}

\newpage

\section{ \large Appendix $A_2$: \\ The histories formalism}  \bigskip \bigskip

\bigskip \bigskip \bigskip \bigskip \bigskip \bigskip \bigskip \bigskip

\footnote{ Due to limitations imposed upon  submissions's size the former
paper, which fully appears  in the thesis submitted to the Bar-Ilan
University, is omitted here. This paper may be  downloaded at 
http://www.arxive.org/abs/quant-ph/0209012}
  
  {\noindent \Large The paper \\ ``Lax-Phillips evolution as an evolution of 
Gell-Mann-Hartle-Griffiths 
 histories  and emergence  of the Schr\"oedinger
equation for a stable history'' \\ 
by D. Bar and L. P. Horwitz \\
 Published in \\ Phys. Lett A, {\bf 303}, 135-139, (2002)}

 \hbox{}
 
\bigskip \bigskip \bigskip \bigskip \bigskip \bigskip \bigskip \vspace{6cm}

 \newpage
   
  \section{ \large Appendix $A_3$: \\ Effect of dense measurement in classical systems}
   
\bigskip \bigskip \bigskip \bigskip \bigskip \bigskip \bigskip \bigskip

\footnote{Due to limitations imposed upon  submissions's size the former
paper, which fully appears  in the thesis submitted to the Bar-Ilan
University, is omitted here. }

{\noindent \Large  The paper \\``Effect of dense measurement in classical systems'' \\ 
by D. Bar  \\ 
Published in \\  Physica A,   {\bf 292}(10), 494-508, (2001).}

 \newpage

  \hbox{} 
  
   \section{ \large   Appendix $A_4$: \\ The space Zeno effect}

 \bigskip \bigskip \bigskip \bigskip \bigskip \bigskip \bigskip \vspace{2cm} 
 
 \footnote{Due to limitations imposed upon  submissions's size the mentioned 
 three sections of the former paper, which  appear  
  in the thesis submitted to the Bar-Ilan
University, are omitted here.}

{\noindent \Large Sections 1-3 of the paper \\``Space Zeno effect'' \\ 
by D. Bar and L. P. Horwitz \\ 
Published in \\ Int. J. Theor. Phys  {\bf 40}(10), 1897-1713, (2001).}

 \newpage

 \section{ \large  Appendix $A_5$: \\ The one-dimensional multi-barrier potential of
   finite range and finite number of barriers}

  \bigskip \bigskip \bigskip \bigskip \bigskip \bigskip \bigskip \bigskip

\footnote{Due to limitations imposed upon  submissions's size 
the mentioned two sections of the former  paper, which appear  
in the thesis submitted to the Bar-Ilan
University, are omitted here.    The whole paper may be  downloaded at 
http://www.arxive.org/abs/quant-ph/0112027} 
   
 {\noindent \Large Sections 1-2 of the paper \\ ``Dynamical effects of a one-dimensional multibarrier potential 
         of finite range'' \\ 
by D. Bar and L. P. Horwitz \\ 
Published in \\ Eur. Phys. J.  B, {\bf 25},  505-518, (2002) }

  \hbox{}  
   
 \bigskip \bigskip \bigskip \bigskip \bigskip \bigskip \bigskip \vspace{6cm} 
 
 \newpage

   \section{ \large   Appendix $A_6$: \\ The one-dimensional multi-barrier potential of
   finite range and infinite number of barriers} 
      
  \bigskip \bigskip \bigskip \bigskip \bigskip \bigskip \bigskip \bigskip

  \footnote{Due to limitations imposed upon  submissions's size the former 
  paper, which fully appears  in the thesis submitted to 
  the Bar-Ilan
University, is omitted here.} 
     
 {\noindent \Large   The paper \\ ``Manifestation of the Zeno effect and chaotic-like effects on a 
         one-dimensional multibarrier potential of finite range'' \\ 
	 by D. Bar and L. P. Horwitz \\ 
	Published in \\  Phys. Lett A, {\bf 296}(6), 265-271, (2002)}

      \hbox{} 
   
  \bigskip \bigskip \bigskip \bigskip \bigskip \bigskip \bigskip \vspace{6cm}
  
  \newpage

\newpage
   
  \section{ \large  Appendix $A_7$: \\  The one-dinensional multitrap system of finite
   range and infinite number of  traps} \bigskip \bigskip

   \bigskip \bigskip \bigskip \bigskip \bigskip \bigskip \bigskip \bigskip  
   
   \footnote{Due to limitations imposed upon  submissions's size 
the former paper, which fully appears  in the thesis submitted 
   to the Bar-Ilan
University, is omitted here.   }
      
    { \noindent \Large  The paper \\ ``Diffusion-limited reaction in the presence of $n$ traps'' \\ 
	 by D. Bar  \\ 
	Published  in \\ Phys. Rev E,  {\bf 64}, 026108/1-10  (2001)}

        \hbox{} 
   
  \bigskip \bigskip \bigskip \bigskip \bigskip \bigskip \bigskip \vspace{6cm}

 \newpage

\section{ \large   Appendix $A_8$: \\  The one-dinensional multitrap system of finite
   range and finite number of  traps} 
   
    \footnote{Due to limitations imposed upon  submissions's size  the former
paper, which which fully appears  in the thesis submitted to the Bar-Ilan
University, is omitted here.  This paper may be  downloaded at 
http://www.arxive.org/abs/physics/0212094. }

   \bigskip \bigskip \bigskip \bigskip \bigskip \bigskip \bigskip \bigskip  
      
    {\noindent  \Large  The paper \\ ``Diffusion-limited reaction for the one-dimensional 
      trap system'' \\ 
	 by D. Bar  \\ 
	 Published  in \\ Phys. Rev E, {\bf 67}, 056123/1-8   (2003) }

        \hbox{} 
   
  \bigskip \bigskip \bigskip \bigskip \bigskip \bigskip \bigskip \vspace{6cm}

   \newpage

 \section{ \large   Appendix $A_9$: \\    List of publications}  
   
    \bigskip  \bigskip

	\bigskip 
	
	\footnote{Some papers, which were  under refereeing process at the time
	of submitting this thesis, were by now accepted for publication (see footnotes
	at pages 28 and 60)}

 \noindent (1)   \hspace{0.5 cm} ``Manifestation of the Zeno effect and chaotic-like effects on a 
         one-dimensional 
	 
	  \hspace{0.73 cm} multibarrier potential of finite
	 range'',

	 \hspace{0.73 cm}  D. Bar and L. P. Horwitz,  
	  
	  \hspace{0.73 cm} Phys. Lett A, 296(6), 265-271, 2002.

 \noindent (2)   \hspace{0.5 cm}  ``Dynamical effects of a one-dimensional multibarrier potential 
         of finite range'',  
	 
	  \hspace{0.73 cm} D. Bar and L. P. Horwitz,

\hspace{0.73 cm}	 Eur. Phys. J.  B, 25, 505-518, 2002.

 \noindent (3)   \hspace{0.5 cm}  ``Space Zeno effect'', 
 
 \hspace{0.73 cm}  D. Bar and L. P. Horwitz, 
 
\hspace{0.73 cm} Int.  J.  
         Theor.  Phys,       40(10), 1897-1713, 2001.

 \noindent (4)    \hspace{0.5 cm}   ``Diffusion-limited reaction in the presence
 of $n$ traps'', 
 
  \hspace{0.73 cm}        D. Bar, 
  
 \hspace{0.73 cm} Phys. Rev. E, 64(2), 026108/1-10, 2001.	

 \noindent 5)    \hspace{0.5 cm}   ``Diffusion-limited reaction for the
 one-dimensional trap system'',  
  
 \hspace{0.73 cm}        D. Bar, 
  
 \hspace{0.73 cm}  Phys. Rev. E	{\bf 67}, 056123/1-8   (2003)

 \noindent (6)   \hspace{0.5 cm}  ``Effect of dense measurement in classical
 systems'', 
 
  \hspace{0.73 cm} D. Bar,

  \hspace{0.73 cm}      Physica A, 292(1-4), 494-508, (2001)

 \noindent (7)   \hspace{0.5 cm}  ``The Zeno effect in the EPR paradox,  in the teleportation process,

   \hspace{0.73 cm}     and in Wheeler's delayed-choice experiment'', 
   
 \hspace{0.73 cm}    D. Bar, 
 
 \hspace{0.73 cm} Found. Phys, 
       Vol: 30, 813-38, (2000)

 \noindent (8)  \hspace{0.5 cm}   ``The Zeno effect for coherent states'',  
 
\hspace{0.73 cm} D. Bar, 

\hspace{0.73 cm} Physica A, 280, 
        374-81, (2000)

 \noindent (9)   \hspace{0.5 cm}  ``The Zeno effect for spins'', 
 
\hspace{0.73 cm}  D. Bar, 

\hspace{0.73 cm} Physica A, 267(3-4), 
        434-442, 1999.

 \noindent (10)    \hspace{0.5 cm}  ``The Feynman path integrals and Everett's
 universal wave function'', 
  
  \hspace{0.73 cm}      D. Bar, 
  
 \hspace{0.73 cm}  Found. Phys, 28(8), 1383-1391, 1998.

 \noindent (11)   \hspace{0.5 cm}   ``Identification of hidden variables through
 Everett`s formalism'', 
 
      \hspace{0.73 cm}      D. Bar, 
      
  \hspace{0.73 cm}    Found. Phys. Lett, 10(1), 99-103, 1997.

 \noindent (12)   \hspace{0.5 cm}  ``Lax-Phillips evolution as an evolution of 
 Gell-Mann-Hartle-Griffiths 
          histories  
	  
	\hspace{0.73 cm}  and emergence  of the Schr\"oedinger equation for a stable 
	  history'',   
	  
\hspace{0.73 cm}	  D. Bar and L. P. Horwitz,

\hspace{0.73 cm}  Phys. Lett A, {\bf 303}, 135-139, (2002). 
	  
 \noindent (13)	\hspace{0.5 cm}``Phase transitions in a one-dimensional multibarrier potential 
         of finite range'', 
	 
\hspace{0.73 cm}	   D. Bar and L. P. 
	 Horwitz, 
	 
\hspace{0.73 cm}	  J. Phys B, {\bf 35}, 4915, (2002) 
         
 \noindent  (14) \hspace{0.5 cm}      ``Quantum field theory and dense measurement'', 
   
 \hspace{0.73 cm}   D. Bar,

  \hspace{0.73 cm}   
  Int. Jour. Theor. Phys {\bf 42}, 443-463  (2003)
	         
\noindent 
         
	  \noindent  (15)   \hspace{0.5 cm}     `` Effect of increasing the 
	  rate of repetitions of 
   classical reactions''

  \hspace{0.73 cm}       by D. Bar,  
  
 \hspace{0.73 cm}  Int. J. Theor. Phys, {\bf 43}, 1169-1190 (2004)

\noindent  (16) \hspace{0.5 cm}	 ``The effect of related experiments'',

	 \hspace{0.73 cm}   D. Bar, 
	 
	\hspace{0.73 cm}   Int. J. Theor. Phys,  {\bf 44}, 1095-1116 (2005)

\noindent  (17) \hspace{0.5 cm}	 ``Internet websites statistics expressed in the
framework of the Ursell-Mayer 

\hspace{0.73 cm} cluster formalism'',

	 \hspace{0.73 cm}   D. Bar, 
	 
	\hspace{0.73 cm}  Found. Phys, {\bf 34}, 1203-1223  (2004) 
	
	 \noindent  (18)  \hspace{0.5 cm}      ``Computer simulations discussed in
  physical terms and terminology''
	 
 \hspace{0.73 cm}	   by D. Bar,  
	   
	 \hspace{0.73 cm}  Archive:   physics/0205010

\newpage
   
 \pagestyle{myheadings}
\markright{ \ \ \ REFERENCES}

\bigskip \bibliographystyle{plain}   
		   				              		       	         							          \bigskip \bigskip \bigskip \bigskip \bigskip \bigskip \bigskip \vspace{100cm}  \bigskip   \bibliographystyle{plain}

\end{document}

%% file: listoffigures.tex
 \pagestyle{empty}

\section{\large List of figures}

\noindent Figure 3.1: \hspace{0.75 cm} Schematic representation of the concentric circular billiard that

	  \hspace{2.2 cm} simulates   the reversible reaction $A+B \leftrightarrow C+D$.
	  .................................................29	   
	   
\noindent Figure 3.2: \hspace{0.75 cm} The three curves represent the activities obtained when all the 
             particles 
	     
	   \hspace{2.2 cm}   are: (1) only in  ``state'' 1.  (2)  only  in ``state'' 2. 
             (3) allowed to pass between 
	     
	   \hspace{2.2 cm}        the  two ``states'' after every 1100 
	     reflections.
	     .........................................................29 

\noindent Figure 3.3: \hspace{0.75 cm} The two curves represent the activities obtained when all the 
             particles are: 
	     
	 \hspace{2.2 cm}     (1) only in  ``state'' 2. (2) allowed to 
	      pass 
	     from ``state'' 1 to 2 after each
	     
	  \hspace{2.2 cm}     reflection and from 2 to 1 after 
	     every 1100 reflections.
	     ...........................................32
 
 \noindent Figure 3.4: \hspace{0.75 cm}The two curves represent the activities obtained when all the 
            particles are: 
	    
	  \hspace{2.2 cm}   (1) only in  ``state'' 1. (2) allowed to pass 
	    from ``state'' 2 to 1 after each 
	    
	\hspace{2.2 cm}     reflection and from 1 to 2 after 
	   every 1100 reflections. ...........................................32  
 
 \noindent Figure 3.5: \hspace{0.75 cm}This Figure   show the activities
 obtained  when the degree of "densing"  is 
 
          \hspace{2.2 cm}   changed slightly from reactioning 
 after each single reflection 
 
 \hspace{2.2 cm}  to doing that after every two reflections (compare
 with Figure 5.3)  ........................33  
 
\noindent  Figure 5.1:\hspace{0.75 cm} Seven representative Feynman paths of states.  The collective 
dense

      \hspace{2.2 cm}        measurement  is done along the middle one.
      ..........................................................44

 \noindent  Figure 5.2:\hspace{0.75 cm} A schematic representation of the physical situation after
              performing  
	      
	\hspace{2.2 cm}      the collective dense measurement shown in Figure
	5.1.  
              The emphasized  
	      
	\hspace{2.2 cm}  path   is  common to all the ensemble.
	.....................................................................46

 \noindent Figure 5.3: \hspace{0.75 cm} A three-dimensional surface of the relative rate of the number 
              of 
	      
	  \hspace{2.2 cm}    observers as function of the number of possible results $K$ 
	      for each  
	      
	   \hspace{2.2 cm}  experiment   and the number $r$ of places occupied by 
              preassigned 
	      
	    \hspace{2.2 cm}  
	   eigenvalues............................................................................................................50

 \noindent Figure 5.4:\hspace{0.75 cm}  The cylinder with the four pistons.
 ........................................................................54

 \noindent  Figure 5.5:\hspace{0.75 cm} A three-dimensional surface  of the entropy per molecule $s$ 
              as a function   
	      
	   \hspace{2.2 cm}  of  the fractions $\frac{n_o}{n}$ and
	   $\frac{n_i}{n}$  of molecules that step out and into 
	      the  
	      
	  \hspace{2.2 cm} interval   $(-x,x)$......................................................................................................57

 \noindent Figure 5.6: \hspace{0.75 cm} The result of perfoming 1000  different 
 experiments of lifting up the 
              pistons 
	      
	   \hspace{2.2 cm}     as a function of the number of observers $N$. 
	  ...........................................................59
 
 \noindent Figure 5.7:\hspace{0.75 cm} The same as Figure 15 except that the values of $(-x_j,x_j)$, 
             $\frac{n_{o_j}}{n}$ and $\frac{n_{i_j}}{n}$ 
	     
	  \hspace{2.2 cm}    are chosen
randomely.............................................................................................59
